\documentclass[fleqn,usenatbib]{mnras}

\usepackage{newtxtext,newtxmath}

\usepackage[T1]{fontenc}
\usepackage{ae,aecompl}

\usepackage{color}

\usepackage{graphicx}	
\usepackage{amsmath}	
\usepackage{amssymb}	

\newcommand{\Ha}{\mbox{H$\alpha$}}
\newcommand{\Hb}{\mbox{H$\beta$}}

\newcommand\hii{H{\sc ii}}

\title[Anti-correlation between SFR and metallicity]{Local anti-correlation between star-formation rate and gas-phase metallicity in disk galaxies}

\author[S\'anchez Almeida et al.]{
J.~S{\'a}nchez Almeida$^{1,2}$\thanks{E-mail: jos@iac.es},
N.~Caon$^{1,2}$, 
C.~Mu{\~n}oz-Tu{\~n}{\'o}n$^{1,2}$, 
M. Filho$^{3,4}$, and
\newauthor 
M.~Cervi\~no$^{1,5}$
\\
$^{1}$Instituto de Astrof{\'\i}sica de Canarias, E-38200 La Laguna,
           Tenerife, Spain\\
$^{2}$Departamento de Astrof{\'\i}sica, Universidad de la Laguna, 
           E-38205 La Laguna, Tenerife, Spain\\
$^3$Center for Astrophysics and Gravitation - CENTRA/SIM, Departamento de F\'\i sica, Instituto Superior T\'ecnico, Universidade de Lisboa,\\
~~~~Av. Rovisco Pais 1, P-1049-001 Lisbon, Portugal\\
$^4$Departamento de Engenharia F\'\i sica, Faculdade de Engenharia, Universidade do Porto, Rua Dr. Roberto Frias, s/n, P-4200-465,\\ 
~~~~Oporto, Portugal\\
$^{5}$Instituto de Astrof\'\i sica de Andaluc\'\i a (IAA-CSIC ), Apdo. 3004, 18008  Granada, Spain
}

\date{Accepted XXX. Received YYY; in original form ZZZ}

\pubyear{2018}

\begin{document}
\label{firstpage}
\pagerange{\pageref{firstpage}--\pageref{lastpage}}
\maketitle

\begin{abstract}
Using a representative sample of 14 star-forming dwarf galaxies in the local Universe, we show the existence of a spaxel-to-spaxel anti-correlation between the index N2\,$\equiv\,\log({\rm[NII]}\lambda 6583/{\rm H}\alpha)$ and the H$\alpha$ flux. These two quantities are commonly employed as proxies for  gas-phase metallicity and star formation rate (SFR), respectively. Thus, the observed N2 to H$\alpha$ relation may reflect the existence of an anti-correlation between the metallicity of the gas forming stars and the SFR it induces. Such an anti-correlation is to be expected if variable external metal-poor gas fuels the star-formation process. Alternatively, it can result from the contamination of the star-forming gas by stellar winds and SNe, provided that intense outflows drive most of the metals out of the star-forming regions. 
We also explore the possibility that the observed anti-correlation is due to variations in the physical conditions of the emitting gas, other than metallicity. Using alternative methods to compute metallicity, as well as  previous observations of \hii\ regions and photoionization models, we conclude that this possibility is unlikely. The radial gradient of metallicity characterizing disk galaxies does not produce the correlation either. 
\end{abstract}

\begin{keywords}
galaxies: star formation -- galaxies: abundances -- galaxies: evolution -- galaxies: dwarf -- galaxies: irregular
\end{keywords}



\section{Introduction}

\citet{2010MNRAS.408.2115M} and \citet{2010A&A...521L..53L} discovered that the  scatter in the well-known mass-metallicity relation (MZR) correlates with the star-formation rate (SFR). Galaxies with a higher SFR  show lower metallicity at a given  stellar mass ($M_\star$). This correlation between $M_\star$, SFR and gas-phase metallicity is called the fundamental metallicity relation (FMR), and its existence had been  hinted at previously \citep{2008ApJ...672L.107E, 2009ApJ...695..259P, 2010A&A...521A..63L}.
The FMR has gone through a close scrutiny during the last years, with some early works questioning its existence 
\citep[e.g.,][]{2013A&A...554A..58S,2014A&A...561A..33I, 2015AJ....149...79D, 2015ApJ...799..138S, 2016ApJ...823L..24K}, 
but with a significantly larger number of confirmations 
\citep[e.g.,][]{2014ApJ...797..126S, 2013ApJ...765..140A, 2012PASJ...64...60Y,
2016MNRAS.457.2929W, 2014ApJ...780..122L, 2015ApJ...805...45L,
2014A&A...568L...8A, 2014MNRAS.440.2300C, 2014ApJ...792....3M,
2016MNRAS.458.1529B}, and with the correlation upholding to redshift 3 \hfill\break
\citep[][]{2014A&A...563A..58T,2017A&A...606A.115H}. 
The FMR is found to evolve with redshift so
that metallicities at higher redshift are systematically lower given $M_\star$ and
SFR \citep[e.g.,][]{2014ApJ...780..122L, 2015ApJ...805...45L}. 
There are also proposals to replace the SFR with the gas mass, which provides a deeper physical insight   
\citep{2013MNRAS.433.1425B, 2016arXiv160604102B, 2016MNRAS.455.1156B},  
and reduces the  dispersion of the relation \citep{2015ApJ...812...98J}. 
\citet{2015MNRAS.446.1449L} claim that stellar age, rather than SFR or gas mass,  is the third parameter in the FMR
\citep[see also][]{2012ApJ...756..163S}.

The importance of the FMR is related to its physical cause. Current interpretations are always made in terms of the star formation feeding from metal-poor gas recently accreted by the galaxy disk. The advent of external gas does not change $M_\star$, but it decreases the mean metallicity of the gas,  while simultaneously triggering star formation. Subsequently, the star formation consumes the gas and produces metals that increase the gas metallicity through stellar winds and SNe, until new metal-poor gas arrives and the cycle repeats. This scenario was already suggested by \citet{2010MNRAS.408.2115M}, and it  has been refined using simple analytical models  
\citep[e.g.,][]{2012ApJ...750..142B, 2013ApJ...772..119L, 2013MNRAS.430.2891D, 2014MNRAS.443..168F, 2014MNRAS.441.1444P}, as well as cosmological numerical
simulations of galaxies with various degrees of sophistication  \citep{2012MNRAS.422..215Y, 2013ApJ...770..155R, 2015MNRAS.452..486D,2016ApJ...826L..11K}.

There are hints in literature that the anti-correlation between SFR and metallicity also occurs locally in galaxy disks, so that regions of high surface SFR are associated with drops in metallicity. The ten extremely metal-poor galaxies of the local Universe studied by \citet{2015ApJ...810L..15S} have a large starburst with a metallicity between 5 and 10 times lower than the underlying host galaxy \citep[see also][]{2013ApJ...767...74S, 2014ApJ...783...45S}.
The same anti-correlation between SFR and metallicity was observed in three redshift 3 galaxies by \citet{2010Natur.467..811C}. 
A large \hii\ region at the edge of the dwarf galaxy GAMA J1411-00 provides most of the SFR of the galaxy, and has an oxygen abundance that is lower than the rest of the galaxy by $\sim$0.2 dex \citep{2014MNRAS.445.1104R}.
The optical emission line analysis of the galaxy HCG~91c reveals that at least three \hii\ regions harbour an oxygen abundance $\sim$0.15~dex lower than expected from their immediate surroundings and from the abundance gradient \citep{2015MNRAS.450.2593V}. 
%
%
\citet{2016MNRAS.456.1549L} find a slight  anti-correlation between gas
metallicity and  SFR at spaxel scales, in the sense that higher SFRs are found in
regions of lower metallicity (see their Fig.~11).
Although the metallicity of IZw~18 is rather homogeneous, there is hint of scatter of approximately 0.2~dex, which anti-correlates with the SFR, as inferred from integrated H$\alpha$ flux \citep[][Figs.~8 and 2]{2016MNRAS.459.2992K}.

This work presents evidence for the existence of such
an anti-correlation between the local  variations of SFR and gas-phase
metallicity in more massive, more metallic star-forming galaxies.  
We use published 2D spectra (IFU\footnote{Integral Field Unit} data)  of blue compact galaxies, and
the  anti-correlation is present in 85\,\% of the cases
(Sect.~\ref{Sect:Data}). 
The observed anti-correlation is analyzed in terms of variable mixtures of 
metal-poor and metal-rich gas, and the closed-box evolution of the original
metal-poor gas. Both possibilities account for the observed relation 
(Sect.~\ref{Section:Analysis}).
The metallicities are inferred using a strong-line ratio (explicitly, ${\rm N2} \equiv \log({\rm[NII]}\lambda 6583/{\rm H}\alpha)$, whereas the SFR is derived from the H$\alpha$ flux. Variations in the properties of the local ionizing flux may induce spurious correlations between H$\alpha$ and N2, which could be incorrectly interpreted as variations in metallicity. 
Section~\ref{sect:hiic} shows how the anti-correlation remains even when these variations are taken into account in the metallicity determination.  Sections~\ref{Sect:sebastian}  and \ref{Sect:photoio} are 
devoted to comparing the observed variations of line ratios with model predictions, to conclude that the observations seem to require variations in metallicity.

%
In Sect.~\ref{Sect:Discussion} we refer to  cosmological numerical simulations showing how this type of local anti-correlation can also result from the accretion of metal-poor gas that triggers the observed star formation bursts \citep{2016MNRAS.457.2605C, 2014MNRAS.442.1830V}. This interpretation is not devoid of uncertainty, since it involves, in those galaxies showing the relation, either inefficient gas mixing in the galaxy disk, or the very recent accretion of metal-poor gas. The gas in the disk is expected to mix in a timescale of the order of the rotational period or smaller \citep[e.g.,][]{2002ApJ...581.1047D, 2012ApJ...758...48Y,2015MNRAS.449.2588P}. Gas accretion, on the other hand, is theoretically predicted to fuel star formation in disk galaxies \citep[e.g.,][]{2009Natur.457..451D,2012RAA....12..917S}, but the observational evidence of this process is still rather indirect \citep{2014A&ARv..22...71S,2017ASSL..430...67S}.

\section{The observational results}\label{Sect:Data}

\subsection{The data}
To carry out this study, we assembled a group of 18 blue star-forming galaxies, for which we have already produced maps of emission-line fluxes, line ratios, extinction, and gas kinematics. Among the group, 10 blue compact dwarf (BCD) galaxies were observed with PMAS-PPAK \citep{2005PASP..117..620R}, with the data published in \cite{Cairos2009-mrk1418}, \cite{Cairos2009-mrk409} and \cite{Cairos2010}, while the other 8 blue compact galaxies were observed with VIMOS \citep{2003SPIE.4841.1670L}, with the data published in \cite{Cairosetal2015}.

The PMAS data cover an area of approximately $16\,\arcsec \times 16\,\arcsec$,  with a spatial sampling of $1\arcsec \times 1\arcsec$, a spectral range from 3590 to 7000~\AA{}, and a spectral resolution of approximately 7\,\AA{}. The VIMOS data cover an area of approximately $27\arcsec \times 27\arcsec$, with a spatial sampling of $0.67\,\arcsec \times 0.67\,\arcsec$. After combining blue grism and orange grism spectra, the spectral range goes from 4150 to 7400\,\AA{}, with a spectral resolution of approximately 2.0\,\AA{}. (More details about the PMAS and the VIMOS data can be found in \citeauthor{Cairos2010}~\citeyear{Cairos2010} and \citeauthor{Cairosetal2015}~\citeyear{Cairosetal2015}, respectively.)

In this paper, we excluded two VIMOS galaxies from the analysis: Haro~15, which was only observed with the blue grism, and has neither \Ha{} nor 
[NII]$\lambda$6583 data, and Mrk~1131, which was only observed with the orange grism, and lacks the \Hb{} data required to correct the \Ha{} fluxes for interstellar extinction. Thus, the final sample is made up of 16 galaxies, two of which will be discarded because they possess an AGN (Active Galactic Nucleus).
Table~\ref{Table:Galaxies} lists the main data characterizing the present sample of galaxies. Distances are employed in the paper to obtain absolute magnitudes and fluxes from the apparent magnitudes and fluxes, as well as to compute the scale in physical units (arcsec kpc$^{-1}$). We use NED (NASA Extragalactic Database) values, which are corrected for local proper motions (more details in the caption of Table~\ref{Table:Galaxies}).  Maps with their H$\alpha$ emission are given in Figs.~\ref{Fig:lgoh12_lgsfr} and \ref{Fig:lgoh12_lgsfr_agn}, left column.

\begin{table*}
\caption{Galaxy sample}
\label{Table:Galaxies}
\begin{center}
\begin{tabular}{ccccccccccc}
\hline
Galaxy           & RA          & DEC        &  m$_{B}$  & D       & $M_{B}$  & Sampling              & Resolution &  SFR &$\log M_\star$              & Other designation  \\
                 &  (2000)     &   (2000)   &  (mag)    & (Mpc)   & (mag)    & (pc\,\arcsec$^{-1})$  & (pc)   & ($M_\odot\,{\rm yr}^{-1}$) & ($\log M_\odot$)     &                    \\
    (1)          &  (2)        &    (3)     &   (4)     &  (5)    &  (6)     &   (7)                 &    (8)     &    (9)  & (10)           & (11)               \\
\hline\\[-5pt]
\multicolumn{8}{l}{VIMOS ---------------} \\
Haro\,11         & 00 36 52 & $-33$ 33 19   & $14.3^{a}$  &  84.0  &  $-20.3$  & 410      & 550    &  19.9 & 9.5  & ESO\,350-IG 038        \\
Haro\,14         & 00 45 46 & $-15$ 35 49   & $13.7^{b}$  &  13.0  &  $-16.9$  & 63       & 84     &   0.9  & 8.4& NGC\,0244              \\
Tol\,0127-39 & 01 29 15 & $-39$ 30 38   & $16.2^{c}$  &  71.3  &  $-18.1$  & 350      & 460    &   1.2 & 9.3  &                        \\
Tol\,1924-41 & 19 27 58 & $-41$ 34 32   & $14.0^{b}$  &  42.4  &  $-19.2$  & 210      & 280    &   6.5 & 9.2  & ESO\,338-IG 004        \\
Tol\,1937-42 & 19 40 58 & $-42$ 15 45   & $15.1^{b}$  &  41.1  &  $-17.9$  & 200      & 270    &   0.39  & 9.2&                        \\
Mrk\,900         & 21 29 59 & $+02$ 24 51   & $14.2^{b}$  &  18.9  &  $-17.2$  & 92       & 120    &   0.17  & 8.6 &NGC\,7077, UGC\,11755  \\
\multicolumn{8}{l}{PMAS ---------------}                                                                                                            \\
Mrk\,1418        & 09 40 27 & $+48$ 20 16   & $13.9^{b}$  &  14.6  &  $-17.0$  & 70       & 140    &   0.11 & 9.3 & UGC\,05151             \\
Mrk\,407         & 09 47 47 & $+39$ 05 04   & $15.4^{a}$  &  27.2  &  $-16.8$  & 130      & 260    &   0.08  & 8.7 &                       \\
Mrk\,409         & 09 49 41 & $+32$ 13 16   & $14.4^{b}$  &  26.3  &  $-17.7$  & 130      & 260    &   0.09  & 8.9 &NGC\,3011, UGC\,5259   \\
Mrk\,32          & 10 27 02 & $+56$ 16 14   & $16.1^{a}$  &  16.4  &  $-14.9$  & 80       & 160    &   0.02  &  7.8&                      \\
Mrk\,750         & 11 50 02 & $+15$ 01 24   & $15.8^{a}$  &  5.2  &  $-13.3$  & 25       & 50     &   0.02  &      7.0  &                \\
Mrk\,206         & 12 24 17 & $+67$ 26 24   & $15.4^{a}$  &  24.3  &  $-16.9$  & 120      & 240    &   0.32  & 9.2&                        \\
Tol\,1434+03 & 14 37 08 & $+03$ 02 50   & $16.9^{a}$  &  29.2  &  $-15.5$  & 140      & 280    &   0.06  & 7.7& SHOC 474                       \\
Mrk\,475         & 14 39 05 & $+36$ 48 22   & $16.4^{a}$  &  11.9  &  $-14.2$  & 58       & 120    &   0.05  & 8.6 &                       \\
I\,Zw\,123       & 15 37 04 & $+55$ 15 48   & $15.4^{a}$  &  15.4  &  $-15.7$  & 75       & 150    &   0.11  & 8.3 &Mrk\,487               \\
I\,Zw\,159       & 16 35 21 & $+52$ 12 53   & $15.7^{a}$  &  43.8  &  $-17.2$  & 210      & 420    &   0.39  & 9.0 & Mrk\,1499              \\
\hline
\end{tabular}
\end{center}
\begin{flushleft}
Cols. (2) and (3): Units of right ascension are hours, minutes, and seconds, and units of declination are degrees, arcmin, and arcsec.\\
Col. (4): $B$-band apparent magnitudes: (a) magnitudes taken from HyperLeda (\url{http://leda.univ-lyon1.fr/}; \citealp{Paturel2003}); 
(b) integrated magnitudes from \cite{GildePaz2003}, corrected for Galactic extinction; 
(c) asymptotic  magnitudes obtained by extrapolating the growth curves, and corrected for Galactic extinction \citep{Cairos2001-II}; 
the asymptotic magnitudes listed in \cite{Cairos2001-II} were corrected for Galactic extinction following \cite{Burstein1982}. 
Here they have been recomputed using the \cite{Schlegel1998} extinction curve.\\
Col. (5): Distance calculated using a Hubble constant of 73\,km\,s$^{-1}$\,Mpc$^{-1}$, and taking into account 
the effect of the Virgo Cluster, the Great Attractor, and the Shapley supercluster (from NED -- \url{http://nedwww.ipac.caltech.edu/}).\\
Col. (6): Absolute magnitudes in the $B$-band, computed from the tabulated $B$ apparent magnitudes and distances.\\
Col. (7): Plate scale.\\ 
Col. (8): Assuming two spaxels per resolution element.\\
Col. (9): SFR computed from the extinction-corrected \Hb{} fluxes for the integrated spectra published in 
\cite{Cairos2009-mrk1418},  \cite{Cairos2010}, and \cite{Cairosetal2015}, assuming $F(\Ha)/F(\Hb)=2.86$ and using Eq.~(\ref{Eqn:HaSFR}). 
For Mrk\,409  \citep{Cairos2009-mrk409}, we used the sum of the fluxes of the 6 SF knots.
As the \Ha{} emission of the PMAS galaxies usually extends beyond the PMAS FOV, the corresponding SFRs are likely lower limits to the actual values.\\
Col.(10): Stellar mass inferred from the absolute magnitude and the colors using the mass-to-light ratios in \citet{2001ApJ...550..212B}.
\end{flushleft}
\end{table*}


\subsubsection{Comments on three sources}\label{Sect:individual}
Haro\,11 and Mrk\,409 represent two well-documented cases of AGNs among the sources in Table~\ref{Table:Galaxies}. They are discarded from our analysis because their emission lines are not excited by star formation, at least not exclusively \citep[e.g.,][]{2013ApJ...774..100K}. However, they remain in some of the plots to evidence differences with the other galaxies. 

Haro\,11 displays  three distinct  components (Fig. \ref{Fig:lgoh12_lgsfr_agn}, first row). The Eastern knot is associated with a young, high mass, X-ray binary  \citep{2015ApJ...812..166P} with ultraviolet emission \citep{2003ApJ...597..263K}. The southern knot shows strong ultraviolet emission \citep{2003ApJ...597..263K}. The central knot is coincident with HI observed in absorption \citep{2014MNRAS.438L..66M}, and harbors a luminous, compact, hard X-ray source, perhaps an intermediate mass black hole binary. The evidence for an accreting black hole is reinforced by the detection of a radio source \citep[][and references therein]{2006ApJ...643..173S}. There is also strong evidence that Haro\,11 is the result of a merger \citep[][and references therein]{Ostlin2015}. The BPT diagram of this galaxy (i.e., O3$\,=\,\log([{\rm OIII}]\lambda 5007/{\rm H}\beta)$ versus N2$\,=\,\log([{\rm NII}]\lambda 6583/{\rm H}\alpha)$; after Baldwin, Phillips, and Terlevich~\citeyear{1981PASP...93....5B}) clearly shows evidence for an AGN exciting the emission-line spectrum of this source.

Mrk\,409 (Fig. \ref{Fig:lgoh12_lgsfr_agn}, second row) shows a central star-forming component and an outer ring of star formation, likely triggered by an expanding starburst-driven superbubble  \citep[][]{Cairos2009-mrk409}.  Although there is no radio nor X-ray confirmation, the central source is associated with HeII $\lambda$4686 emission, and with a Seyfert 2-type AGN \citep[][]{2014AJ....148..136M}. The BPT diagram of this galaxy provides evidence of an AGN exciting the emission-line spectrum.

Only one galaxy in Table~\ref{Table:Galaxies} presents  signs of current interaction (Tol\,1924-416). It is still included in the following analysis because its companion is quite distant (see below), and its emission lines behave as the other isolated galaxies.  Tol\,1924-416 appears to be composed of at least three star-formation knots (Fig.~1, third row). The source is documented as interacting with ESO\,338-IGO4B, 72 kpc away \citep{2004ApJ...608..768C}. However, the distance is large compared with the physical size of the galaxy ($\sim$ 5\,kpc effective radius), and so the companion is not expected to exert a significant tidal force on the galaxy \citep[e.g.,][Sect.~3.4]{2015ApJ...802...82F}.    
\medskip

\subsection{The anti-correlation N2 versus H$\alpha$}
\label{Sect:anticorr}

Here we analyze the relationship between SFR, as measured from the \Ha{} flux, and metallicity, as parameterized by the N2 line  ratio, 
\begin{equation}
{\rm N2}=\log({\rm[NII]}\lambda6583/{\rm H}\alpha). 
\label{eq:defN2}
\end{equation}
We first converted \Ha{} fluxes to SFRs by means of the formula \citep{Kennicutt1998},
\begin{equation}
\label{Eqn:HaSFR}
{\rm SFR}/(M_\odot\;\mathrm{yr}^{-1}) = 7.9 \times 10^{-42} L(\Ha)/(\,\mathrm{ergs}\; \mathrm{s}^{-1}),
\end{equation}
where the \Ha{} luminosity, $L(\Ha)$, is corrected for extinction. The correction factor
was computed, spaxel-by-spaxel, from the observed \Ha/\Hb{} ratio, adopting the usual Case B low-density limit \citep{Osterbrock2006},
and the \citet{Cardelli1989} extinction law.
As for the relationship between N2 and metallicity, we adopted the calibration,
\begin{equation}
\label{Eqn:HaNIImet}
12+\log{\rm (O/H)} = 9.07 + 0.79 \times {\rm N2},
\end{equation}
proposed by \citet{PerezMonteroContini2009}. The calibration we use for the SFR and the metallicity can be replaced by other similar calibrations in literature  \citep[e.g.,][]{2004MNRAS.348L..59P,2012ARA&A..50..531K,2014ApJ...797...81M} without modifying the main results in the paper. 

%
Figure~\ref{Fig:lgoh12_lgsfr} show maps of \Ha{} flux (left column) and N2 (central column) for all the galaxies in our sample. Each row corresponds to one of the galaxies as labeled.  A simple inspection shows that the peaks in \Ha{} flux tend to coincide with dips in N2, and vice-versa. 
\begin{figure*}
\centerline{
\includegraphics[width=0.62\textwidth]{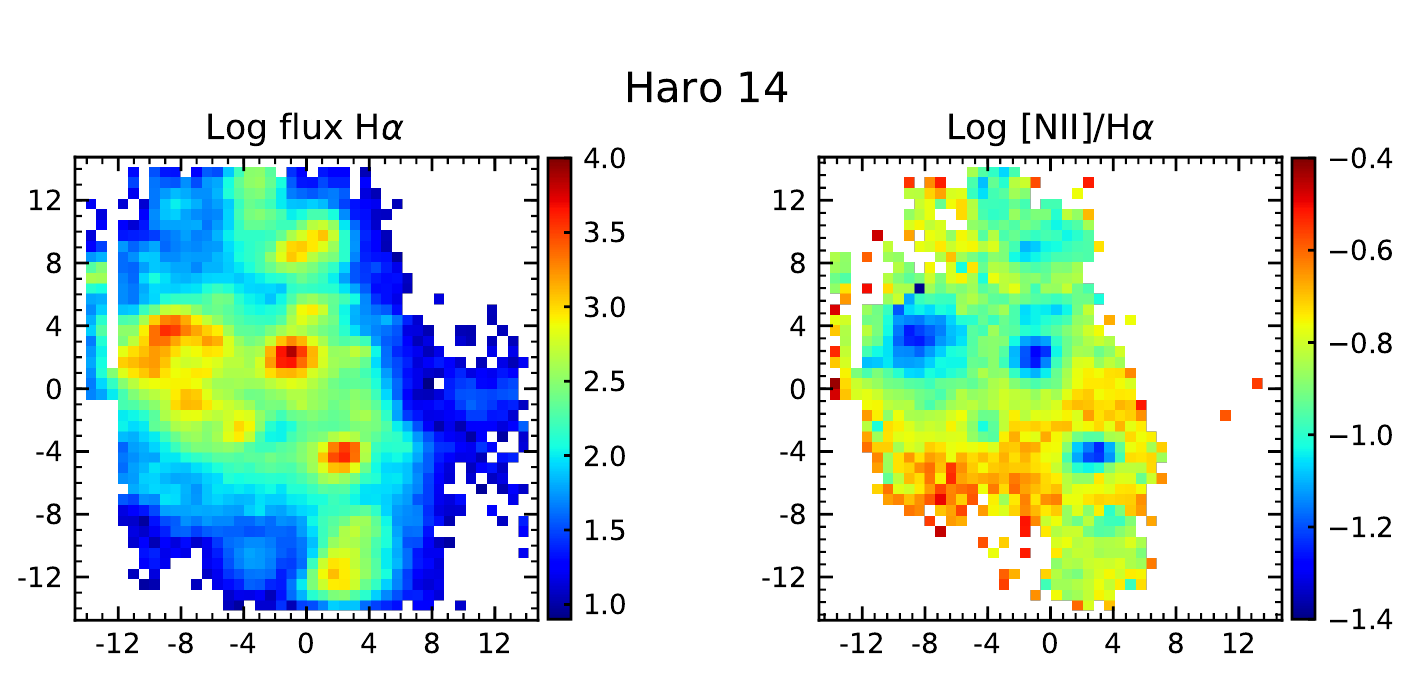} \hspace{5pt}
\includegraphics[width=0.35\textwidth]{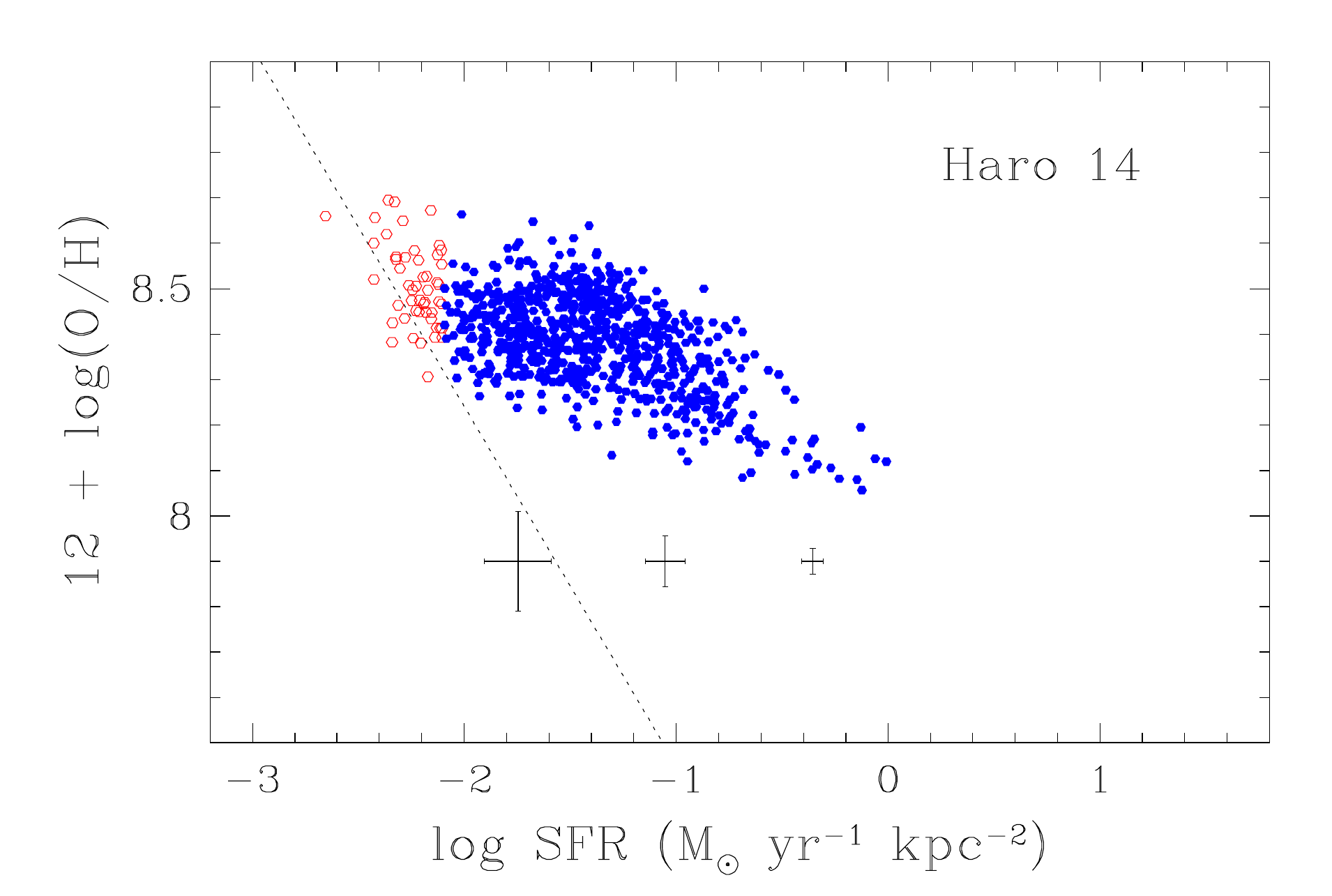}
}
\centerline{
\includegraphics[width=0.62\textwidth]{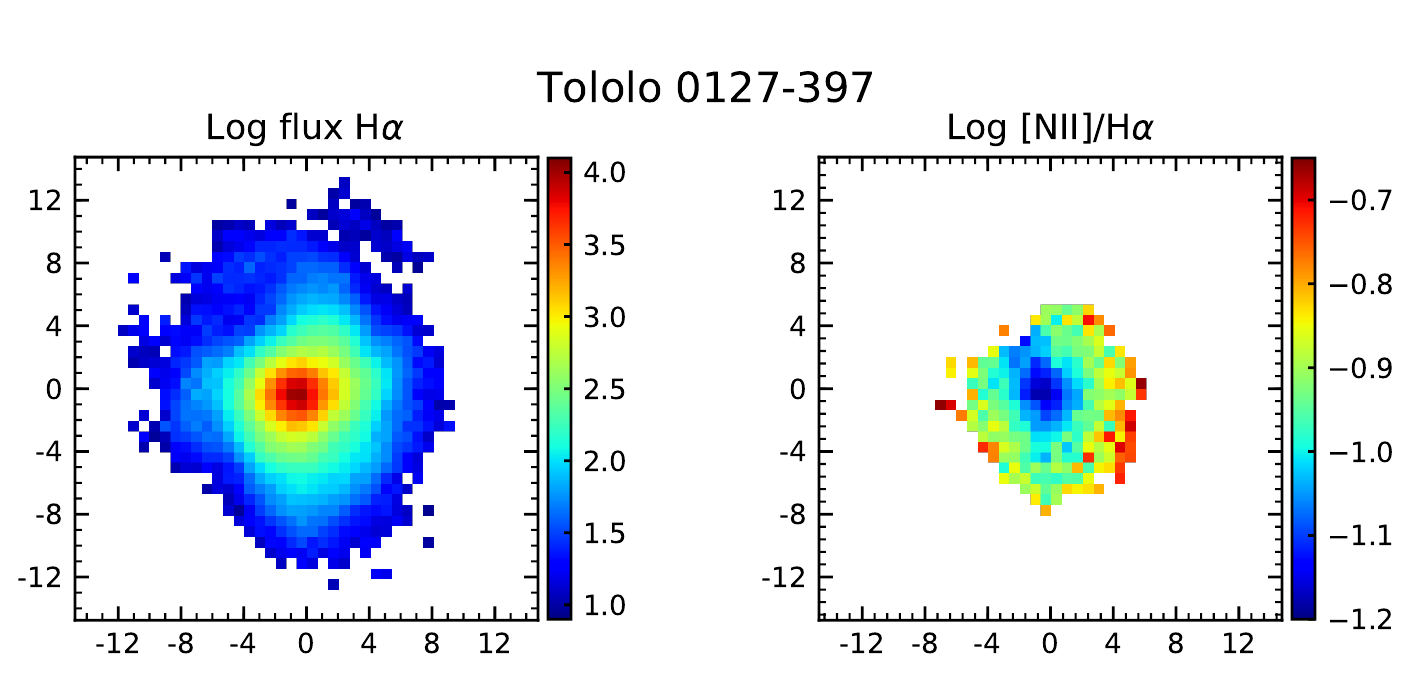} \hspace{5pt}
\includegraphics[width=0.35\textwidth]{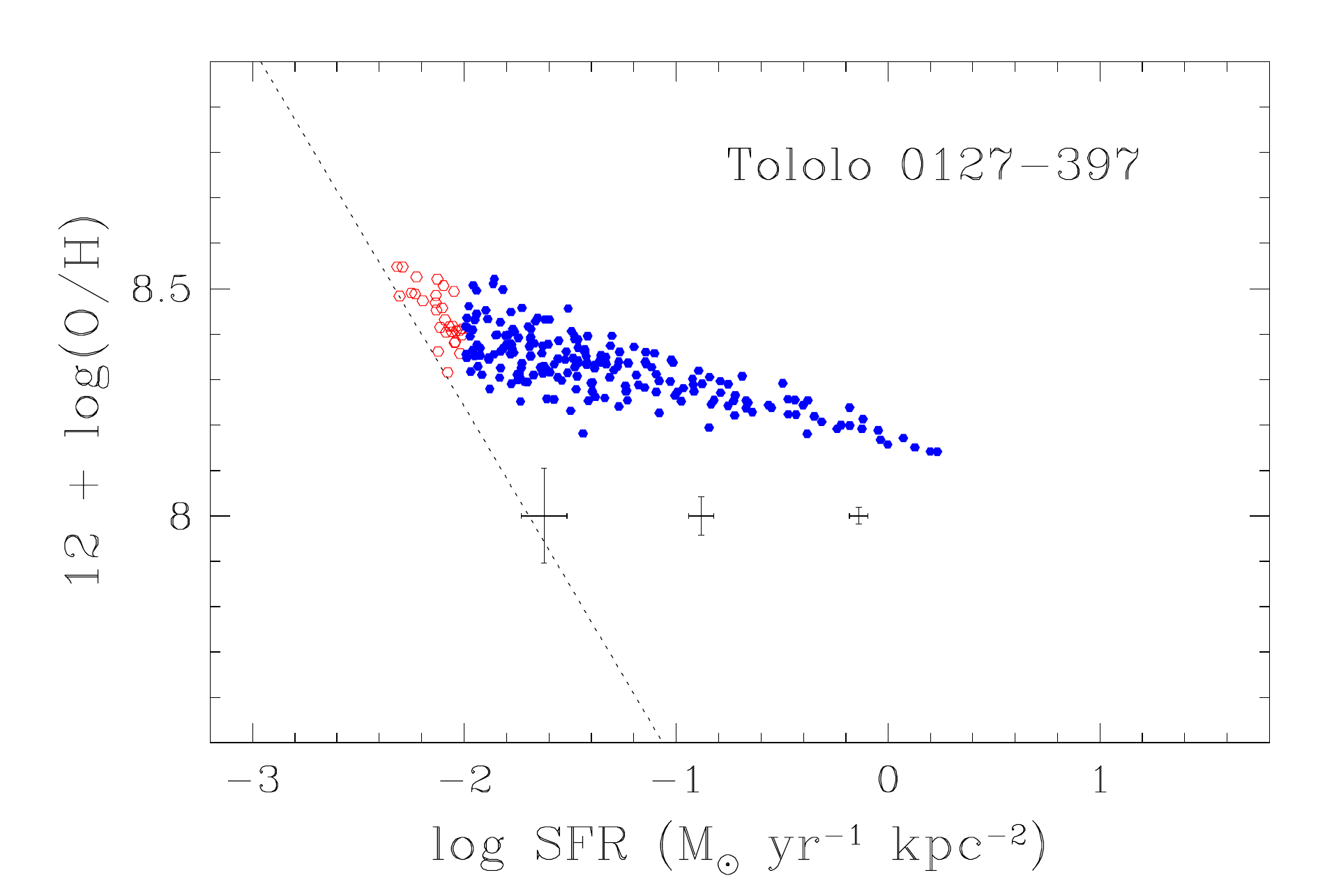}
}
\centerline{
\includegraphics[width=0.62\textwidth]{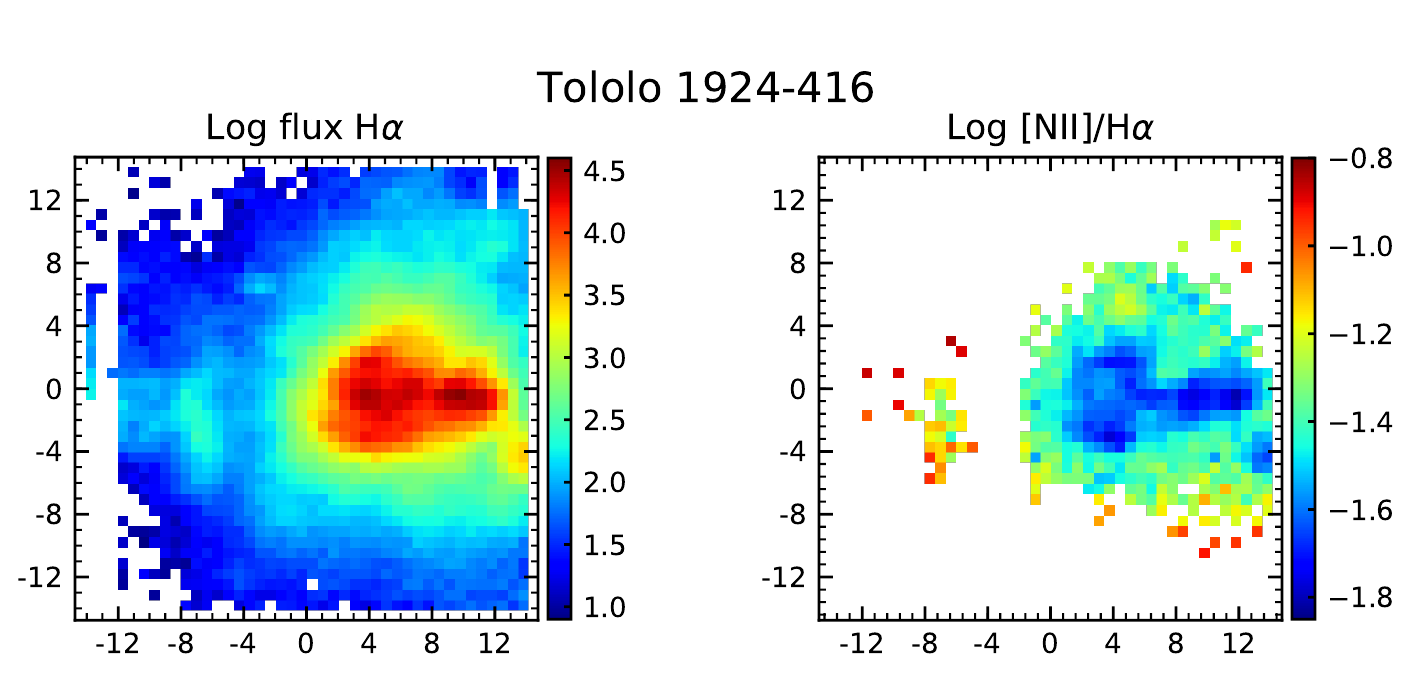} \hspace{5pt}
\includegraphics[width=0.35\textwidth]{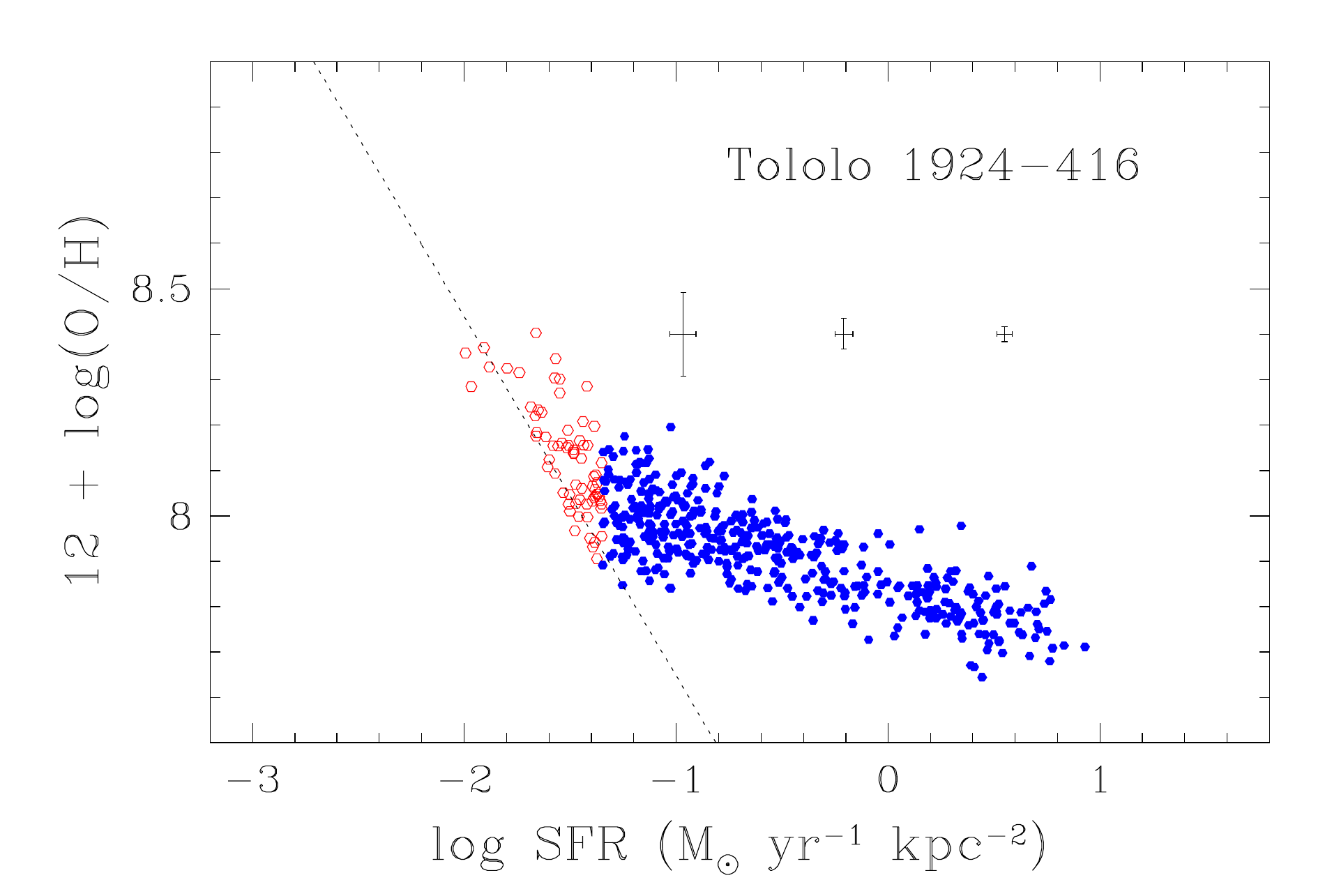}
}
\centerline{
\includegraphics[width=0.62\textwidth]{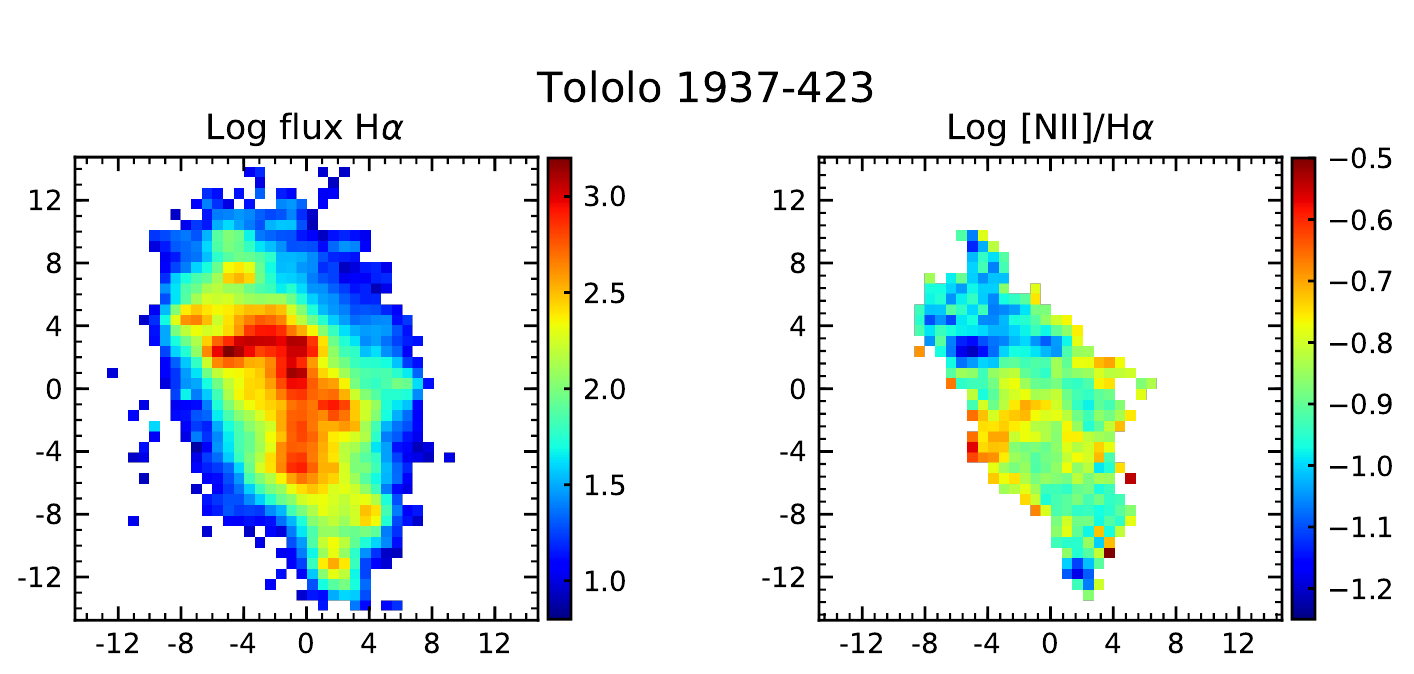} \hspace{5pt}
\includegraphics[width=0.35\textwidth]{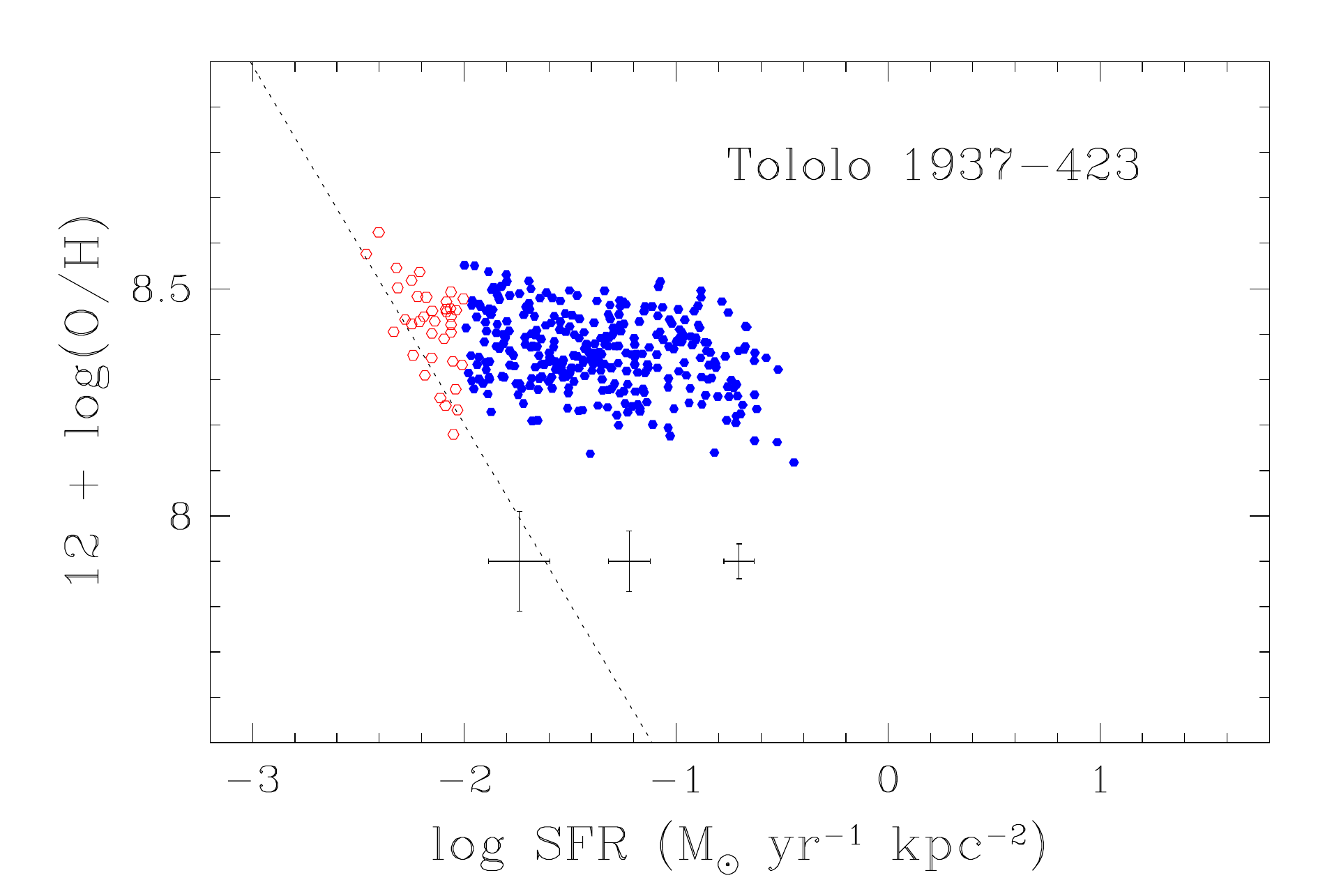}
}
\caption{
Left and central columns: maps of \Ha{} (left) and [NII]$\lambda$6583/\Ha{} (center), both in logarithmic scale, for  the galaxies studied in the paper. Each row corresponds to one galaxy as labelled. The VIMOS data cover an area of approximately $27\arcsec \times 27\arcsec$ with a spatial sampling of $0.67\arcsec \times 0.67\arcsec$. The area covered by PMAS is $16\arcsec \times 16\arcsec$ with a sampling of $1\arcsec \times 1\arcsec$.  Note the anti-correlation between the peaks and dips in the two maps.  The fluxes in H$\alpha$ are given in units of $10^{-16}\,{\rm erg\,s}^{-1}\,{\rm cm}^{-2}$. 
Right column: scatter plots metallicity versus surface SFR.  Noise in the spectra may induce an artificial correlation, represented as dotted lines in the individual panels (see main text). Such artifact does not explain the observed correlation except perhaps for the red points, which have been excluded from the  analysis. The typical error bars for 12+log(O/H) and SFR at three representative SFRs are also shown in each panel.These error bars do not include the error introduced by the calibration to infer 12+log(O/H) from N2, which is probably significantly larger than the bars on display. The consequences of the calibration error on the relation are discussed in the main text.
}
\label{Fig:lgoh12_lgsfr}
\end{figure*}

\addtocounter{figure}{-1}
\begin{figure*}
\centerline{
\includegraphics[width=0.62\textwidth]{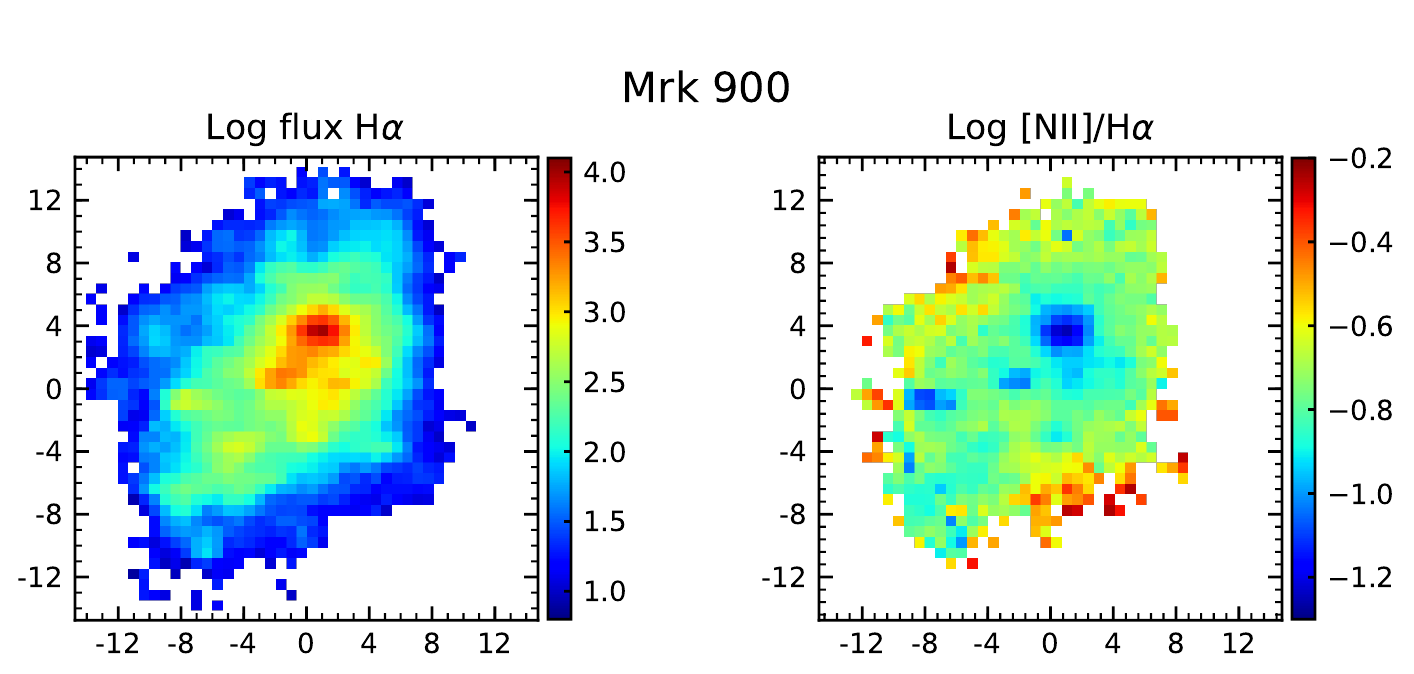} \hspace{5pt}
\includegraphics[width=0.35\textwidth]{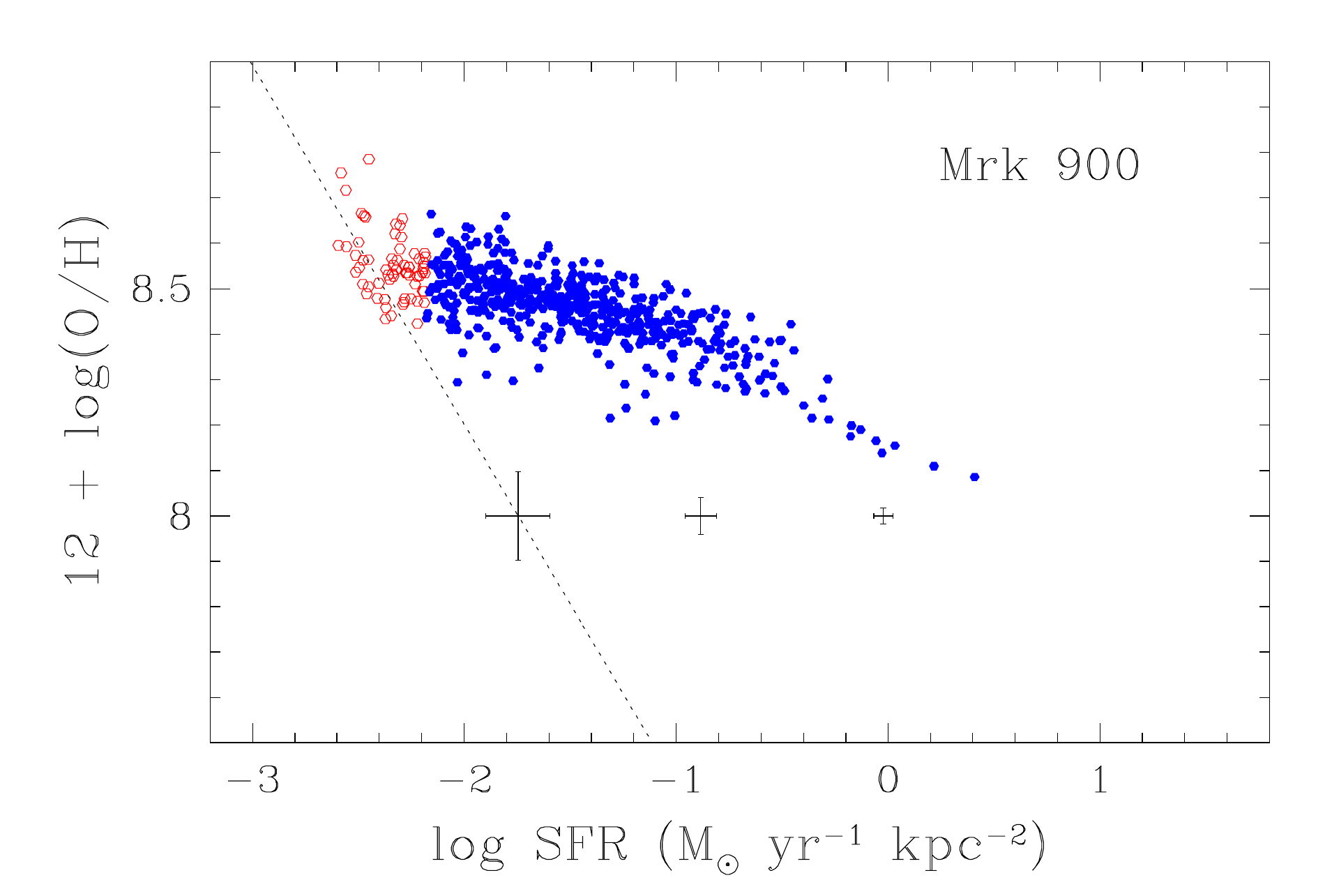}
}
\centerline{
\includegraphics[width=0.62\textwidth]{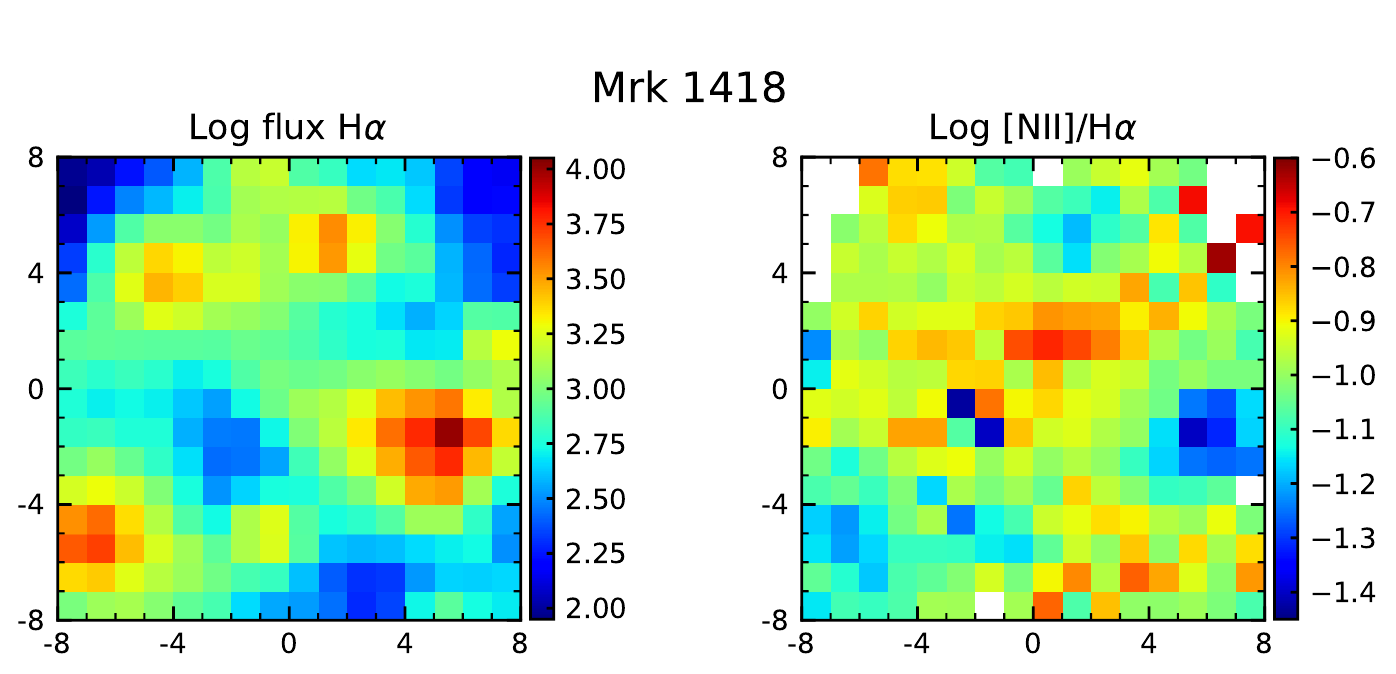} \hspace{5pt}
\includegraphics[width=0.35\textwidth]{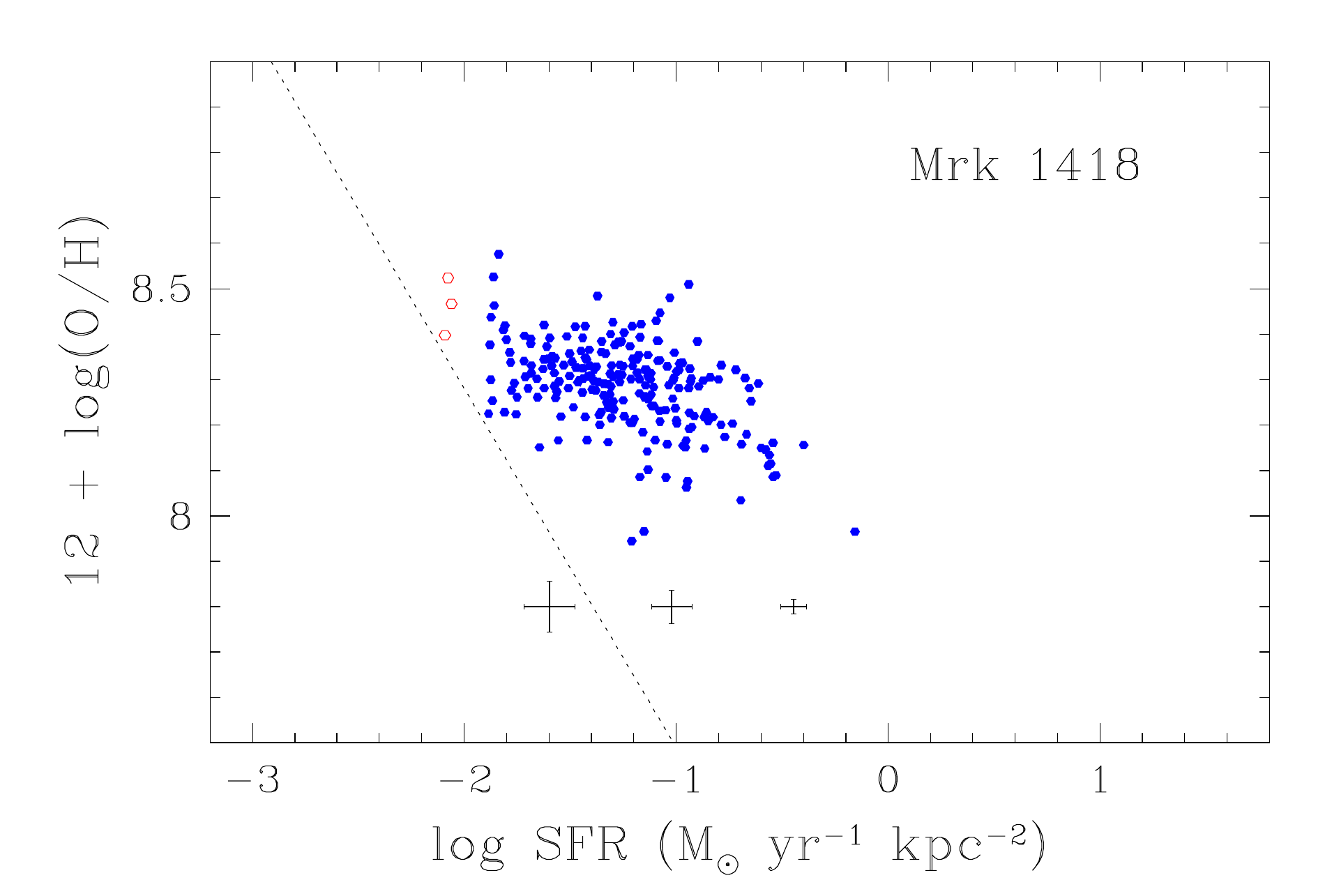}
}
\centerline{
\includegraphics[width=0.62\textwidth]{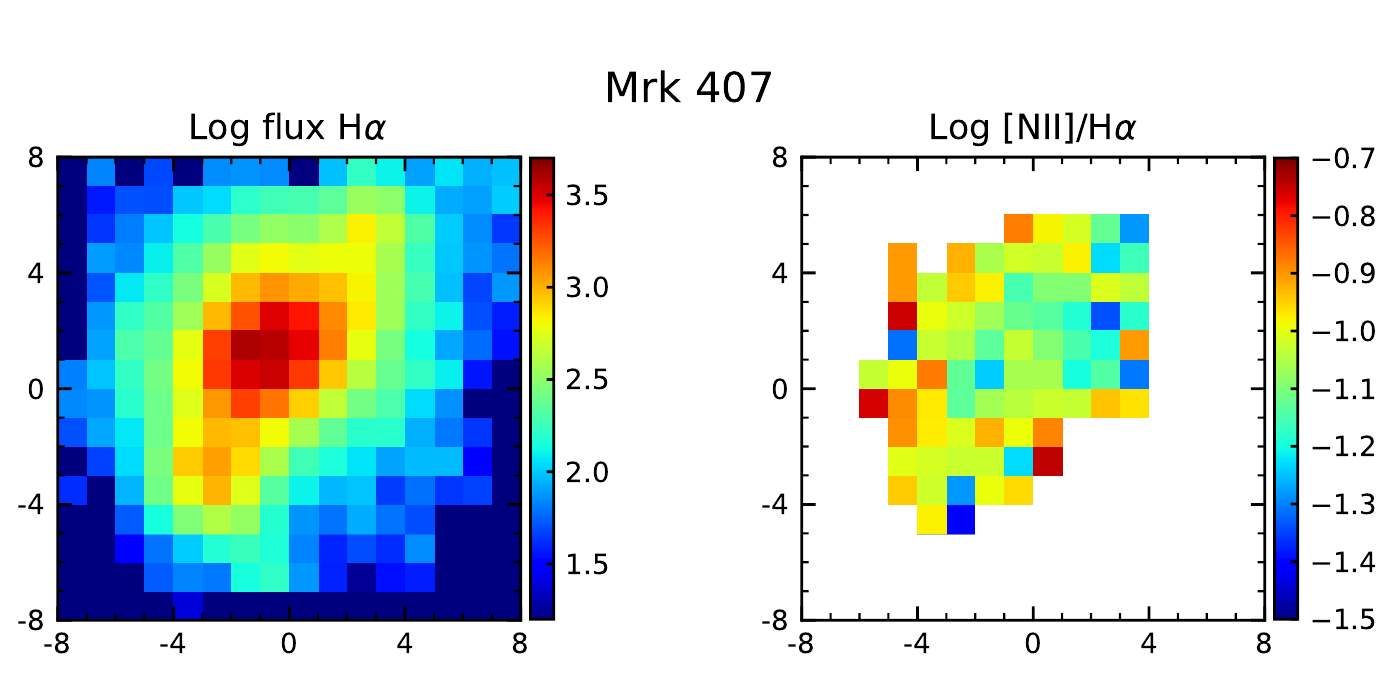} \hspace{5pt}
\includegraphics[width=0.35\textwidth]{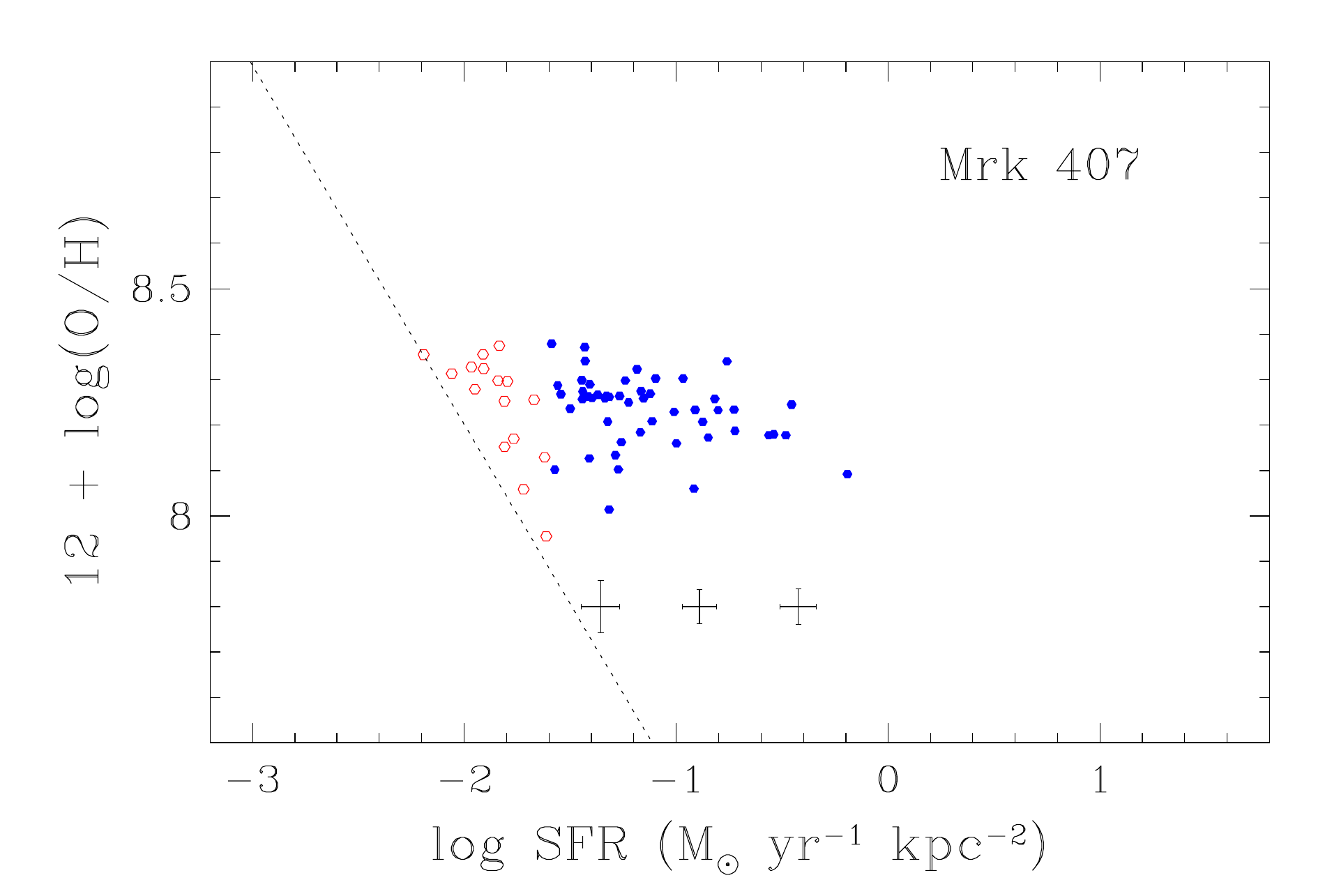}
}
\centerline{
\includegraphics[width=0.62\textwidth]{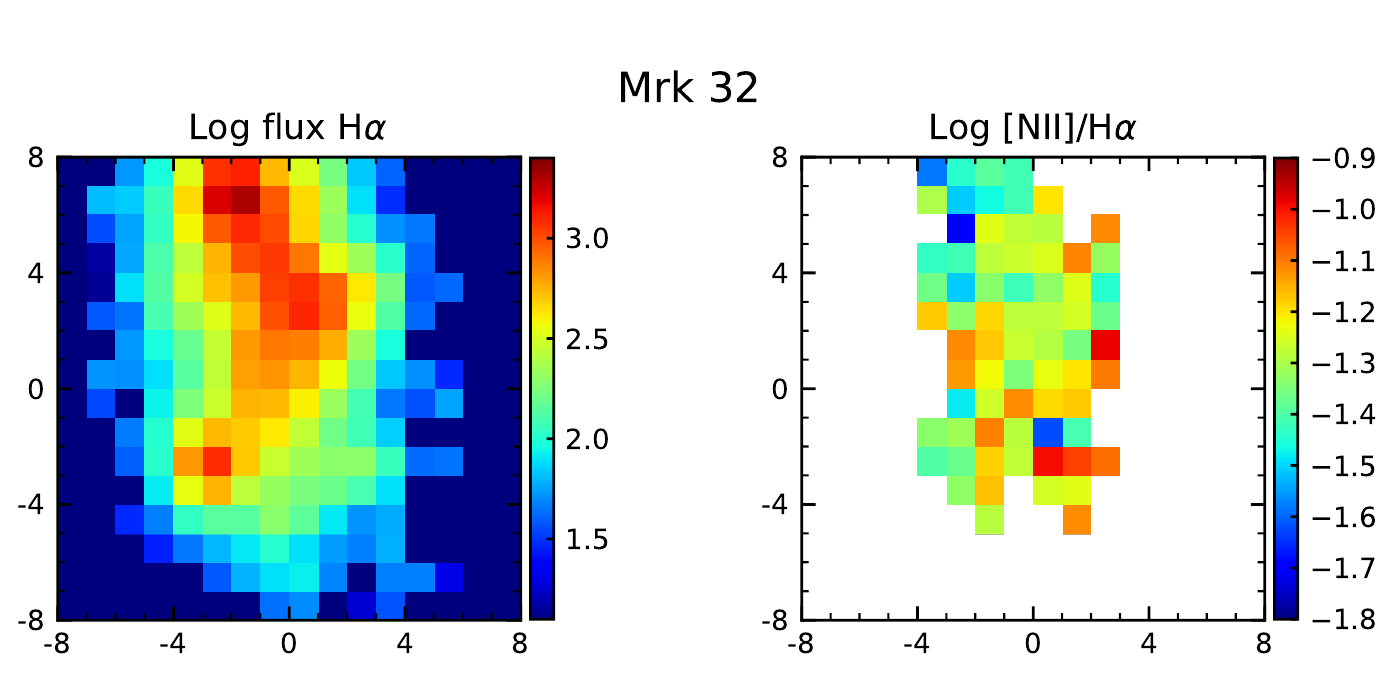} \hspace{5pt}
\includegraphics[width=0.35\textwidth]{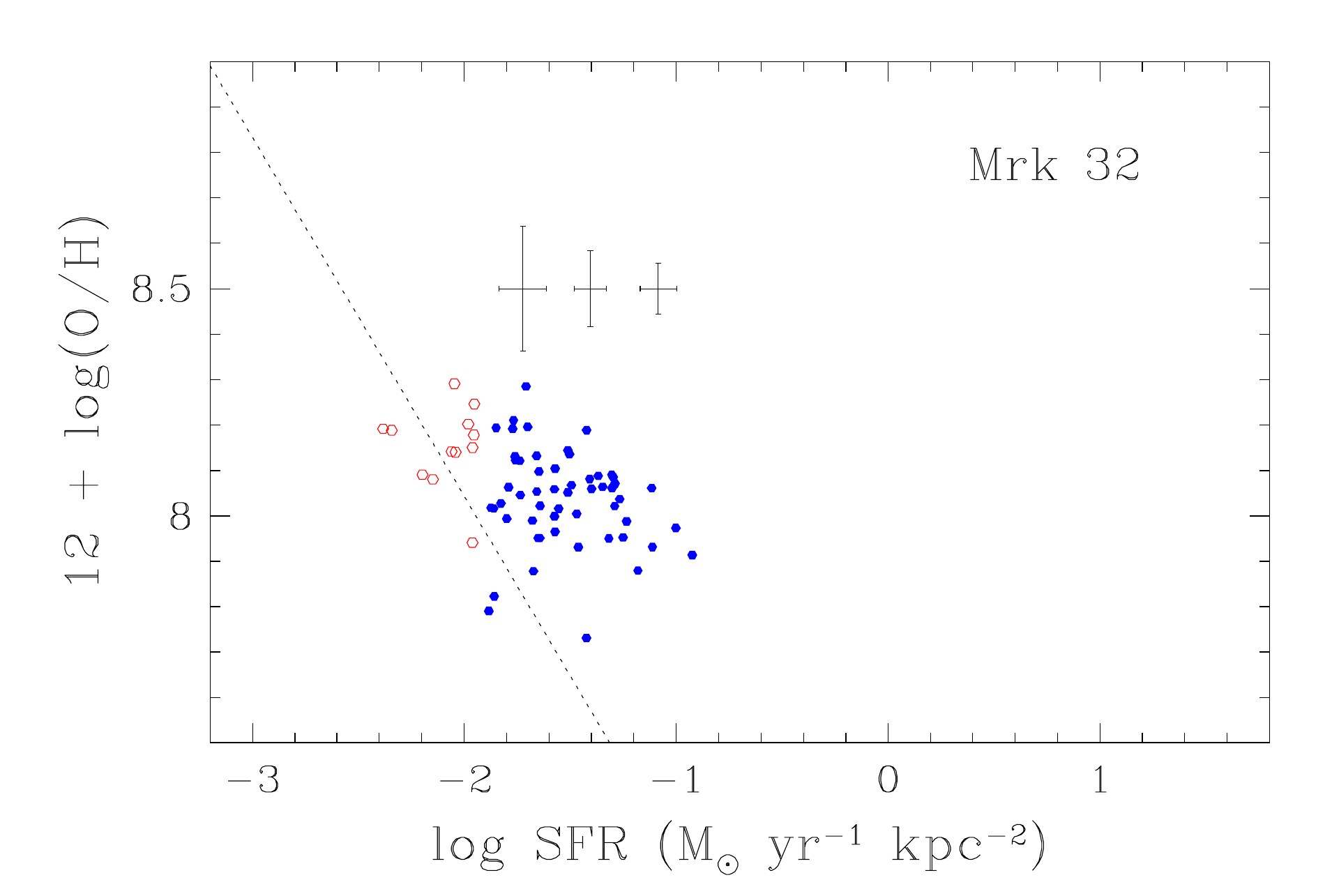}
}
\label{Fig:HavsNII_VIMOS2}
\caption{Continued.}
\end{figure*}

\addtocounter{figure}{-1}
\begin{figure*}
\centerline{
\includegraphics[width=0.62\textwidth]{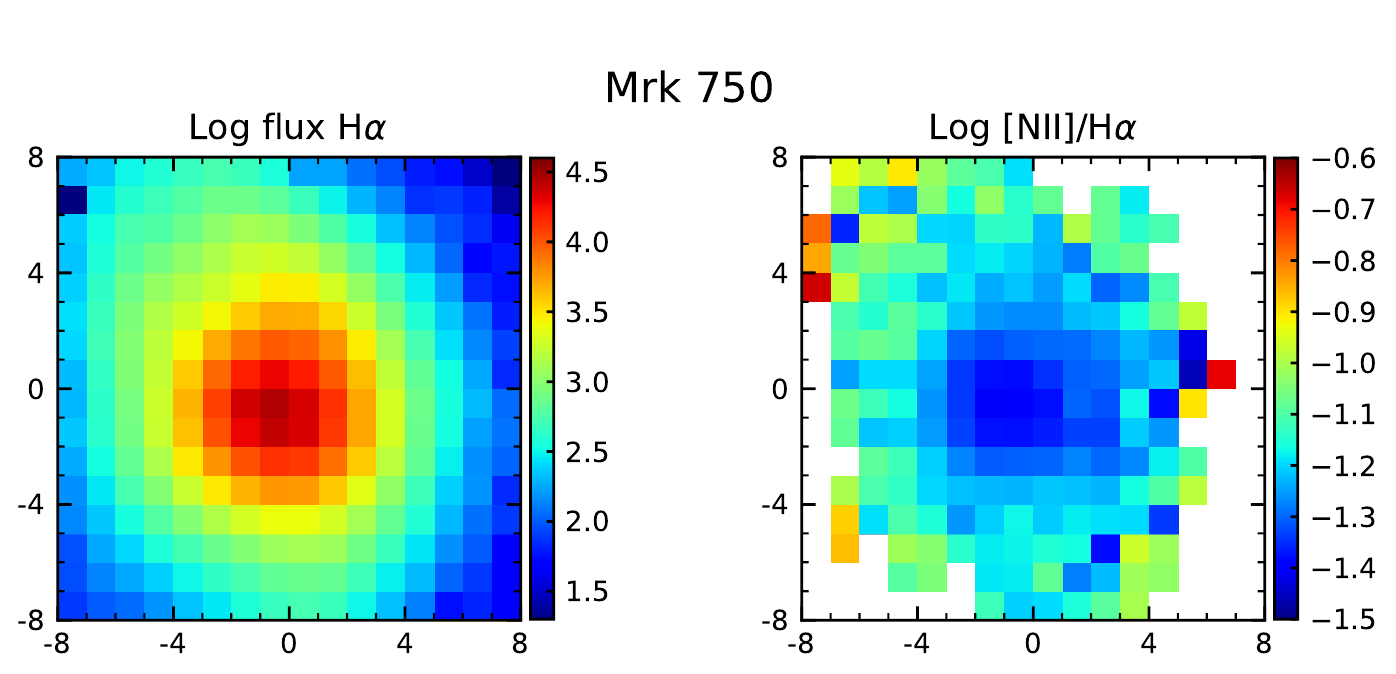} \hspace{5pt}
\includegraphics[width=0.35\textwidth]{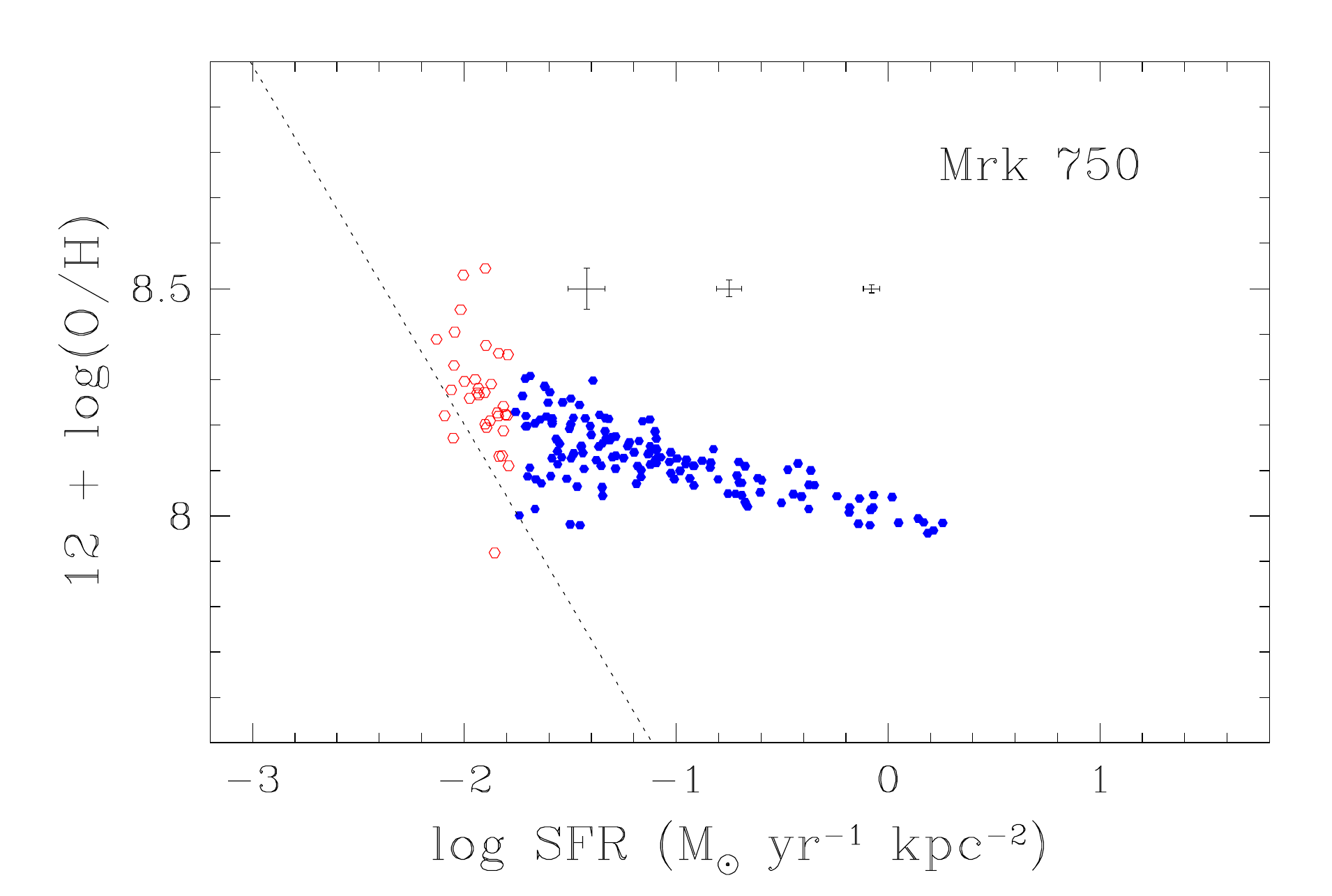}
}
\centerline{
\includegraphics[width=0.62\textwidth]{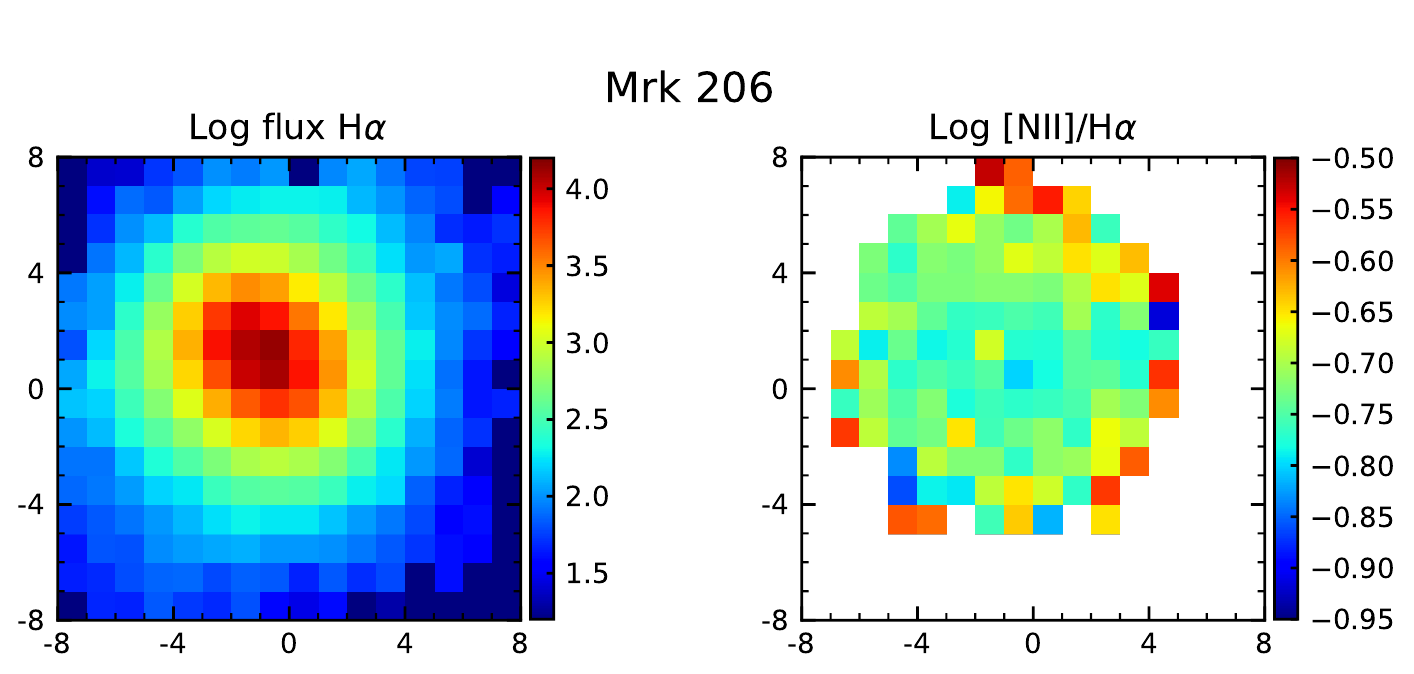} \hspace{5pt}
\includegraphics[width=0.35\textwidth]{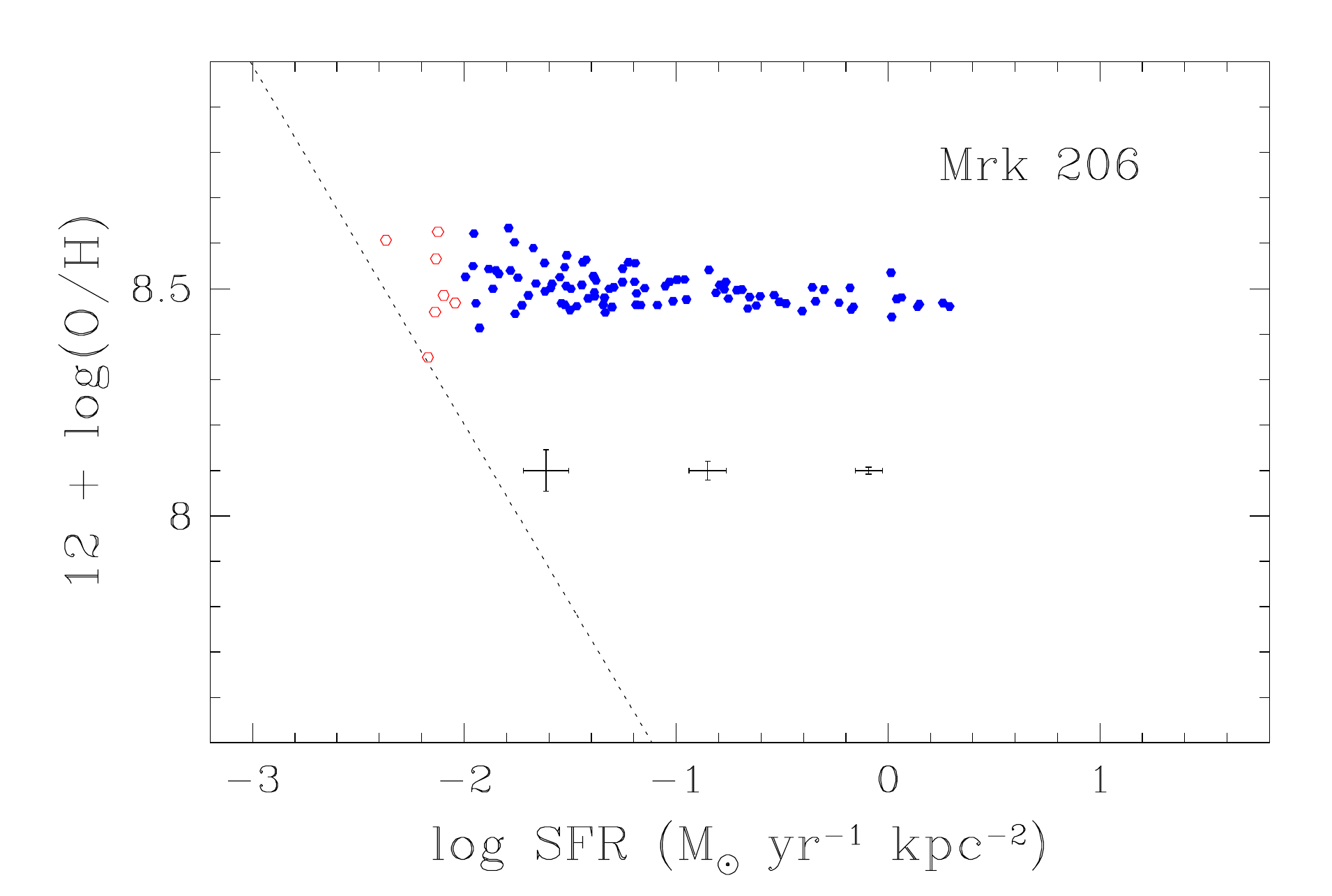}
}
\centerline{
\includegraphics[width=0.62\textwidth]{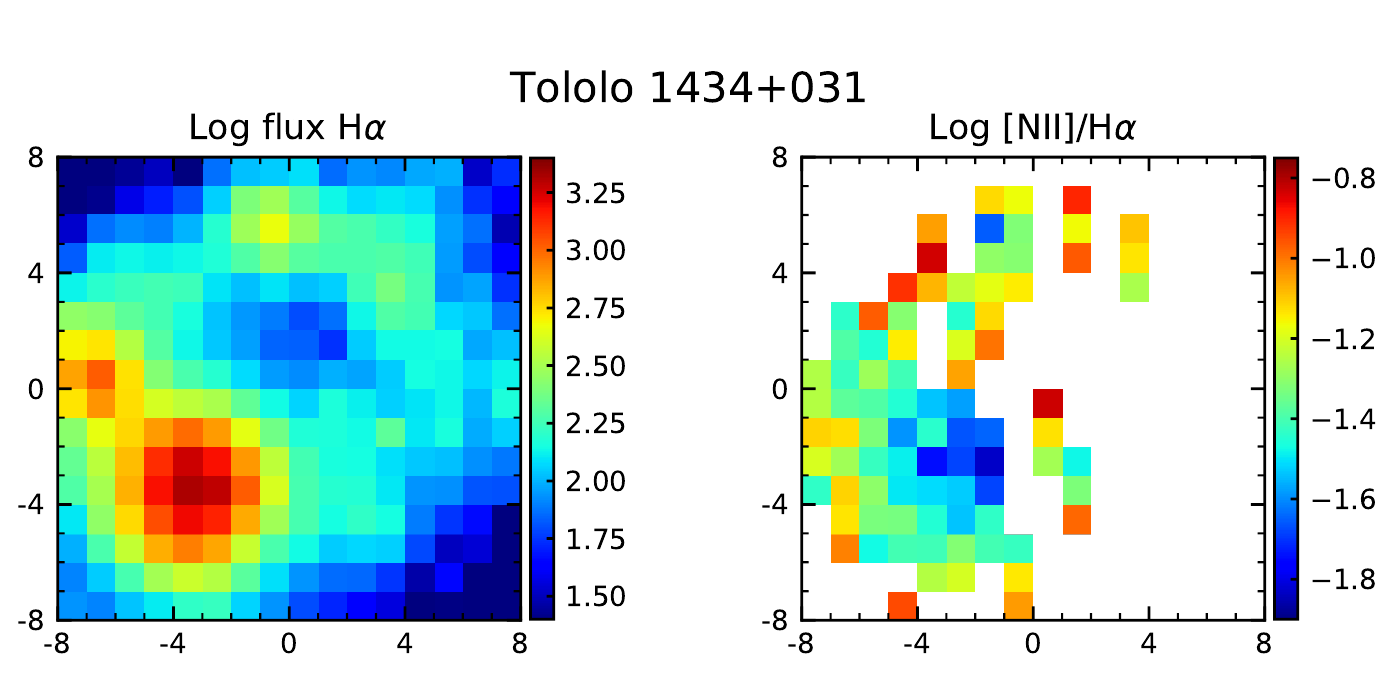} \hspace{5pt}
\includegraphics[width=0.35\textwidth]{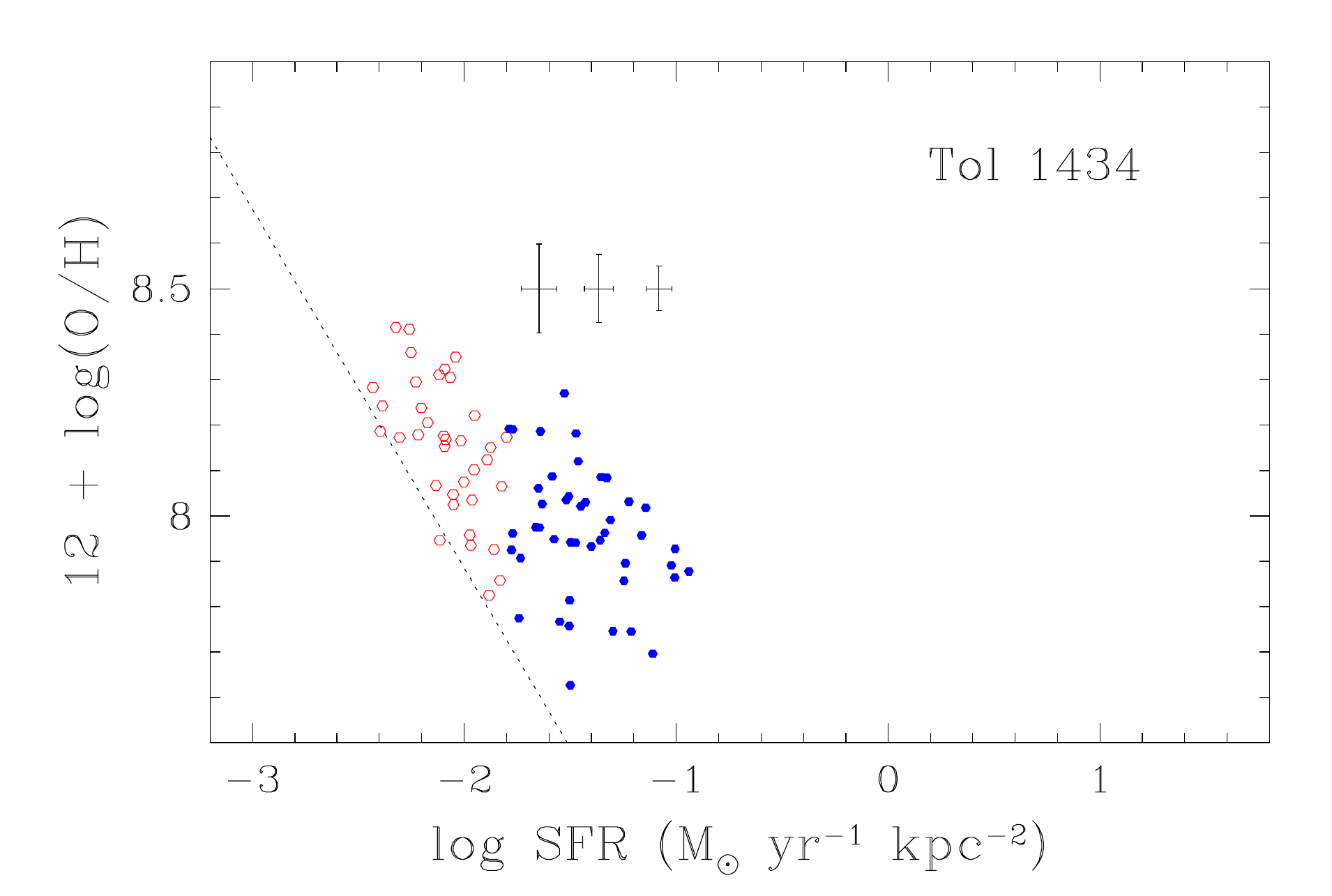}
}
\centerline{
\includegraphics[width=0.62\textwidth]{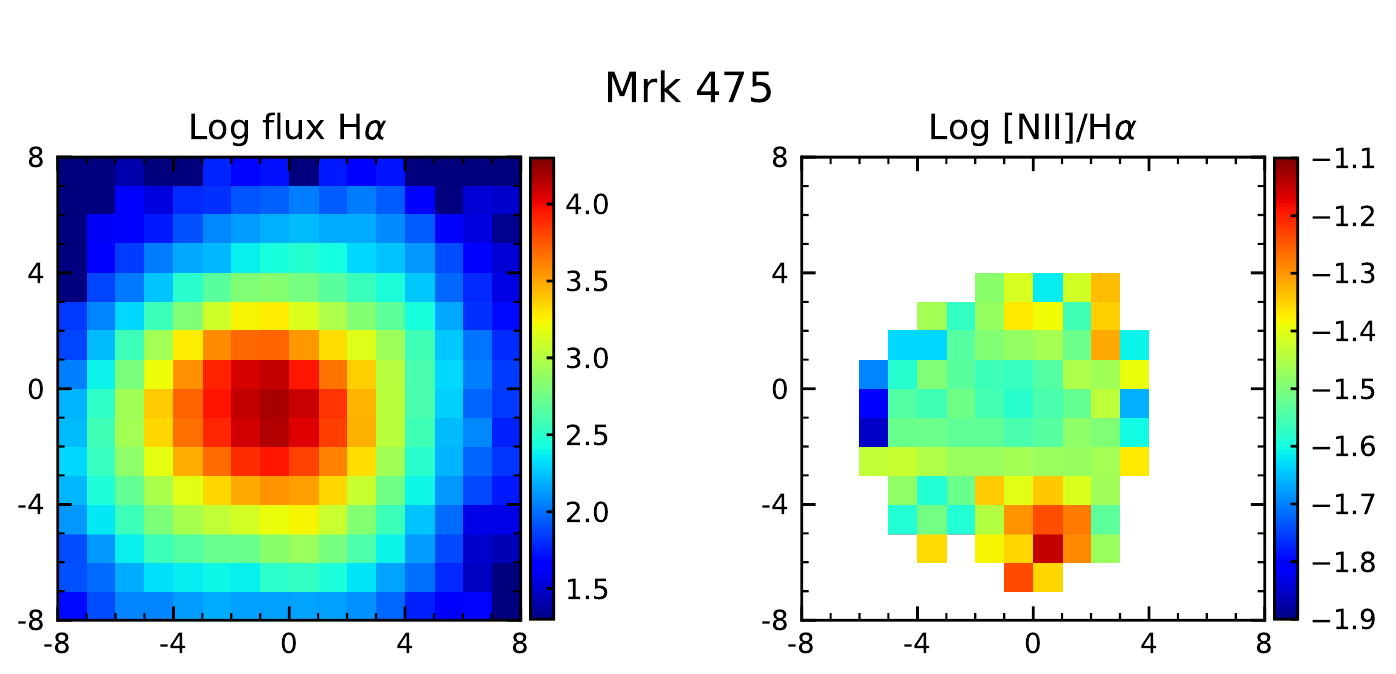} \hspace{5pt}
\includegraphics[width=0.35\textwidth]{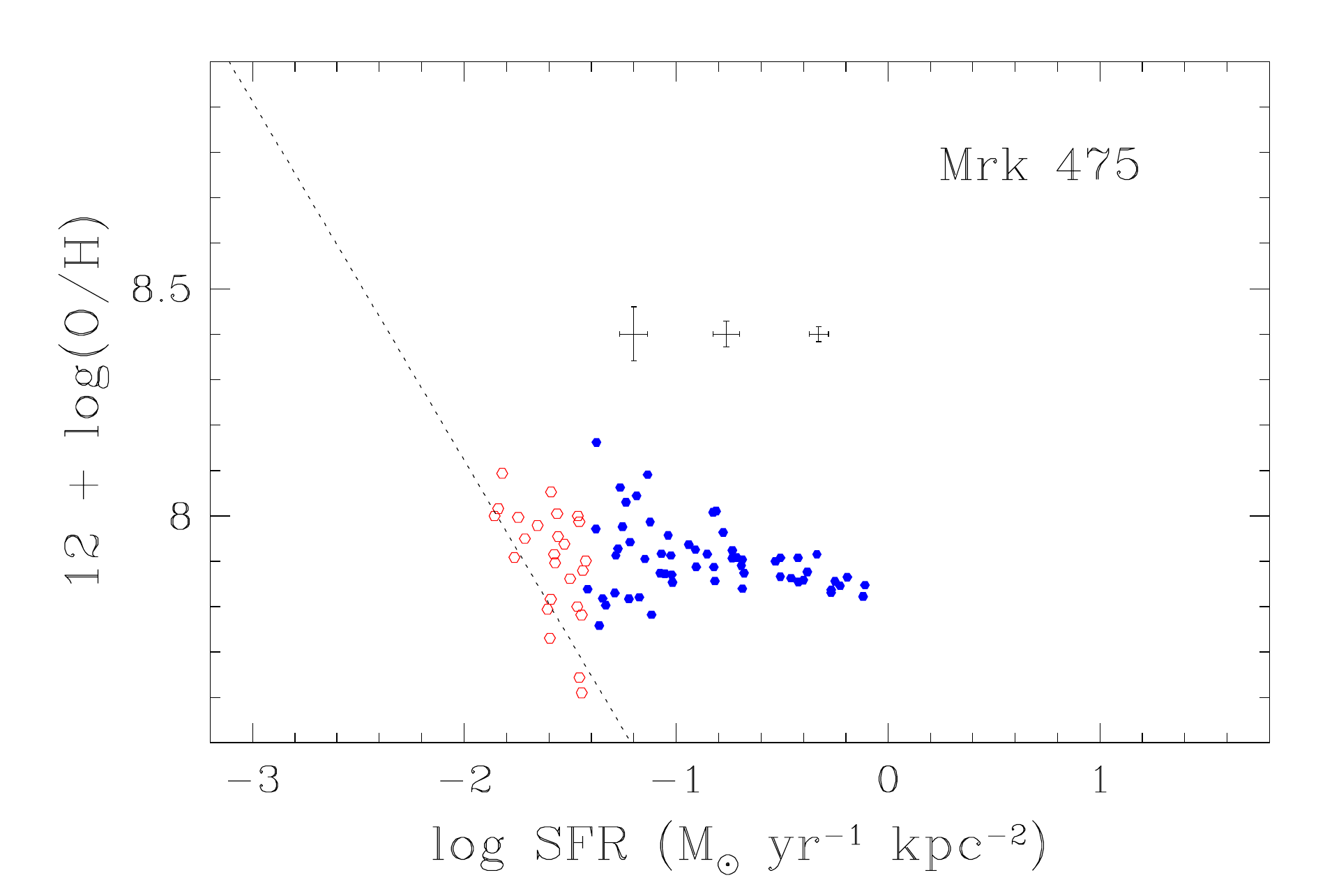}
}
\caption{Continued.}
\end{figure*}

\addtocounter{figure}{-1}
\begin{figure*}
\centerline{
\includegraphics[width=0.62\textwidth]{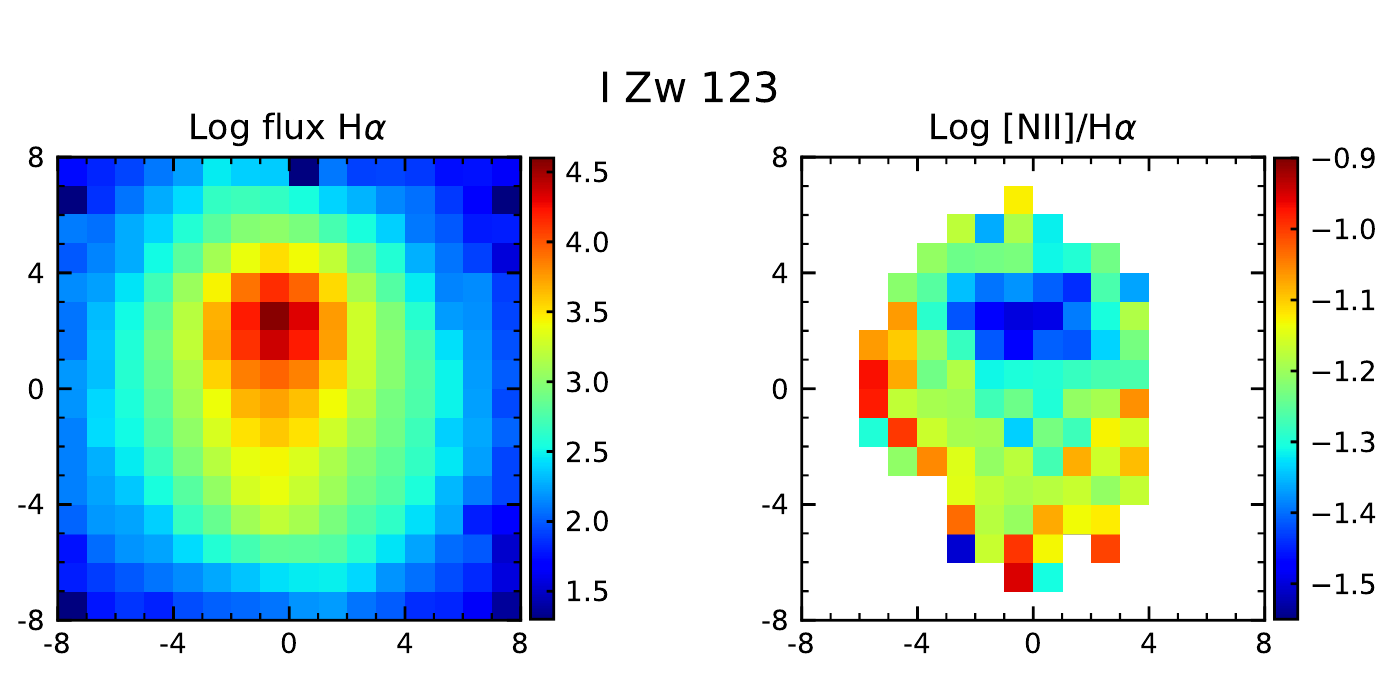} \hspace{5pt}
\includegraphics[width=0.35\textwidth]{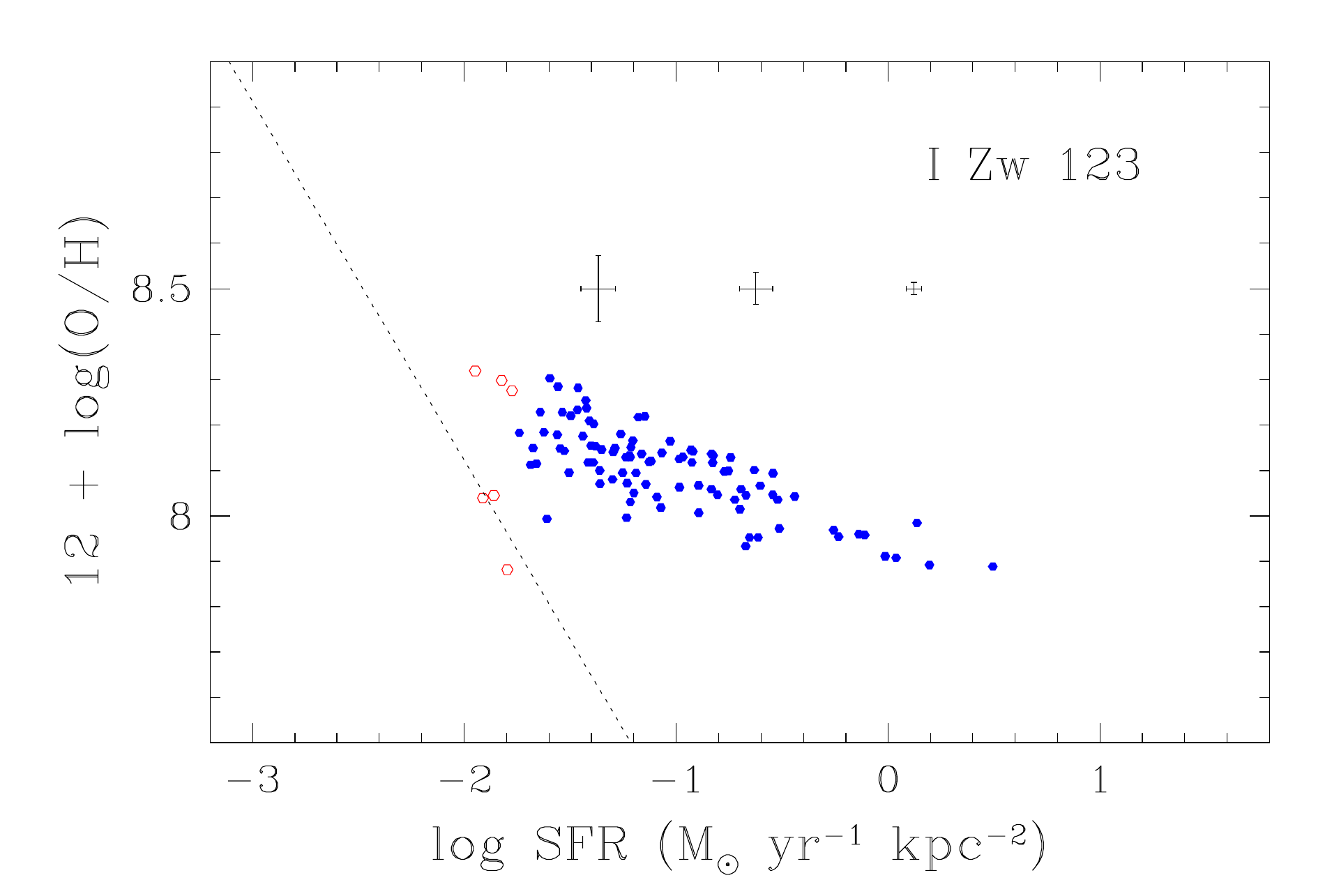}
}
\centerline{
\includegraphics[width=0.62\textwidth]{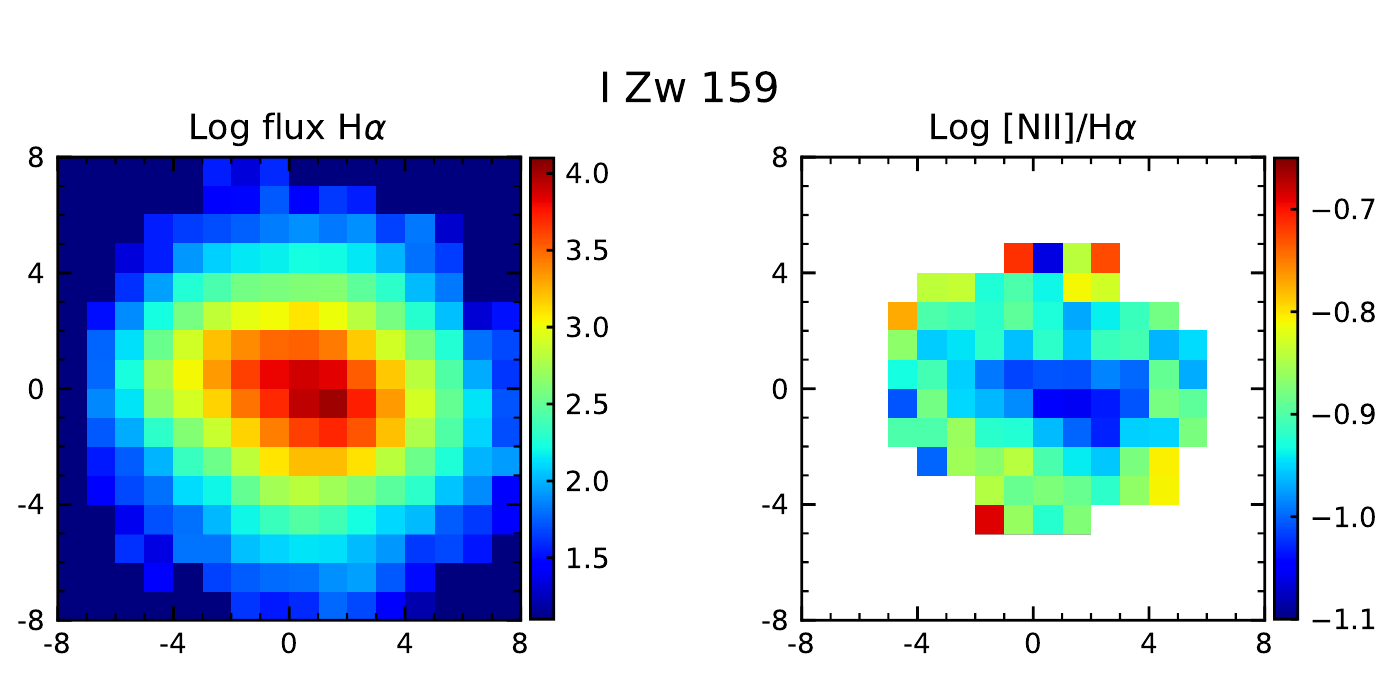} \hspace{5pt}
\includegraphics[width=0.35\textwidth]{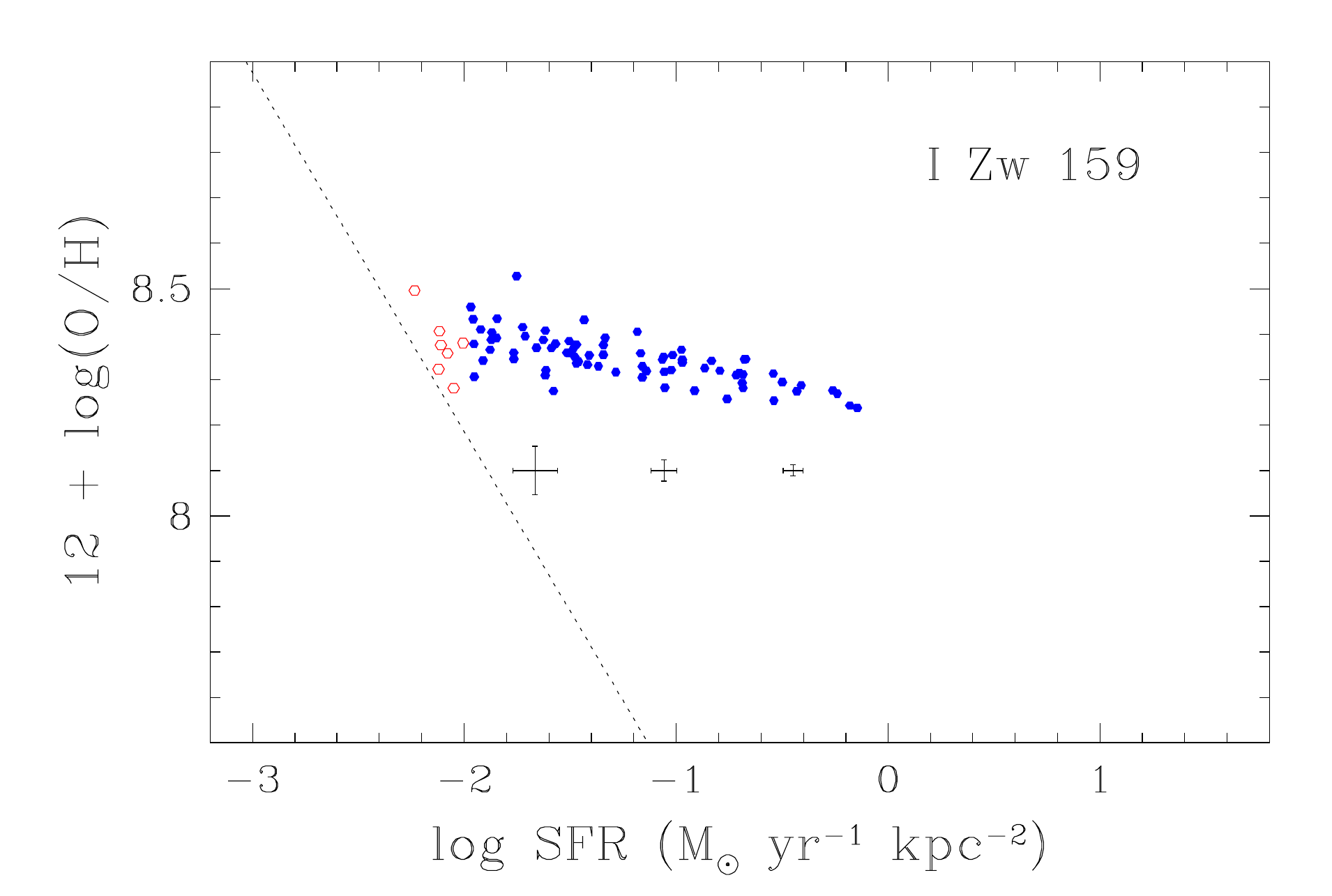}
}
\caption{Continued.}
\end{figure*}

The anti-correlation is more apparent in the spaxel-by-spaxel scatter plot metallicity versus SFR. The third panel of each row in Fig.~\ref{Fig:lgoh12_lgsfr} contains the corresponding scatter plot. As the galaxies have different spatial samplings, we bring them to a common reference by plotting the surface SFR, in units of $M_\odot\,{\rm yr}^{-1}\,{\rm kpc}^{-2}$. As can be inferred from Fig.~\ref{Fig:lgoh12_lgsfr}, right column, the anti-correlation between $12+\log({\rm O/H})$ and SFR goes from mild (e.g., Mrk~206) to very clear (e.g., Tololo~1924-416).
%

%
\begin{figure*}
\centerline{
\includegraphics[width=0.62\textwidth]{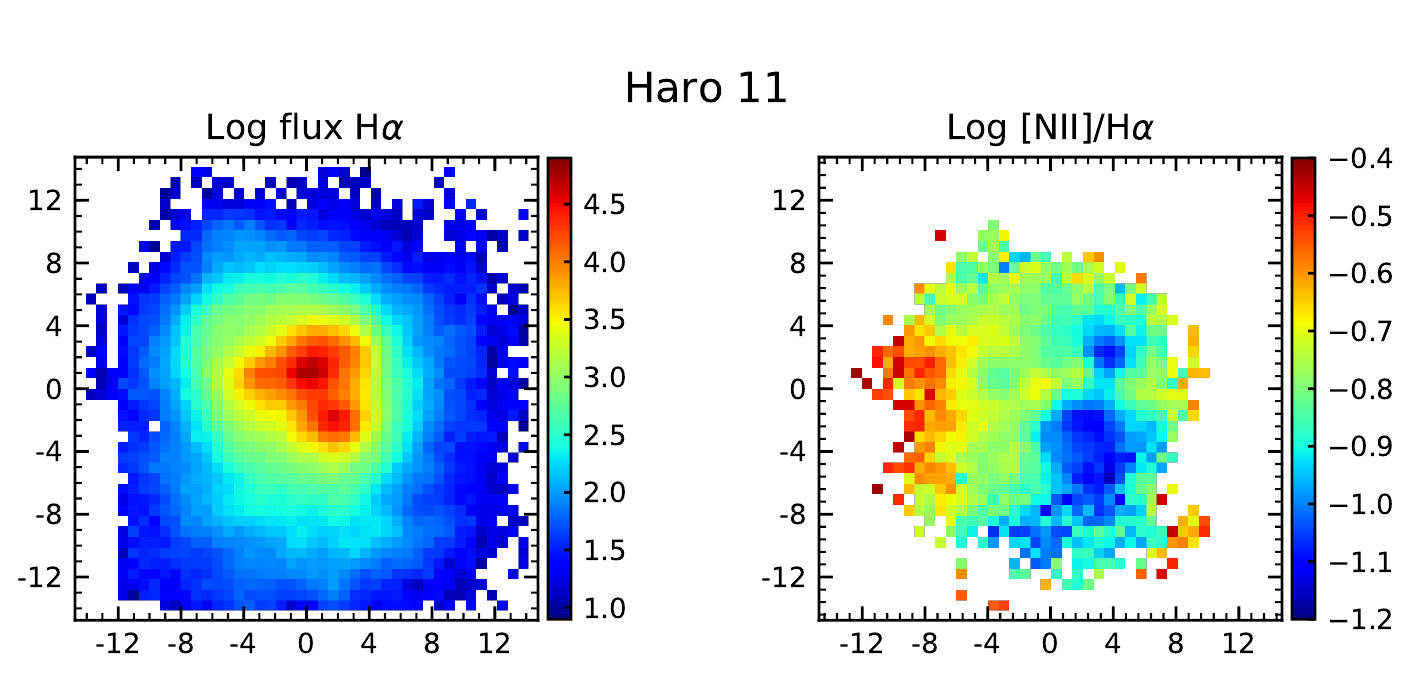} \hspace{5pt}
\includegraphics[width=0.35\textwidth]{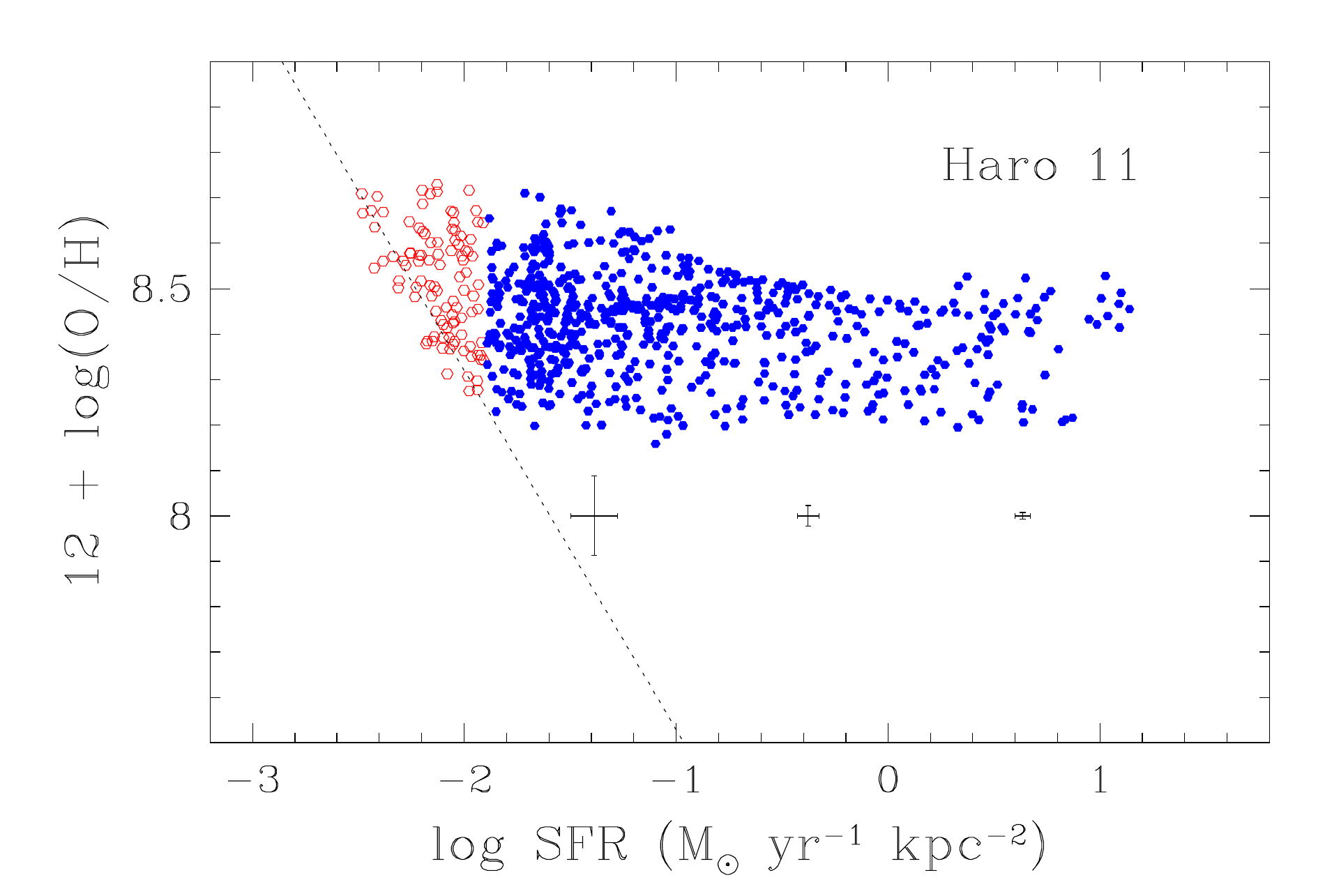}
}
\centerline{
\includegraphics[width=0.62\textwidth]{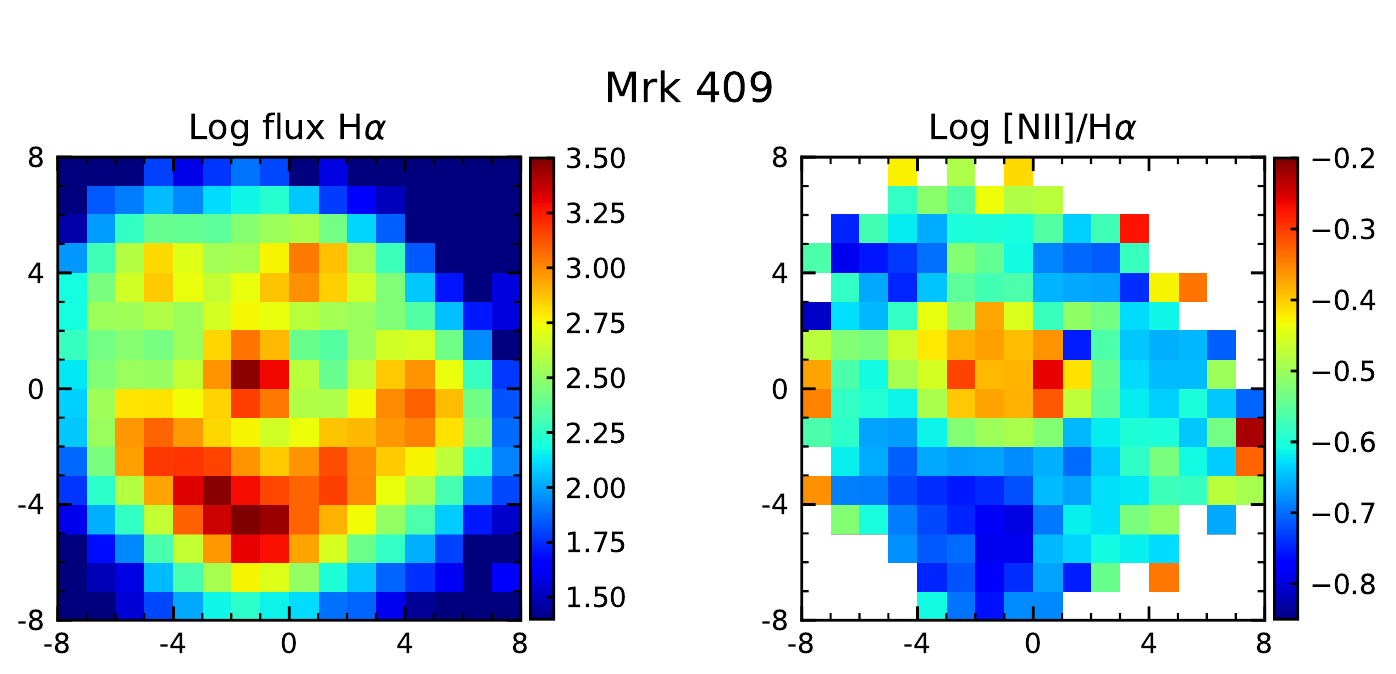} \hspace{5pt}
\includegraphics[width=0.35\textwidth]{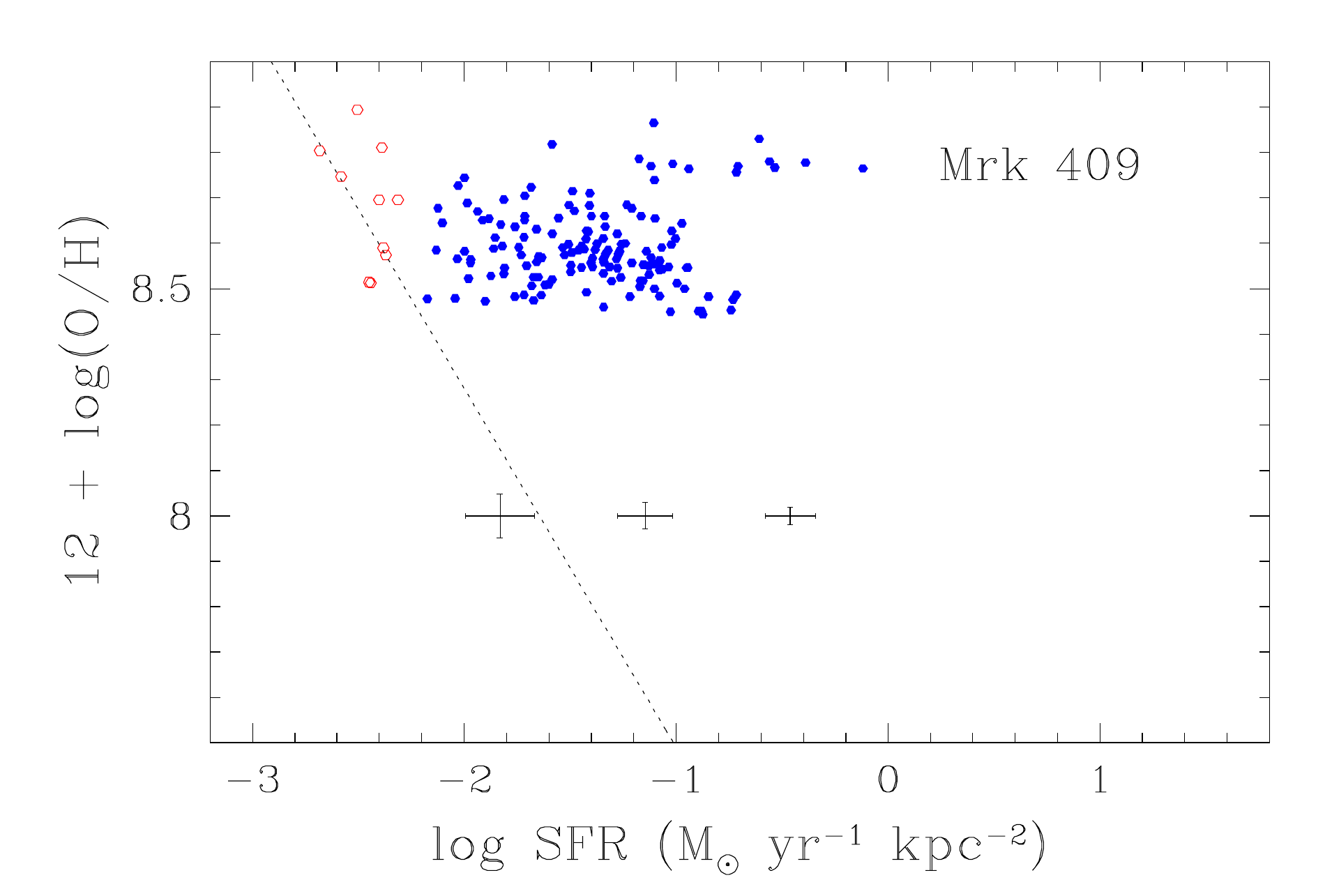}
}
\caption{
Same as Fig.~\ref{Fig:lgoh12_lgsfr} with the two well-documented cases of AGNs. They are not included in our analysis. Note that the scatter plots metallicity versus surface SFR  (third column) show no obvious trend, in contrast with many of the star-forming galaxies in Fig.~\ref{Fig:lgoh12_lgsfr}.  
}
\label{Fig:lgoh12_lgsfr_agn}
\end{figure*}


In order to quantify the correlation, we fit straight lines to the data points of the individual galaxies,
\begin{equation}
12+\log({\rm O/H})=m\times {\rm \log SFR}+k,
\label{Eq.:line}
\end{equation}
where $k$ and $m$ are obtained  using a variant of the least square fit method that takes into account uncertainties on both coordinates, as implemented in the FITEXY routine of the IDL\footnote{Interactive Data Language} astro library. The values for the slope $m$ and the intercept $k$ are listed in Table~\ref{table:slopes}.
Figure~\ref{Fig:lgoh12_lgsfr}, right column, shows the uncertainty of the individual values of $12+\log({\rm O/H})$ and SFR used by the fitting routine. To avoid  overcrowding the figure,  only errors at three representative SFRs are shown. They include measurement errors of the emission line fluxes, and the error of \Ha{}/\Hb{} due to calibration uncertainties, which may be as large as 10\,\%\ for the VIMOS data \citep[see][for details]{Cairosetal2015}.
Excluding Haro~11 and Mrk~409, because they host an AGN (see Sect.~\ref{Sect:individual}), we find 
\begin{eqnarray}
m=&-0.15\pm 0.07,\label{eq:slope} \\ 
k=&8.04\pm 0.23,\nonumber 
\end{eqnarray}
%
where the error bars represent the standard deviation among the fits for the different galaxies. We note a slight trend for the lower metallicity galaxies (smaller intercept) to show slightly more negative slopes.
The error bars for the $m$ and $k$ of the individual galaxies are given in Table~\ref{table:slopes}. They have been estimated using {\em bootstrapping}, which randomly resamples the original data and repeats the fit to evaluate the variance of the estimated parameters \citep[e.g.,][]{mooreetal03,Bradley+94}.
\begin{table*}
\begin{center}
\caption{Parameters that characterize the anti-correlation 12+log(OH) versus SFR.}
\begin{tabular}{ccccccccc}
\hline
Galaxy&$m~^a$ & $k~^a$ & $m~^b$&$k~^b$& $Z_1/Z_\odot~^c$ & $1/N_{eff}~^c$ & $Z_1/Z_\odot~^d$ & $Y_{eff}/N/Z_\odot~^e$ \\
\hline
Haro\,11         & $+0.023 \pm 0.042$ & $8.402 \pm 0.011$&$-0.031 \pm 0.005$& $8.390 \pm 0.061$& \multicolumn{4}{c}{No correlation}  \\
Haro\,14         & $-0.232 \pm 0.009$ & $8.079 \pm 0.010$ &$-0.147\pm 0.009$&$8.181\pm 0.012$& $0.000$           & $0.205 \pm 0.009$ & $0.283 \pm 0.011^\star$  & $0.087 \pm 0.004$ \\
Tol\,0127& $-0.117 \pm 0.005$ & $8.176 \pm 0.008$ &$-0.108\pm0.005$&$8.180\pm 0.005$& $0.000$           & $0.113 \pm 0.005$ & $0.303 \pm 0.006^\star$  & $0.042 \pm 0.002$ \\
Tol\,1924& $-0.143 \pm 0.008$ & $7.854 \pm 0.004$ &$-0.138\pm 0.007$&$7.846\pm 0.005$& $0.054 \pm 0.031$ & $0.251 \pm 0.070$ & $0.115 \pm 0.002$        & $0.020 \pm 0.001$ \\
Tol\,1937& $-0.101 \pm 0.018$ & $8.234 \pm 0.023$ &$-0.063\pm 0.012$&$8.274\pm 0.018$& $0.000$           & $0.094 \pm 0.065$ & $0.381 \pm 0.013$        & $0.045 \pm 0.009$ \\
Mrk\,900         & $-0.190 \pm 0.008$ & $8.195 \pm 0.011$ &$-0.136\pm 0.007$&$8.252\pm 0.010$& $0.000$           & $0.183 \pm 0.008$ & $0.332 \pm 0.019^\star$  & $0.083 \pm 0.003$ \\
Mrk\,1418        & $-0.238 \pm 0.019$ & $8.008 \pm 0.023$ &$-0.138 \pm 0.017$&$8.114\pm 0.023$& $0.000$           & $0.197 \pm 0.031$ & $0.248 \pm 0.021$        & $0.073 \pm 0.007$ \\
Mrk\,407         & $-0.079 \pm 0.026$ & $8.150 \pm 0.028$ &$-0.069\pm 0.036$&$8.153\pm 0.038$& $0.000$           & $0.057 \pm 1.466$ & $0.311 \pm 0.015$        & $0.021 \pm 0.009$ \\
Mrk\,409         & $-0.344 \pm 0.284$ & $8.148 \pm 0.336$ &$+0.026\pm 0.025$&$8.630\pm 0.038$& \multicolumn{4}{c}{No correlation}  \\
Mrk\,32          & $-0.185 \pm 0.060$ & $7.804 \pm 0.089$ &$-0.097\pm 0.061$&$7.891\pm 0.088$& $0.000$           & $0.116 \pm 0.075$ & $0.182 \pm 0.020$        & $0.027 \pm 0.013$ \\
Mrk\,750         & $-0.121 \pm 0.005$ & $8.001 \pm 0.005$ &$-0.106\pm0.009$&$8.008\pm 0.007$& $0.024 \pm 0.199$ & $0.124 \pm 0.049$ & $0.197 \pm 0.003$        & $0.027 \pm 0.002$ \\
Mrk\,206         & $-0.025 \pm 0.008$ & $8.473 \pm 0.007$ &$-0.028\pm 0.006$&$8.470\pm 0.008$& $0.592 \pm 0.430$ & $0.465 \pm 0.328$ & $0.600 \pm 0.011$        & $0.014 \pm 0.005$ \\
Tol\,1434& $-0.294 \pm 0.059$ & $7.579 \pm 0.081$ &$-0.205\pm 0.081$&$7.670\pm 0.114$& $0.140 \pm 0.044$ & $0.794 \pm 0.728$ & $0.128 \pm 0.017$        & $0.036 \pm 0.016$ \\
Mrk\,475         & $-0.119 \pm 0.016$ & $7.821 \pm 0.008$ &$-0.061\pm 0.026$&$7.851\pm 0.016$& $0.000$           & $0.088 \pm 0.068$ & $0.139 \pm 0.004$        & $0.015 \pm 0.002$ \\
I\,Zw\,123       & $-0.161 \pm 0.012$ & $7.944 \pm 0.010$ &$-0.148\pm 0.012$&$7.947\pm 0.011$& $0.087 \pm 0.192$ & $0.262 \pm 0.070$ & $0.160 \pm 0.006^\star$  & $0.032 \pm 0.004$ \\
I\,Zw\,159       & $-0.088 \pm 0.007$ & $8.240 \pm 0.007$ &$-0.076\pm 0.009$&$8.251\pm 0.009$& $0.000$           & $0.084 \pm 0.033$ & $0.370 \pm 0.007^\star$  & $0.036 \pm 0.003$ \\
\hline
\end{tabular}

\raggedright
\begin{tabular}{l}
$^a$\,Slope and intercept defined in Eq.~(\ref{Eq.:line}), computed considering measurement errors in the emission line fluxes.\\
$^b$\,Slope and intercept defined in Eq.~(\ref{Eq.:line}), computed adding 0.34\,dex to the measurement errors in 12+log(O/H).\\
$^c$\,Metallicity of the metal-poor gas and inverse index of the Kennicutt-Schmidt relation; Eqs.~(\ref{Eqn:formulaJorge}) and (\ref{eq:neff}).\\ 
$^d$\,Based on Eq.~(\ref{Eqn:formulaJorge}) with $N=1.4$. The superscript ($^\star$) points out when 
the fit worsens with respect to the case where $1/N_{eff}$ is inferred from the fit.\\
$^e$\,Effective yield from Eqs.~(\ref{eq:openbox}) and (\ref{eq:yeff}).\\
\end{tabular}
\label{table:slopes}
\end{center}
\end{table*}



Several sources of random noise affecting both SFR and $12+\log({\rm O/H})$ were not considered when estimating $m$ and $k$ in Eq.~(\ref{eq:slope}). However, the anti-correlation between SFR and metallicity remains in place even when they are included.  Firstly, we analyze the  fact that both SFR and  $12+\log({\rm O/H})$ depend on the  observed H$\alpha$ flux and so their errors are partly correlated.  [NII]$\lambda$6583 is a weak line, therefore, when its observed flux is dominated by noise, of roughly constant value in each map, then 
${\rm N2} ={\rm constant} -\log{\rm H}\alpha$ (see Eq.~[\ref{eq:defN2}]). In this case
N2 drops when H$\alpha$ increases, giving rise to an artificial anti-correlation between SFR and $12+\log({\rm O/H})$ qualitatively similar to the observed one. The dotted lines in Fig.~\ref{Fig:lgoh12_lgsfr}, right column, show the expected false correlation given the measured noise level in each [NII]$\lambda$6583 map. As one can see, the artificial correlation is clearly unrelated to the observed correlation. To fully discard any influence of this bias, the points around the artificial correlation, indicated by red open symbols in the figures, have been excluded from all analyses.
Secondly, the calibration in Eq.~(\ref{Eqn:HaNIImet}) assumes a one-to-one relation between N2 and  $12+\log({\rm O/H})$. However, this relation presents a significant intrinsic scatter due to differences in the properties of the \hii\ regions \citep[ionizing UV flux, ionizing parameter, and N to O ratio; see, e.g.,][]{2005MNRAS.361.1063P}. \citet{PerezMonteroContini2009} find a scatter of 0.34 dex given N2\footnote{Part of this scatter is not applicable to the within-galaxy variation studied in this paper.  \citeauthor{PerezMonteroContini2009} consider galaxies with large differences in N/O which are not present in a single galaxy (see Sect.~\ref{sect:hiic}). They also use line ratios from very heterogeneous bibliographic sources, so that part of the scatter is not intrinsic to the calibration but attributable to observational errors in the employed data.  Thus, other similar calibrations based on N2 present much smaller scatter, like the 0.18~dex quoted by \citet{2004MNRAS.348L..59P}. Therefore, this 0.34 dex scatter should be regarded as an upper limit to the intrinsic error of the N2 calibration.}.  We carried out a Monte Carlo simulation to estimate whether such scatter could hinder the detection of the observed correlation between SFR and $12+\log({\rm O/H})$. Mock observations were produced starting from a uniform random distribution of SFRs in the range of the observed values (from 0.01 to 10\,$M_\odot\,{\rm kpc}^{-2}\,{\rm yr}^{-1}$), and then computing the associated metallicity as $12+\log({\rm O/H})=m\times {\rm SFR}+k$, with $m$ and $k$ given by  Eq.~(\ref{eq:slope}). Random gaussian noise with a standard deviation of 0.34 dex was added to $12+\log({\rm O/H})$. 
One thousand mock observations were produced assuming 200 spaxels per galaxy, which is a realistic number for the spaxels in the VIMOS fields. 
These mock observations were analyzed as real observations, fitting straight lines to retrieve 1000 different values for $m$ and $k$. The mean slope turns out to be $-0.15\pm 0.03$,  with the error bar representing the dispersion among the 1000 individual estimates of $m$.If 100 spaxels per galaxy are considered, the dispersion increases only moderately, with $m=-0.150\pm 0.04$. The dispersion inferred from the Monte Carlo simulation is significantly smaller than the observed value, therefore, a random noise of 0.34 dex in metallicity does not impede finding and characterizing the slope of a correlation between SFR and $12+\log({\rm O/H})$ like the one we obtain (Eq.~[\ref{eq:slope}]). 
We completed this Monte Carlo simulation with a direct estimate of the effect on the fit of having a 0.34~dex extra error in $12+\log({\rm O/H})$. For every single galaxy, the linear fit was repeated adding in quadrature 0.34~dex to the error in $12+\log({\rm O/H})$. The inferred values are consistent with the values given in Eq.~(\ref{eq:slope}), explicitly, $m=-0.11\pm 0.05$ and $k=8.08\pm 0.21$, with the error bars representing the scatter among the galaxies in our sample. Table~\ref{table:slopes} includes the slope and intercept ($m$ and $k$) once the 0.34 dex extra error is included in the least squares fitting procedure. The list also includes formal error bars which have been evaluated using bootstrapping, as we did with the original estimate.


\subsection{Spatial resolution compared with the physical size of the \hii\ regions}\label{Sect:sizes}
The spatial resolution of the maps is listed in Table~\ref{Table:Galaxies}. We estimate the resolution as twice the sampling interval, following the Shannon sampling theorem. The observed values span from 50 to 500\,pc, with the typical value around 200\,pc. These sizes are significantly larger than the expected size of an \hii\ region, and comparable to the size of giant \hii\ regions. For example, the Str\"omgren radius of the model \hii\ regions are all smaller than 200 pc, provided that the mass of the ionizing stellar cluster is smaller than $10^{6}\,M_\odot$ \citep[e.g.,][]{2010A&A...517A..93V}. Similarly, the observed \hii\ regions have diameters consistent with this range in sizes \citep[e.g.,][]{2000AJ....120..752F,2009A&A...507.1327H,2012MNRAS.422.3339W,2016A&A...592A.122H,2017ApJ...834..181O}. The fact that the spatial resolution is coarser than the typical \hii\ region size implies that, to first order, the individual resolution elements in our maps spatially integrate the signal coming from full \hii\ regions. This fact justifies the analysis made Sect. 4, where the properties observed in each spaxel are compared with the properties of integrated \hii\ regions.


\subsection{Correlation after subtracting the running mean average}\label{Sect:nolarge}

\begin{figure}
\includegraphics[width=0.48\textwidth]{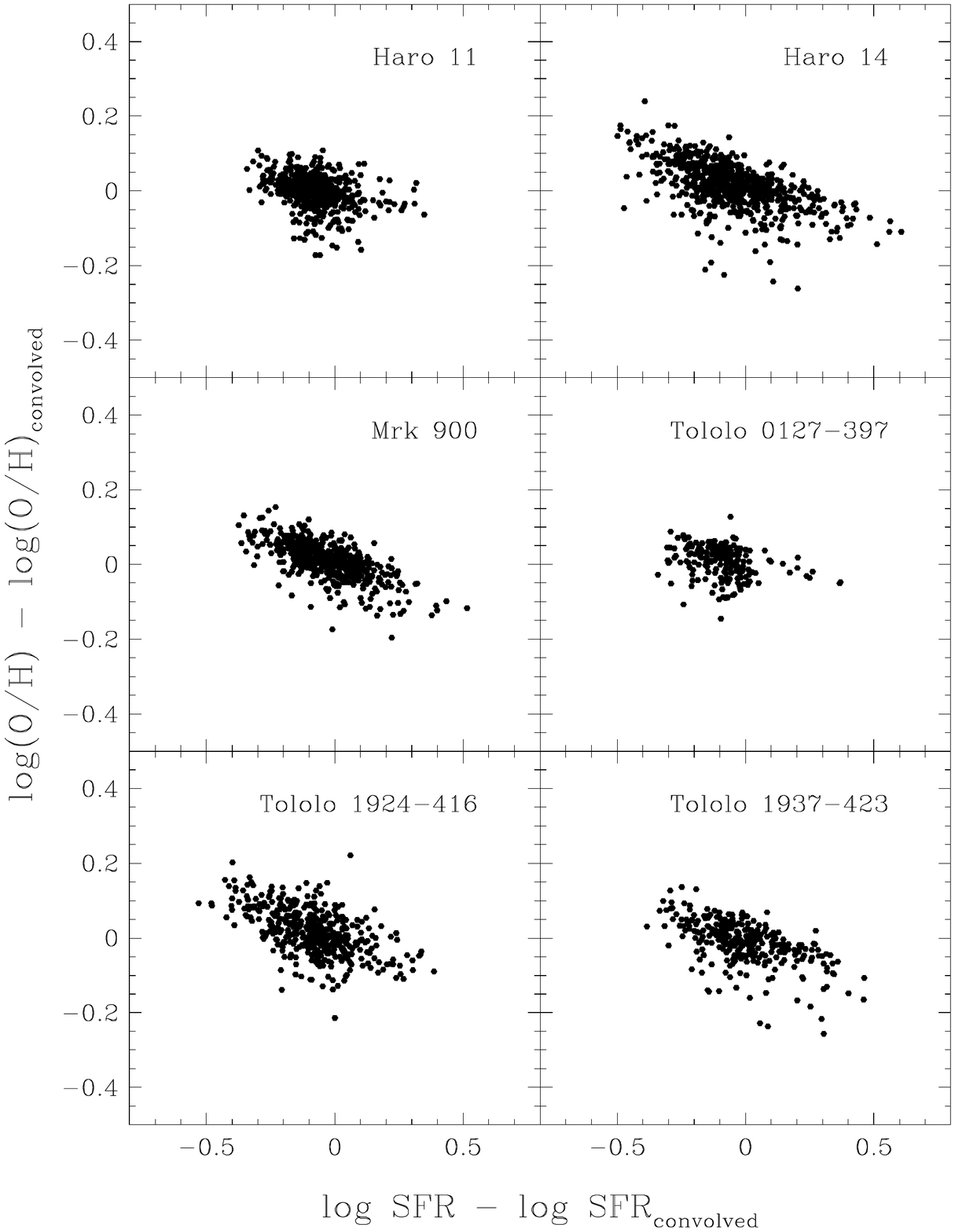}
\caption{Similar to the panels on the right column of Fig.~\ref{Fig:lgoh12_lgsfr}, except that running-mean average images have been subtracted from the original 12+log(O/H) and SFR images. The anti-correlation between the excess of metallicity, 12+log(O/H)$-$12+log(O/H)$_{\rm convolved}$ and the excess of SFR, log\,SFR$-$log\,SFR$_{\rm convolved}$, remains with an amplitude similar to that of the correlation of the original images (cf. Fig.~\ref{Fig:lgoh12_lgsfr}). The units of the axes are the same as in Fig.~\ref{Fig:lgoh12_lgsfr}.}
\label{Fig:corr_mean_sobtracted}
\end{figure}

The correlation described in the Sect.~\ref{Sect:anticorr} is dominated by {\em local variations} in N2 and H$\alpha$. It is not due to large-scale variations, like the radial gradients of metallicity typical of massive spirals, including the Milky Way \citep[e.g.,][]{1981ARA&A..19...77P,2014A&A...563A..49S}. In order to show this result, we have repeated the plot 12+log(O/H) versus SFR in Fig.~\ref{Fig:lgoh12_lgsfr}, this time removing a running-mean local value from the original signal. This high-pass spatial filtering leaves only the small scale variations in the images of N2 and H$\alpha$. The result for the VIMOS galaxies is
shown in Fig.~\ref{Fig:corr_mean_sobtracted}. The anti-correlation between N2 and H$\alpha$ remains with an amplitude similar to the amplitude in the original images (cf. Figs.~\ref{Fig:lgoh12_lgsfr} and \ref{Fig:corr_mean_sobtracted}). The subtracted running mean images were produced using a box of 5$\,\times\,$5 spaxels. The fact that the correlation remains after removing a running-mean average also holds for the galaxies in the PMAS sample. 

\section{Analysis in terms of a correlation between metallicity and SFR}
\label{Section:Analysis}

Here we explore possible explanations for the observed relation between N2 and H$\alpha$ in terms of a local correlation between metallicity and SFR. We analyze the mixing of pre-existing gas with variable amounts of metal-poor gas, as expected from external metal-poor gas accretion (Sect.~\ref{sect:mixing}), and the self-enrichment due to local variations of the star-formation history (Sect.~\ref{sect:selfcont}). 
Variations  of the physical conditions in the emitting nebula other than metallicity are treated in Sect.~\ref{Sect:bias}. This last option would forge a relation between 12+log(O/H) and SFR. As we describe in Sect.~\ref{Sect:bias}, this possibility is unlikely, although it cannot be completely ruled out.  

\subsection{Mixing of metal-rich with metal-poor gas}\label{sect:mixing}

One possibility to explain the observed correlation is assuming that the mass of gas producing stars, $M_g$, is a mixture of pre-existing 
metal-rich gas of mass $M_{g2}$, blended together with variable masses of metal-poor gas, $M_{g1}$, i.e.,
\begin{displaymath}
M_g = M_{g1} + M_{g2},
\end{displaymath}
\begin{equation}
M_g\,Z = M_{g1}\,Z_1 + M_{g2}\, Z_2,
\label{eq:massgas}
\end{equation}
with $Z_1$, $Z_2$, and $Z$ the metallicity of the metal-poor gas, the metallicity of the metal-rich gas, and the metallicity of the mixture, respectively. 
This scenario is to be expected if the star formation in each galaxy is triggered by the accretion of pristine or nearly-pristine gas falling onto the disk. 
Depending on the mass of gas that ends up in each particular location on the disk, starbursts with different combinations of metallicity and SFR are produced. The metallicity is given by Eq.~(\ref{eq:massgas}), whereas the (surface density) SFR is given by the (surface density) gas mass through the so-called Kennicutt-Schmidt relation
\citep[e.g.,][]{Kennicutt1998},
\begin{equation}
{\rm SFR} \propto M_g^N, 
\label{eq:kslaw}
\end{equation}
with $N\simeq 1.4$ \citep{2012ARA&A..50..531K}. 
Using Eqs. (\ref{eq:massgas}) and (\ref{eq:kslaw}), one obtains  
\begin{equation}
\label{Eqn:formulaJorge}
Z = Z_1 + (Z_2-Z_1) \times ({\rm SFR}_2/{\rm SFR})^{1/N},
\end{equation}
where ${\rm SFR}_2$ represents the SFR to be observed if there is no external gas infall ($M_{g1}=0$).

Equation~(\ref{Eqn:formulaJorge}) predicts an anti-correlation between the metallicity $Z$ and the SFR that accounts for the observed trend (Sect.~\ref{Sect:Data}) once the free parameters in the relation are tunned.
If all the spaxels start off with the same gas mass, $M_{g2}$, then Eq.~(\ref{Eqn:formulaJorge}) predicts that $Z$ will vary as SFR$^{1/N}$, with $1/N\sim 0.71$ \citep{2012ARA&A..50..531K}. On the other hand, if different spaxels begin with different $M_{g2}$, then $M_{g}$ and $M_{g2}$ are correlated, so that
\begin{equation}
M_{g2}/M_{g}\simeq (\langle M_{g2}\rangle /M_g)^\delta,
\label{eq:deltapprox}
\end{equation}
with $\delta \ll 1$ (Appendix~\ref{appa}). The parameter $\langle M_{g2}\rangle$ stands for an appropriate constant scaling factor. In this case, Eq.~(\ref{Eqn:formulaJorge}) still holds, replacing ${\rm SFR}_2$ with  $\langle{\rm SFR}_2\rangle$, and replacing
 $1/N$ with the exponent $1/N_{eff}$ given by  
\begin{equation}
1/N_{eff}=\delta/N << 1/N. \label{eq:neff}
\end{equation}

\begin{figure}
\includegraphics[width=0.5\textwidth]{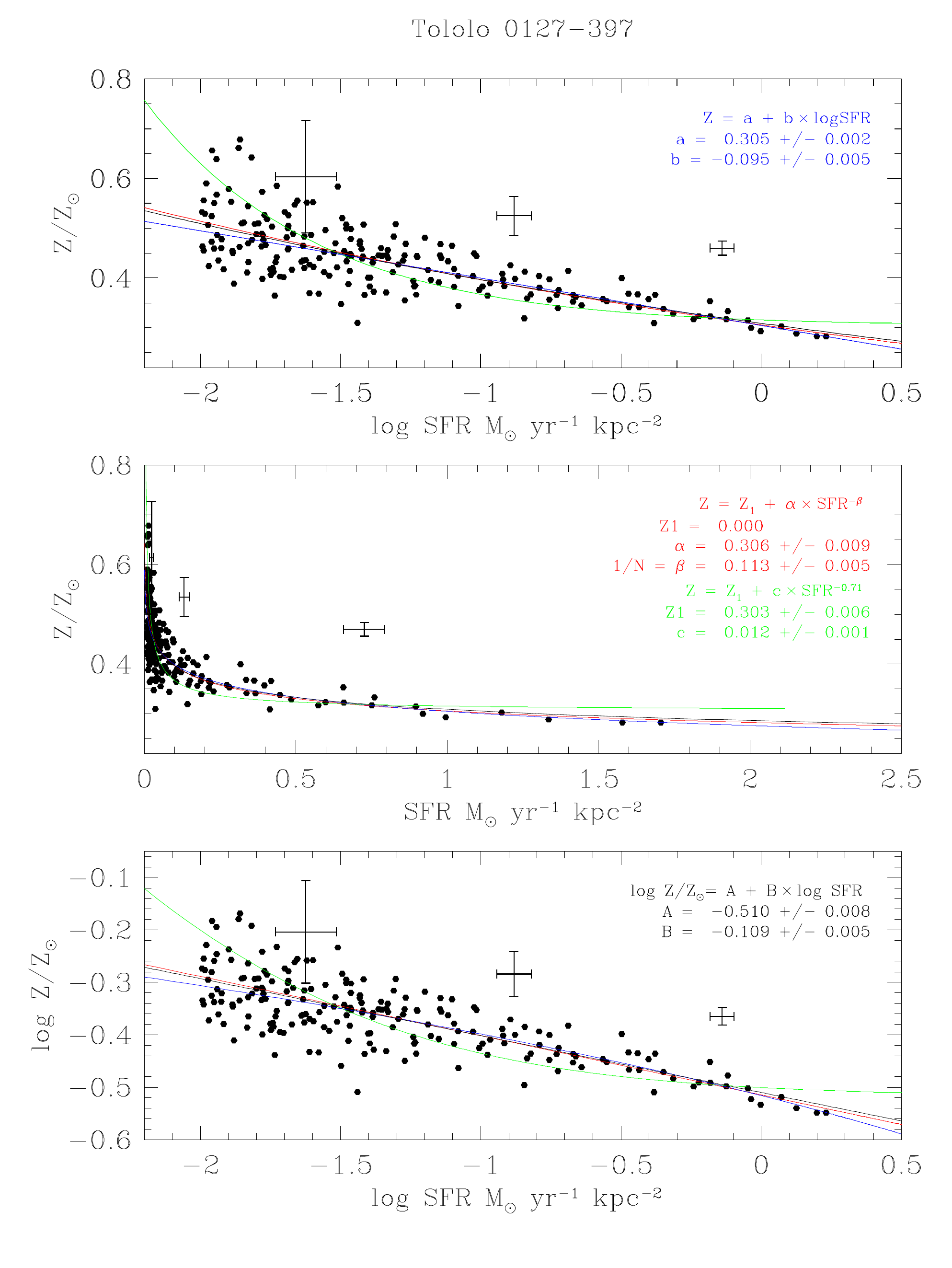}
\caption{
Metallicity ($Z$) versus SFR for Tololo\,0127-397 (black solid circles). The three 
panels show the same points and curves in three different representations;
$Z$ versus $\log{\rm SFR}$ (top), $Z$ versus SFR (middle) and 
$\log Z$ versus $\log{\rm SFR}$ (bottom). 
The red line is the best fit to Eq.~(\ref{Eqn:formulaJorge}), 
obtained by leaving the exponent, and the slope free 
(see the inset in the middle panel), but forcing $Z_1 \geq 0$. 
The green line also represents a fit to Eq.~(\ref{Eqn:formulaJorge}), but setting $N=1.4$ (corresponding to $1/N=0.71$).
The cyan line is the best linear fit to Eq.~(\ref{eq:openbox}) (see inset in the top panel).
Finally, the black line represents a linear fit in the log-log plane
as parameterized in Eq.~(\ref{Eq.:line}).
To convert 12+log(O/H) to metallicity in solar units, we adopted 
$12+\log(\mathrm{O/H})_\odot = 8.69$, from \citet{2009ARA&A..47..481A}. 
The value of the best-fitting parameters, together with
the formal uncertainties from the fitting algorithm, are also shown.
The green line provides a fit that it clearly worse than the 
fit represented by the other three lines. This happens only with some of the 
galaxies; often, the four mathematical models are indistinguishable (see
Fig.~\ref{Figure:fitjorgeMrk206}).
As in Fig.~\ref{Fig:lgoh12_lgsfr}, typical error bars (vertically shifted for clarity) are
shown at three representative SFR values.
}
\label{Figure:fitjorgeTol0127}
\end{figure}

\begin{figure}
\includegraphics[width=0.5\textwidth]{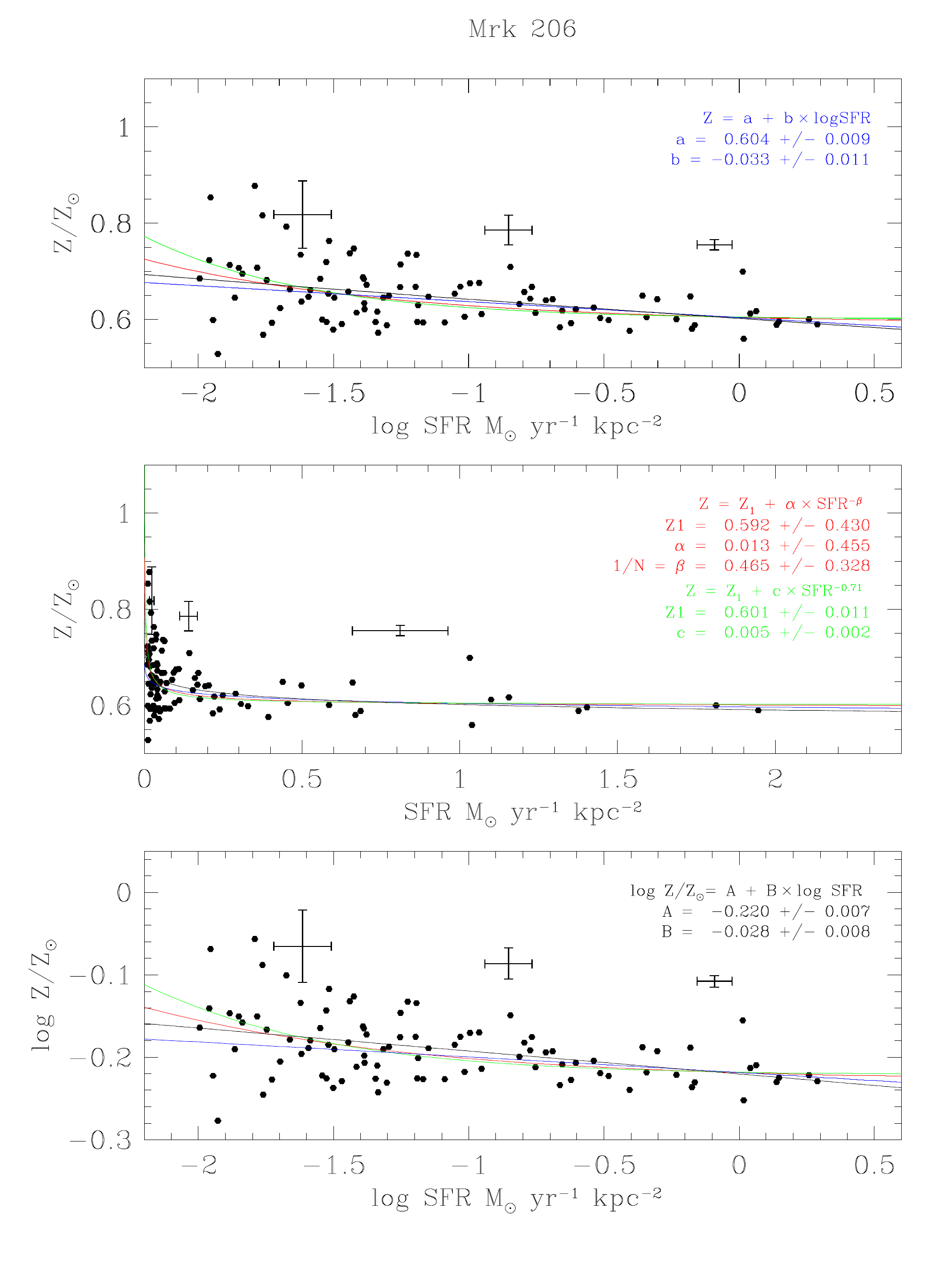}
\caption{Same as Fig.~\ref{Figure:fitjorgeTol0127} for Mrk\,206. 
In this case, all four mathematical models fit the data points satisfactorily.}
\label{Figure:fitjorgeMrk206}
\end{figure}

Examples of the fits to Eq.~(\ref{Eqn:formulaJorge}) are shown in Figures~\ref{Figure:fitjorgeTol0127} and \ref{Figure:fitjorgeMrk206}. They were performed using the MPFIT\footnote{A non-linear least-squares minimization procedure described by \citet{2009ASPC..411..251M}.} package implemented in  IDL. The fits were constrained to have $Z_1 \ge 0$. 
Table~\ref{table:slopes} contains the free parameters inferred from the fits for these and the remaining galaxies.
If $1/N$ is assumed to be a free parameter, the fits are excellent (see the red solid lines in Figs ~\ref{Figure:fitjorgeTol0127} and \ref{Figure:fitjorgeMrk206}). However, one obtains a value for $1/N$ very different from the expected value $1/N\simeq 0.71$ ($N=1.4$). If, on the other hand, $1/N$ is set to 0.71, the fit often worsens (compare the red and the green lines in Fig.~\ref{Figure:fitjorgeTol0127}; 
the 5 galaxies where this fit worsens are marked with an asterisk in Table~\ref{table:slopes}). Table~\ref{table:slopes} summarizes the values for $1/N_{eff}$ obtained in the various fits. Its mean and standard deviation are given by
\begin{equation}
1/N_{eff}=0.20\pm 0.19,
\end{equation}
which, assuming $N=1.4$, renders a value of $\delta= 0.28\pm 0.26$. This low value is consistent with the Monte Carlo simulation given in Appendix~\ref{appa}, which was carried out assuming that the two masses, $M_{g1}$  and $M_{g2}$, are similar ($M_{g1}$ only 50\,\%\ larger than $M_{g2}$), with their spaxel-to-spaxel variation being also similar. 

As one can see in Table~\ref{table:slopes}, the metallicity of the external gas, $Z_1$, tends to be zero, i.e., at the limit forced by the constrained fit. We interpret this fact as an indication that the observed correlation is consistent with the infall of very metal-poor gas, although the actual metallicity remains unconstrained.

%
%
%
%
%
%


\subsection{Self-contamination in open-box evolution}\label{sect:selfcont}

Another possibility that qualitatively explains the observed anti-correlation between metallicity and SFR is self-contamination due to star formation. If the whole galaxy disk starts off with similar amounts of metal-poor gas per unit surface, those regions where the stellar evolution proceeds faster will consume the gas earlier, and will become richer in metals. The differences between different regions can be due to the stochastic nature of the star-formation process. Under these hypotheses,  a simple close-box evolution model \citep[e.g.,][]{1980FCPh....5..287T} and the Kennicutt-Schmidt relation (Eq.~[\ref{eq:kslaw}]) render,
\begin{equation}
Z=Z_i-Y\,{{\ln 10}\over{N}}\log({\rm SFR}/{\rm SFR_i}),
\label{eq:openbox}
\end{equation}
where $Y$ designates the oxygen yield \citep[e.g.,][]{2002A&A...390..561M}, and $Z_i$ and ${\rm SFR_i}$ represent the initial metallicity and SFR, respectively. The above equation can be generalized to an open-box evolution, where part of the metal-enriched gas escapes from the region by star-formation-driven outflows. In this case, one must replace the yield $Y$ with the {\em effective} yield $Y_{eff}$
\citep[e.g.,][]{2015ApJ...810L..15S},
\begin{equation}
Y_{eff}=Y\,{{1-R}\over{1-R+W}},
\label{eq:yeff}
\end{equation}
where $R$ represents the so-called return fraction (i.e., the fraction of gas returned to the interstellar medium by supernova explosions and stellar winds per unit mass locked into stars), and $W$ is the mass loading factor, i.e., the outflow rate in units of the SFR. In practice, $Y_{eff} \ll Y$ since $R \ll 1$ and $W$ is often large (see the paragraph below).
The oxygen yield is believed to be $Y\simeq 0.004$ for metallicities smaller than the solar metallicity \citep{2002A&A...390..561M}. Then, using the value for the solar metallicity from \citet{2009ARA&A..47..481A}, $Y/Z_\odot\simeq 0.70$.

Examples of fits to the observed relation $Z$ versus SFR using Eq.~(\ref{eq:openbox}) are shown as the solid blue lines in Figs.~\ref{Figure:fitjorgeTol0127} and \ref{Figure:fitjorgeMrk206}. Given the scatter of the data, the fits are as good as the fits to Eq.~(\ref{Eqn:formulaJorge}) (the red solid lines) and the generic straight line in Eq.~(\ref{Eq.:line}) (the black solid lines). This statement holds for all the galaxies, with the free parameters of the fit given in Table~\ref{table:slopes}. The main observational constraint set by the parameters resulting from fitting Eq.~(\ref{eq:openbox}) is the need for intense  outflows, capable of carrying away most of the metals produced in each starburst. Specifically, the mean and standard deviation of $Y_{eff}/N$ is 
\begin{equation}
Y_{eff}/N= (0.040 \pm 0.025)\,Z_\odot,
\label{eq:yeffconstraint}
\end{equation}
so that using typical values for $R=0.3$ 
\citep[e.g.,][]{2014A&ARv..22...71S}, $Y/Z_\odot\simeq 0.70$, and $N=1.4$,
Eq.~(\ref{eq:yeff}) renders,
\begin{equation}
W=11.3\pm 6.7.
\label{eq:massloading}
\end{equation}
This value for $W$ is large, but it is consistent with some of the large values found in local dwarf galaxies 
\citep[see ][and references therein]{2005ARA&A..43..769V,2017ApJ...834..181O}. Moreover, $W$ refers to the mass expelled from each spaxel, which does not imply that the gas returns back all the way to the inter-galactic medium. If it only reaches the circum-galactic medium, it may fall back onto the galaxy disk as part of a galactic fountain \citep[e.g.,][]{2017ASSL..430..323F,2017ASSL..430...95R,2017MNRAS.468.4170M}. Stellar feedback processes of the required magnitude should be accompanied by shells and bubbles of swept material visible in emission lines \citep[e.g.,][]{2012MNRAS.424.2442S,2017ApJ...834..181O}.

\section{Are metallicity variations needed to explain the anti-correlation?}\label{Sect:bias}

The metallicities are inferred from the N2 index, whereas the SFR is derived from the H$\alpha$ flux (Sect.~\ref{Sect:Data}). The two quantities are proxies for the true metallicity and the true SFR, respectively, and they are subject to systematic errors. Thus, spatial  variations in the  properties of the local ionizing flux may induce a correlation between N2 and H$\alpha$ \citep[e.g.,][]{2014ApJ...797...81M}, which could be (mis)interpreted as a correlation between metallicity and SFR. This section is devoted to arguing  that the observed spatial variation of line fluxes do require a change in metallicity, so that interpretations such as those provided in Sects.~\ref{sect:mixing} and \ref{sect:selfcont} are needed. In other words, they cannot be produced only by systematic errors resulting from the variation in mass and age of the stellar cluster that ionize each spaxel, and the change in ionization parameter that results.

As we explain in Sect.~\ref{Sect:sizes}, the spatial resolution of the IFU maps used in our work is too coarse to resolve individual \hii\ regions. Therefore, each typical spaxel integrates the spectrum of one or several full \hii\ regions. This hypothesis is underlying the following discussion.

\subsection{Alternative estimate of the metallicities}\label{sect:hiic}
The use of the N2 index as a proxy for metallicity does not consider differences in physical parameters between \hii\ regions with the same metallicity. In order to check whether the existing differences can mimic an anti-correlation between SFR and O/H, we have used another alternative method that considers differences in N/O and ionization parameters to infer the oxygen abundance. The code HII-CHI-MISTRY (HIIC) by \citet{2014MNRAS.441.2663P} compares line ratios of a number of selected emission lines with the predictions of a grid of CLOUDY \citep{2013RMxAA..49..137F} photoionization models having different physical conditions, including different N/O and ionization.  Changes in N/O account for peculiarities in the chemical composition of the \hii\ region, whereas the ionization parameter is related to the age and mass of the ionizing stellar cluster as well as to the geometry of the gas. HIIC has been thoroughly tested, providing results identical to the direct method within around 0.1 dex in a large variety of objects \citep{2014MNRAS.441.2663P}, including those with SFR and metallicity similar to our galaxies \citep[e.g.,][Fig.~3]{2016ApJ...819..110S}.

\begin{figure}
\includegraphics[width=0.48\textwidth]{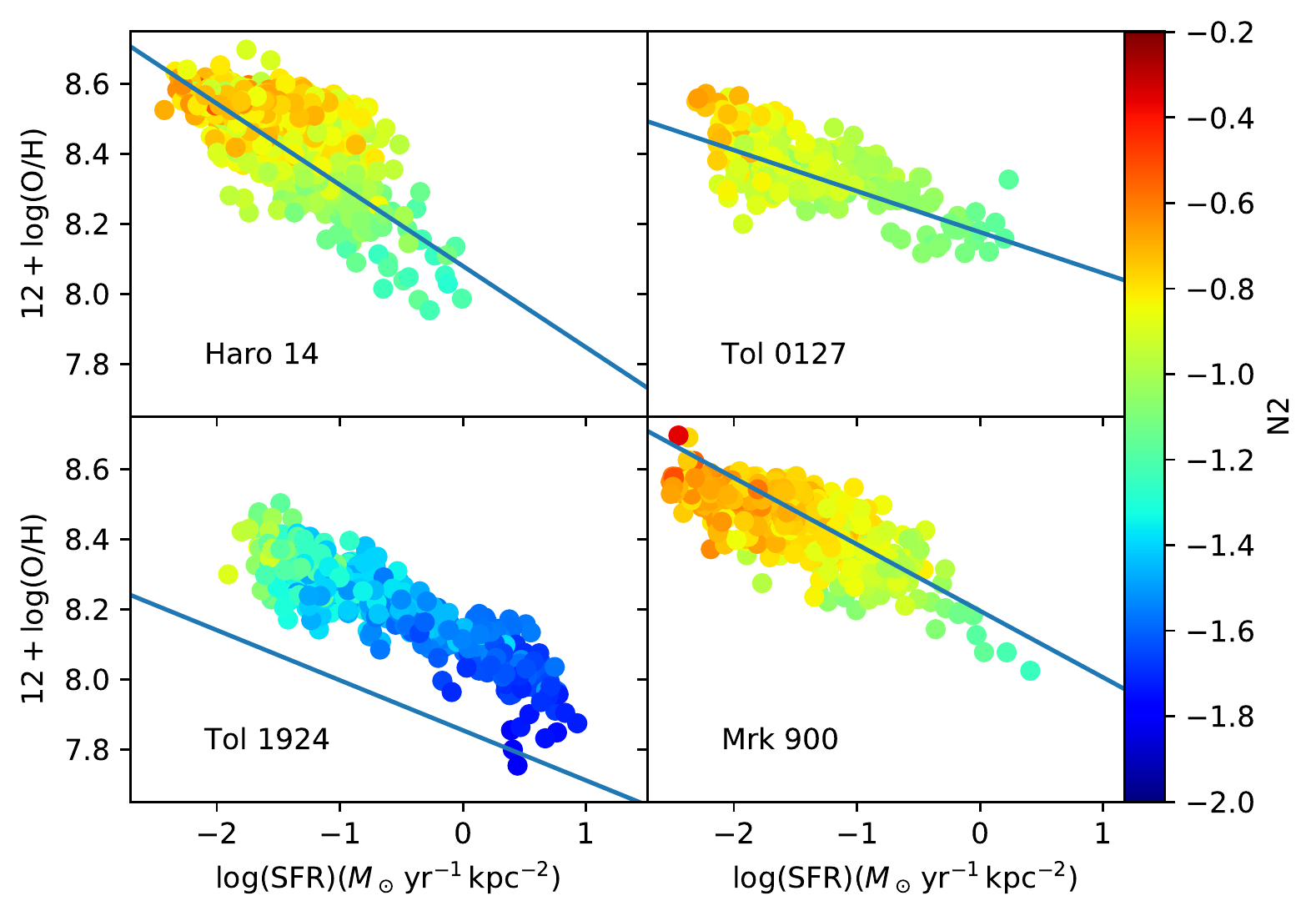} 
\caption{
12+log(O/H) versus SFR for  four VIMOS galaxies (see the labels), with the oxygen abundance provided by HIIC \citep{2014MNRAS.441.2663P}. HIIC considers changes in N/O and ionization parameter within the galaxies. The anti-correlation between O/H and SFR that appears in Fig.~\ref{Fig:lgoh12_lgsfr} remains here. The solid lines show the linear fits to the anti-correlation in Fig.~\ref{Fig:lgoh12_lgsfr}   ($k$ and $m$ in Table~\ref{table:slopes}). The different spaxels are color-coded according to N2, as indicated by the color bar.
The formal error bars for 12+log(O/H) turn out to be around 0.04\,dex, which are inferred by HIIC from the difference between the model and the observed line fluxes. 
}
\label{fig:HIICM_Ha1}
\end{figure}

HIIC can use different emission lines depending on availability. In the case of the galaxies observed with PMAS, the list of lines includes [OII]$\lambda$3727,  [OIII]$\lambda$5007, [NII]$\lambda$6583, [SII]$\lambda$6716,6731, as well as H$\alpha$ and H$\beta$. The VIMOS spectra do not include [OII], and this particular line cannot not be used in the metallicity estimate. Figure~\ref{fig:HIICM_Ha1} shows the scatter plot 12+log(O/H) versus SFR for the four VIMOS galaxies chosen to compute O/H with HIIC. The four examples were selected because they show a clear anti-correlation when O/H is computed using N2 (Fig.~\ref{Fig:lgoh12_lgsfr}).  The same anti-correlation is maintained even when this other more accurate technique for measuring  O/H  is used. Figure~\ref{fig:HIICM_Ha1} includes the linear fit to the anti-correlation obtained when  $12+\log({\rm O/H})$ is based on N2 ($k$ and $m$ in Table~\ref{table:slopes}). The slopes of the anti-correlation inferred from N2 agree remarkably well with the distribution of points obtained using HIIC (Fig.~\ref{fig:HIICM_Ha1}).

\begin{figure}
\includegraphics[width=0.48\textwidth]{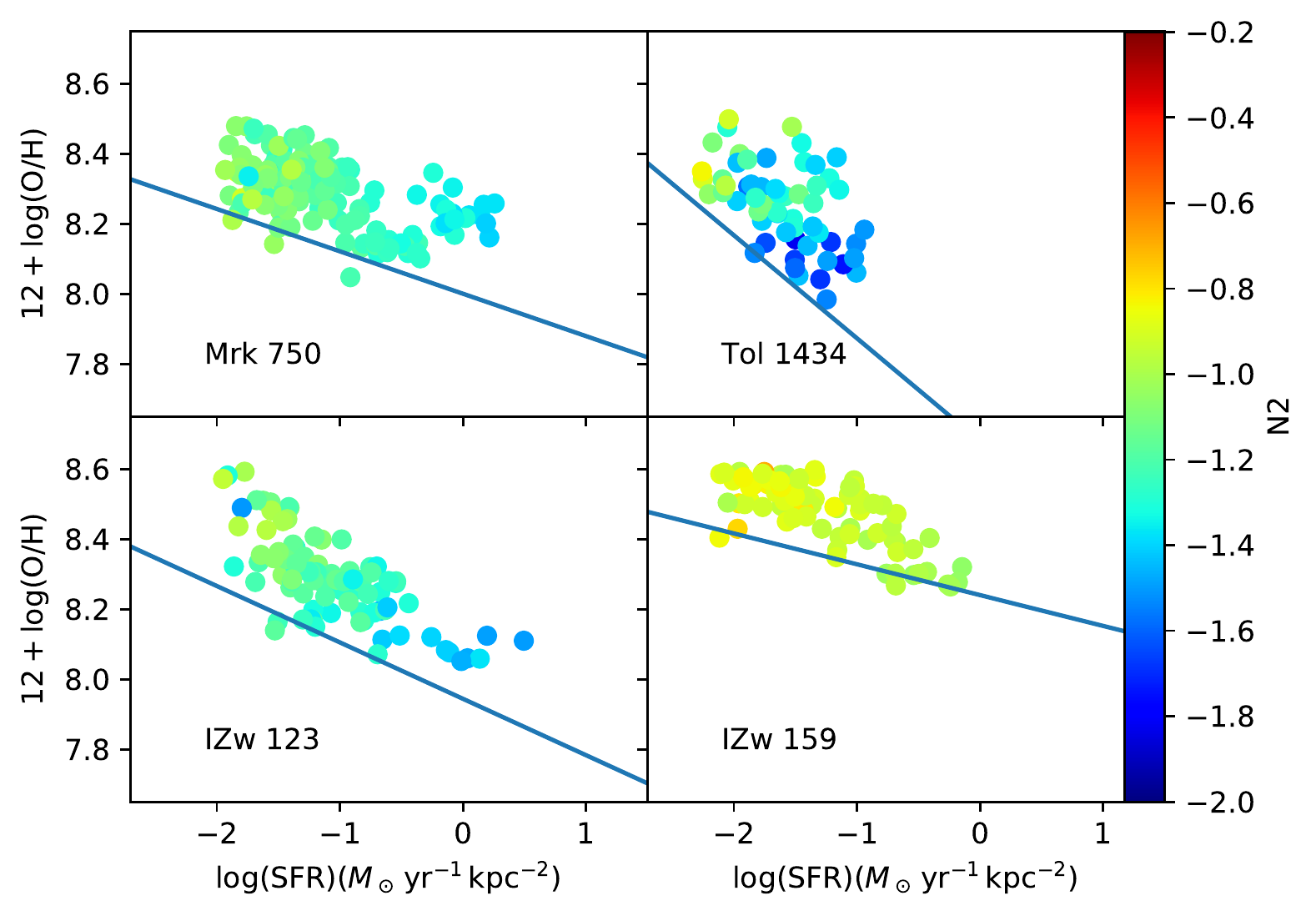} 
\caption{
Same as Fig.~\ref{fig:HIICM_Ha1}, for four PMAS galaxies (see the labels).  In this case the line [OII]$\lambda$3727 is included in the O/H determination. The anti-correlation between O/H and SFR that appears in Fig.~\ref{Fig:lgoh12_lgsfr} remains here. The solid lines show the linear fits to the anti-correlation in Fig.~\ref{Fig:lgoh12_lgsfr}   ($k$ and $m$ in Table~\ref{table:slopes}).
The formal error bars provided by HIIC turn out to be around 0.05\,dex.
}
\label{fig:HIICM_Ha2}
\end{figure}

Similar agreement between N2-based and HIIC-based metallicity is found when comparing PMAS galaxies  (Fig.~\ref{fig:HIICM_Ha2}), which have less spatial resolution but include [OII]. In the case of the PMAS galaxies, HIIC has a direct observational constraint on the ionization through the ratio [OII]/[OIII].  As for the VIMOS galaxies, these examples were chosen to apply HIIC because they show a clear anti-correlation when O/H is computed using N2.
The symbols in Figs.~\ref{fig:HIICM_Ha1} and \ref{fig:HIICM_Ha2} have been color-coded according to N2, which clearly  shows an increase with increasing metallicity, thus explaining why the two methods of estimating O/H depict the same anti-correlation between metallicity and SFR.  

\begin{figure}
\includegraphics[width=0.48\textwidth]{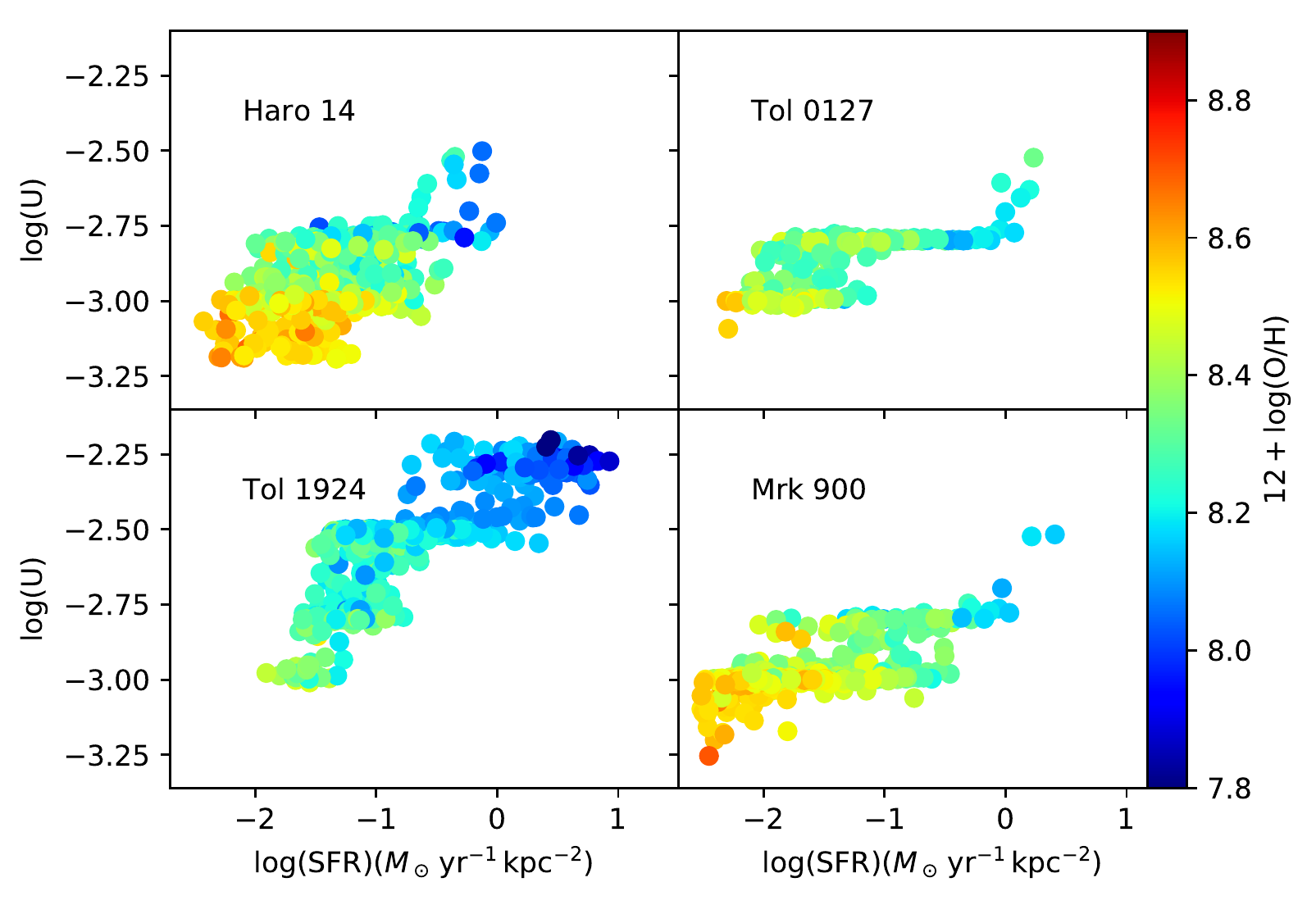} 
\includegraphics[width=0.48\textwidth]{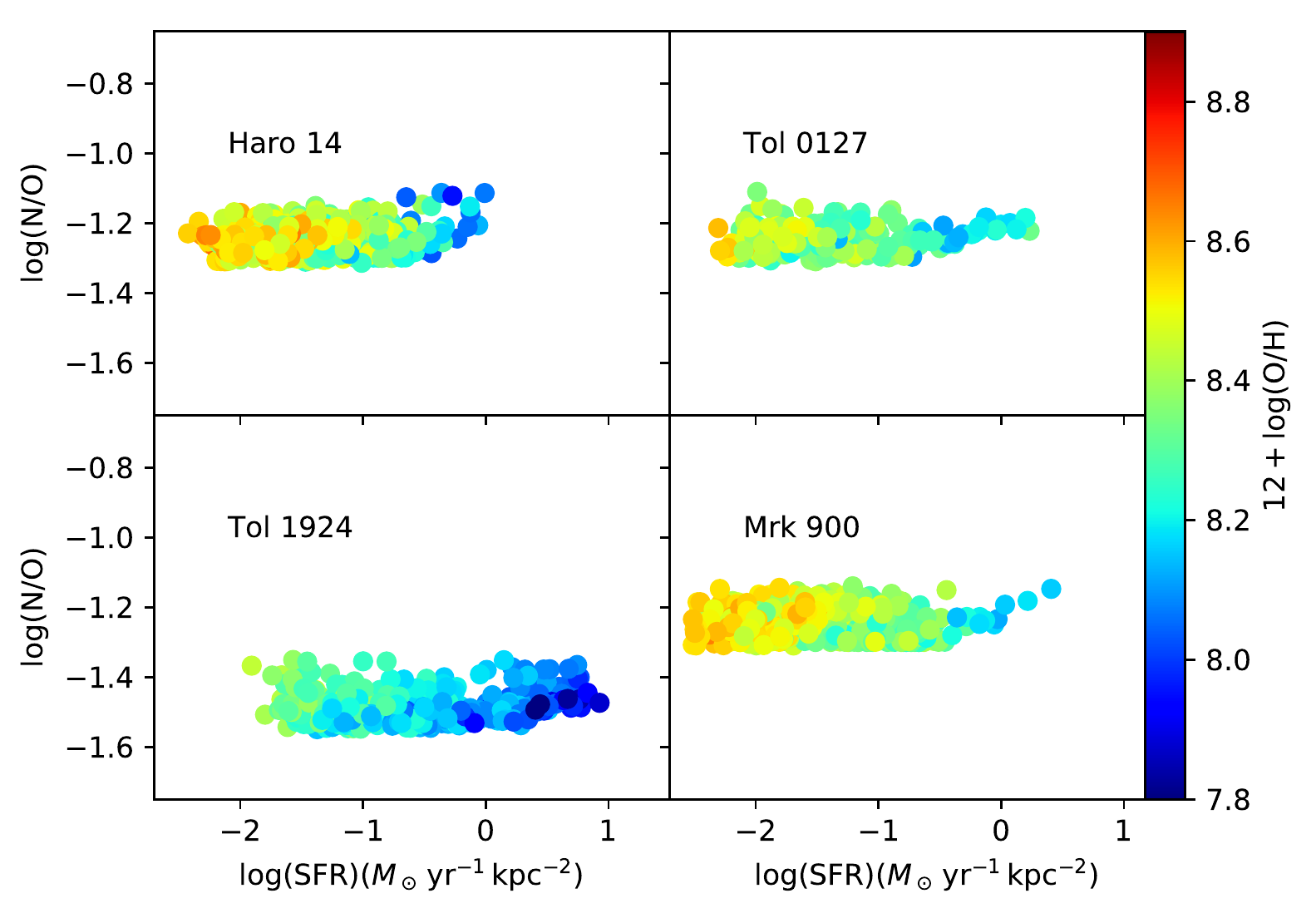} 
\caption{
Top panels: Variation of the ionization parameter (U) with SFR as inferred  from HIIC for the galaxies in Fig.~\ref{fig:HIICM_Ha1}. There is a significant increase of U with increasing SFR. 
Bottom panels: Variation of N/O with SFR as inferred  from HIIC. The ratio remains rather constant across the galaxy.
The different spaxels are color-coded according to their $12+\log({\rm O/H})$ from HIIC, as indicated by the color bar. The range of ordinates corresponds to 1.1dex, which is the same range used in Figs.~\ref{fig:HIICM_Ha1} and \ref{fig:HIICM_Ha2}.
The  formal error bars provided by HIIC for $\log({\rm U})$ and $\log({\rm N/O})$  turn out to be around 0.01\,dex and 0.03\,dex, respectively.
}
\label{fig:HIICM_NO}
\end{figure}

Figure~\ref{fig:HIICM_NO} illustrates how the ionization parameter (U) and N/O vary with SFR according to HIIC. The galaxies show a significant increase of U with increasing SFR, whereas N/O remains rather constant across the galaxy. Figure~\ref{fig:HIICM_NO} includes VIMOS galaxies, but the behavior is very similar for PMAS objects.

To summarize, even when the within-galaxy spread in N/O and ionization parameter is considered, the anti-correlation between metallicity and SFR remains.

\subsection{Comparison with observed \hii\ regions}\label{Sect:sebastian} 

The argumentation will be based on the position of the individual spaxels of the analyzed galaxies in the  BPT diagram \citep[O3 versus N2; ][]{1981PASP...93....5B}. The position of the data points in this plane is very sensitive to both the ionizing spectrum and the metallicity of the emitting gas \citep[e.g.,][]{2003MNRAS.346.1055K,2004MNRAS.348L..59P,2010MNRAS.403.1036C}. On the other hand, the location of data points on the BPT diagram is almost insensitive to reddening or mis-calibration, since the lines in each ratio are very close in wavelength and any multiplicative factor affecting the observed fluxes cancels out. 
We will argue that our galaxies occupy a region of the BPT plane consistent with a spread in metallicity, both comparing them with previous observations (this section) 
and with photoionization models (Sect.~\ref{Sect:photoio}).

\begin{figure*}
\begin{center}
\includegraphics[width=0.9\textwidth]{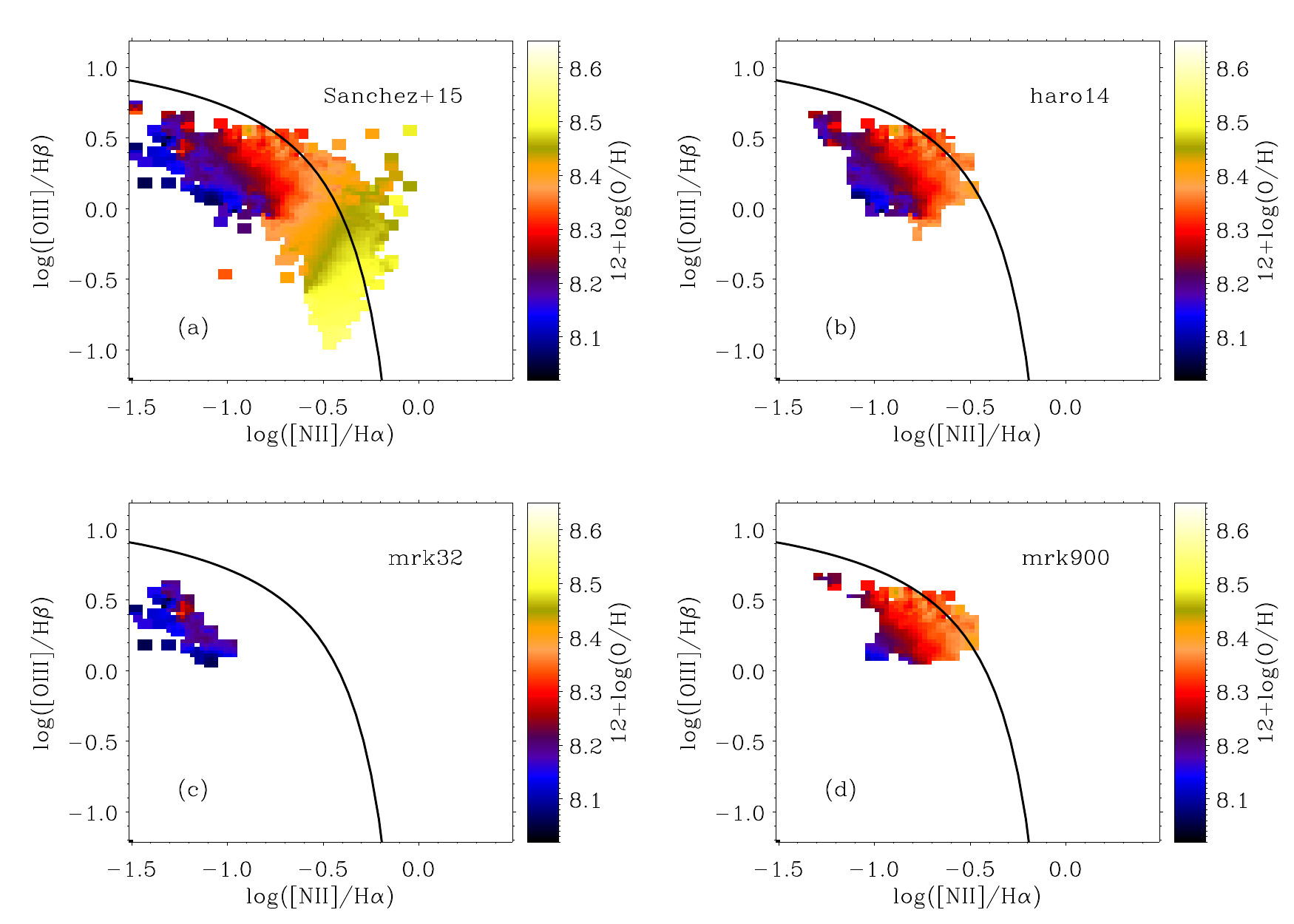}
\end{center}
\caption{
(a) Mean metallicity of the \hii\ regions corresponding to each point in the BPT 
plane (log([OIII]$\lambda$5007/H$\beta$) versus log([NII]$\lambda$6583/H$\alpha$)),
as measured from $\sim$5000 \hii\ regions from the CALIFA survey \citep{2015A&A...574A..47S}.
The color bar  gives the equivalence between color
and metallicity, with each single color corresponding to a range of metallicity of $\sim\,$0.05~dex.
(b) Same as (a) but showing only the area where the spaxels of Haro~14 appear. Note 
that the area shows several colors, and so it covers several metallicities.
(c) Same as (b) for Mrk~32. In this case, all the spaxels of the galaxy are concentrated 
in an area with a single color, implying homogeneity in metallicity. However, many of the 
spaxels of this particular galaxy lie outside the range of log([OIII]$\lambda$5007/H$\beta$) 
and log([NII]$\lambda$6583/H$\alpha$) sampled by \citet[][]{2015A&A...574A..47S}. 
(d) Same as (b) for Mrk~900, which also shows metallicity variations.
The solid line, shown for reference, is the same in all the panels. It corresponds to 
the divide worked out by \citet{2003MNRAS.346.1055K} to separate star-forming galaxies 
from galaxies with AGN-like activity.   
}
\label{Fig:bpt_sanchez}
\end{figure*}

\citet{2015A&A...574A..47S} use a set of 5000 \hii\ regions from the CALIFA survey \citep[which includes galaxies of all masses and Hubble types;][]{2012A&A...538A...8S}, to explore the distribution of observed \hii\ regions on the BPT diagram. Figure~\ref{Fig:bpt_sanchez}a shows the area of the BPT plane covered by these \hii\ regions, and it is color-coded according to the mean metallicity of the regions that appear in that particular location of the plane. One can see a systematic variation which, among other things, indicates that the N2 index does indeed increase with increasing metallicity (as in Eq.~[\ref{Eqn:HaNIImet}]). The metallicities in  \citeauthor{2015A&A...574A..47S} were computed using the strong-line method by \citet{2012MNRAS.424.2316P}, which compares a number of emission lines with those observed in \hii\ regions with metallicities determined through the direct method. Their values were also cross-checked with the O3N2 calibrator \citep[e.g.,][]{2004MNRAS.348L..59P}. The two methods are independent of the N2 method used in our metallicity estimate.

Using the \hii\ regions in CALIFA as reference, we find that the individual spaxels of most of our galaxies cover a region of the BPT plane characterized by more than one metallicity. Examples are shown in Fig.~\ref{Fig:bpt_sanchez}b and \ref{Fig:bpt_sanchez}d. These two panels display the metallicity mapping of the BPT by \citeauthor[][i.e., the same as Fig.~\ref{Fig:bpt_sanchez}a]{2015A&A...574A..47S},  but only in the region containing the spaxels of our galaxies. It is clear how the observed region corresponds to a range of metallicities. Figure~\ref{Fig:bpt_sanchez}c shows another galaxy (Mrk~32) where all the spaxels are concentrated in a small region of the BPT plane, and so they likely correspond to a single metallicity in the mapping. However, even in this second case, one cannot discard more than one metallicity, since many of the galaxy's spaxels appear outside the BPT region covered by the CALIFA sample. From our set of 16 galaxies, 12 galaxies show a clear spread in metallicity  (Haro~11, Haro~14, IZw~123, IZw~159, Mrk~206, Mrk~407, Mrk~409, Mrk~750, Mrk~900, Mrk~1418, Tol~127, and Tol~1937), whereas the remaining 4 objects do not (Mrk~32, Mrk~475, Tol~1434, and Tol~1924). 

\begin{figure}
\includegraphics[width=0.45\textwidth]{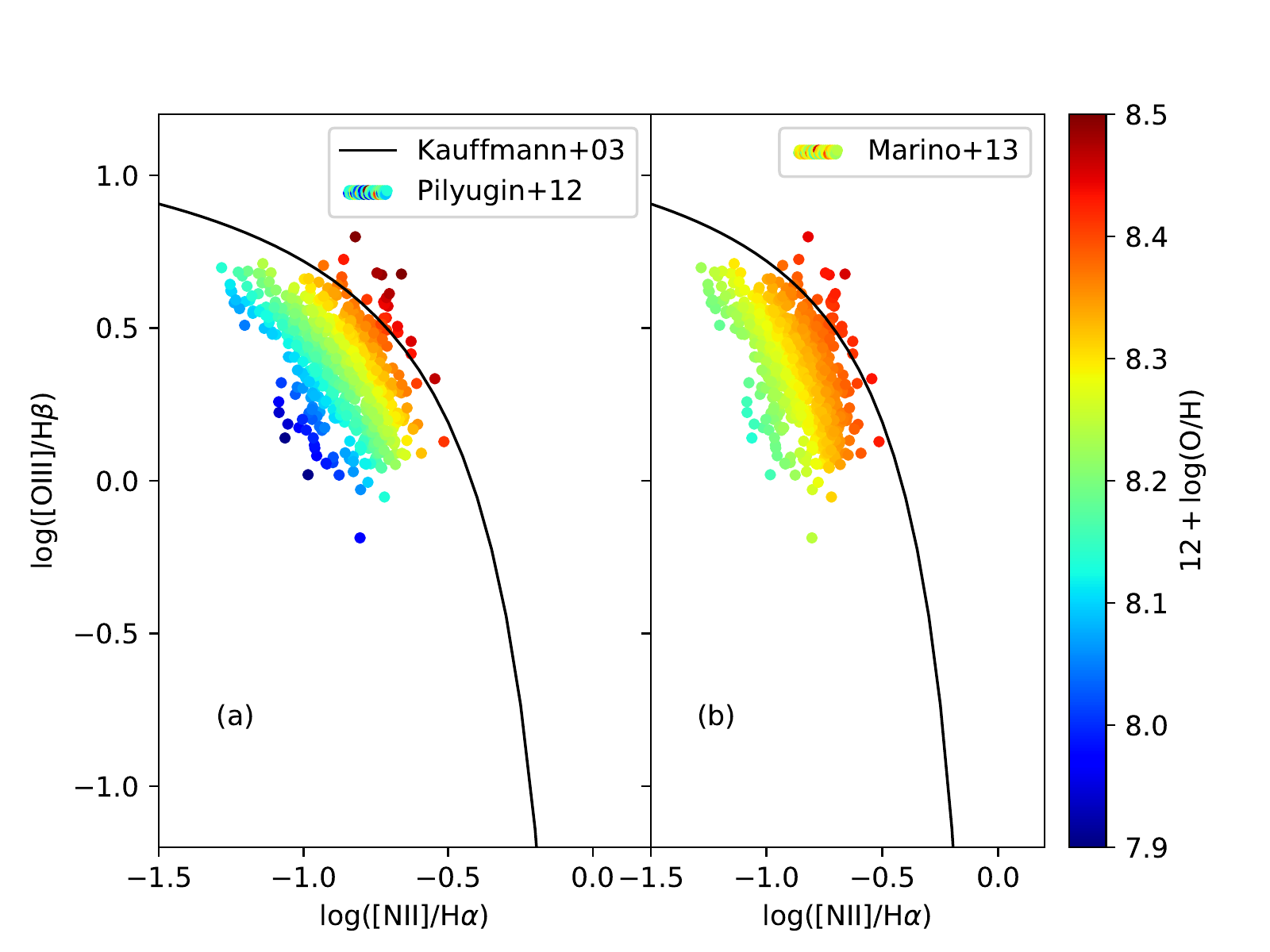}
\caption{
(a) BPT diagram of Haro~14 color-coded with the metallicity of galaxies from \citet{2012MNRAS.424.2316P} having the same N2 and O3 as the individual spaxels in Haro~14. Note the need for a range of metallicities to explain the observed range of N2 and O3. (b) Same as (a) with the metallicity obtained from the galaxies used by \citet{2013A&A...559A.114M}. The solid line corresponds to the divide below which star-forming galaxies appear \citep{2003MNRAS.346.1055K}.
}
\label{fig:pilyugin}
\end{figure}

The above exercise was repeated using other empirical datasets to assign metallicities to each point of the BPT plane. Specifically, the galaxies employed for calibration by \citet{2012MNRAS.424.2316P} allowed us to fit  with a two-dimensional polynomial the scatter plot $12+\log({\rm O/H})$ versus N2 and O3. The fit assigns a metallicity to every pair N2 O3, which we then apply to the values observed in our galaxies to infer the expected $12+\log({\rm O/H})$. The result of the exercise for Haro~14 is included in Fig.~\ref{fig:pilyugin}a, which shows how its N2 and O3 cover a region spanning a range of metallicities. The same conclusion is drawn when fitting the two-dimensional polynomial to the galaxies employed by \citet{2013A&A...559A.114M}; see Fig.~\ref{fig:pilyugin}b.  It is important to realize that the data used for calibration are based on the direct method where differences of physical parameters (electron density and temperature) and chemical composition (N/O) are self-consistently taken into account in the determination of metallicity. 

In short, most of the galaxies analyzed in the paper have log([OIII]$\lambda$5007/H$\beta$) and  log([NII]$\lambda$6583/H$\alpha$) values spreading over an area of the BPT diagram that corresponds to \hii\ regions having a range in metallicities.

\subsection{Comparison with numerical models}\label{Sect:photoio}

In the previous section we compare the observed line ratios with observed \hii\ regions. Here we compare them with model \hii\ regions. 
We wanted to check whether the line ratios observed in the galaxies are compatible with  model \hii\ regions having a realistic range of physical parameters but a single metallicity. The photoionization models by \citet{2013ApJ...774..100K} were meant to explain the evolution of the galaxies on the BPT diagram when the redshift changes from 0 to 3. These models provide a fair reference to compare with since they scan a wide range of physical parameters which (a) reproduce the sequence followed by the local star-forming galaxies, and (b) include the extreme  conditions that characterize high redshift objects. High redshifts galaxies are systematically displaced in the BPT with respect to local galaxies \citep[e.g.,][]{2014ApJ...795..165S,2017ApJ...835...88K}, and the need for the models to reach this region of the BPT plane guarantees sampling a large range of physical conditions.   
The position of the model galaxies on the BPT depends on the ionizing radiation field, the geometry of the emitting gas, the electron temperature and density and, finally, on the metallicity. The radiation field itself depends on the IMF (Initial Mass Function) and on the age and metallicity of the stellar population.  \citeauthor{2013ApJ...774..100K} use the Mappings photoionization code \citep[e.g.,][]{1985ApJ...297..476B,1993ApJS...88..253S} to model O3 and N2 in the \hii\ regions surrounding a large set of ionizing stellar clusters with different properties. They assume a Salpeter IMF with an upper mass limit of 100\,$M_\odot$, but this choice of IMF has negligible impact on the emission line ratios. The stellar model atmospheres include winds of massive stars and WR \citep{2001A&A...375..161P,1998ApJ...496..407H},  and Geneva evolutionary tracks \citep{1994A&AS..103...97M}. The clusters have continuous star formation extending over 4 Myr. Dust in the gas-phase is included, metals are partly depleted into the dust grains, and $\alpha$-element abundances are assumed to scale with metallicity. The nebulae are radiation bounded and isobaric, with an electron density that varies from 10 to $10^3$\,cm$^{-3}$. The variation of the metallicity range with redshift is taken from cosmological numerical simulations of galaxy formation \citep{2011MNRAS.415...11D,2011MNRAS.416.1354D}.
The photoionization models thus constructed have been thoroughly tested, and they reproduce the properties of  the local star-forming galaxies (e.g., Kewley et al. 2001a; Levesque et al. 2010).    

\begin{figure}
\includegraphics[width=0.45\textwidth]{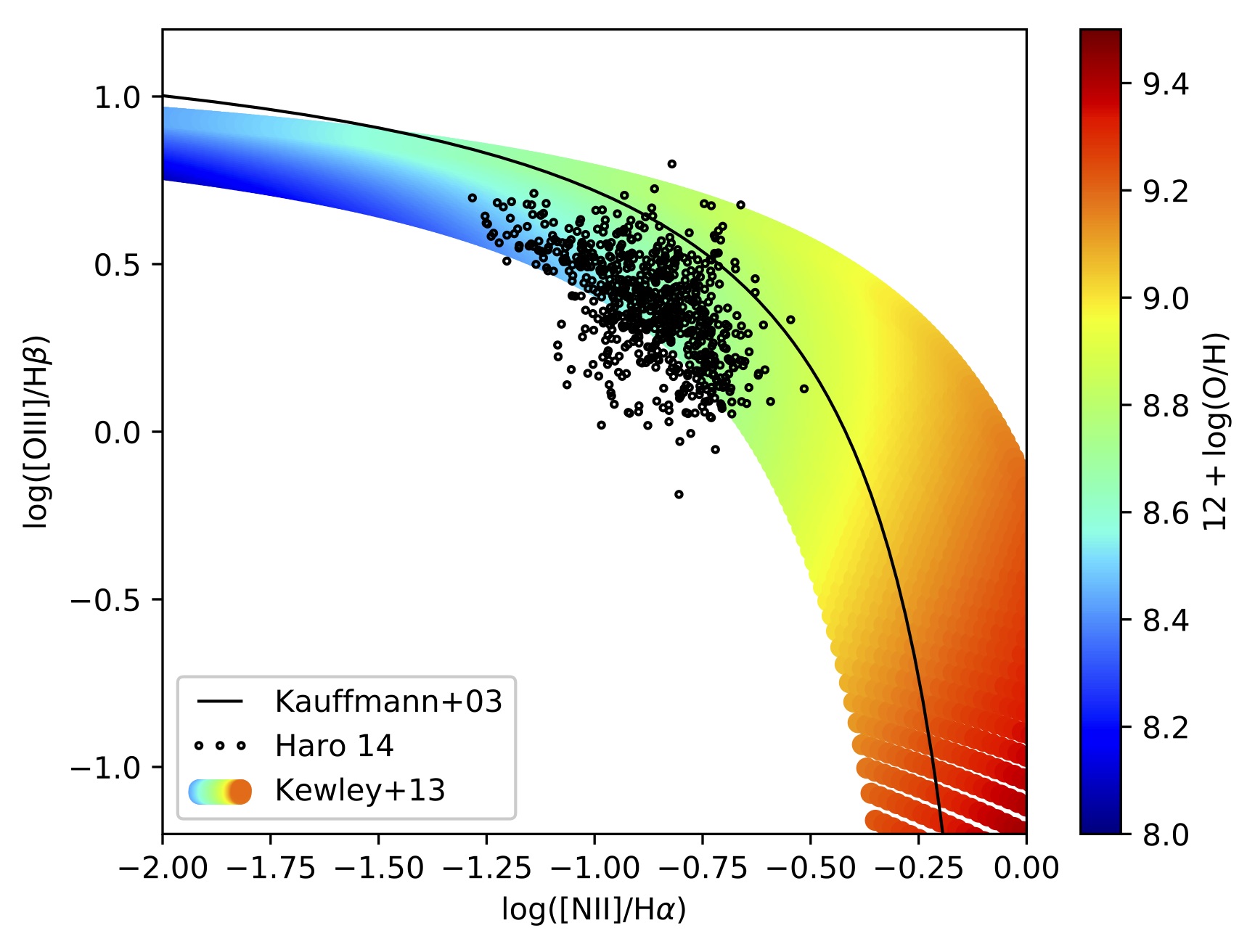}
\caption{
BPT diagram for Haro~14 (the open circles) overlaid on the numerical photoionization models by \citet[][]{2013ApJ...774..100K}.  The models are color-coded according to their gas-phase metallicity, as indicated by  the color bar.  The observed points occupy a region of the model BPT plane characterized by a range of metallicities of around 0.3~dex. The solid line separares star-forming galaxies from galaxies with AGN activity, according to \citet{2003MNRAS.346.1055K}. It is shown for reference and is the same as in Figs.~\ref{Fig:bpt_sanchez} and \ref{fig:pilyugin}.  
}
\label{fig:bpt3}
\end{figure}

At each redshift $z$,
the models by \citeauthor{2013ApJ...774..100K} predict a  mean relation between O3 and N2 given by,
\begin{equation}
{\rm O3}=1.1+0.003\,z+\frac{0.61}{{\rm N2}+0.08-0.1833\,z},
\label{eq:kewley1}
\end{equation}
with the metallicity varying along the sequence as,
\begin{equation}
12+\log({\rm O/H})=8.97+0.0663\,z-({\rm O3-N2})\,(0.32-0.025\,z).
\label{eq:kewley2}
\end{equation}
Thus, Eqs.~(\ref{eq:kewley1}) and (\ref{eq:kewley2}) provide an implicit relation that gives $12+\log({\rm O/H})$ as a function of N2 and O3.  Specifically, the relation gives the metallicity to be expected in a typical galaxy having specific values for N2 and O3.  When the redshift varies from 0 to 3, the relation covers a substantial fraction of the BPT plane, shown as the colored region in Fig.~\ref{fig:bpt3}. 

Figure~\ref{fig:bpt3} shows the position of the spaxels of Haro~14 overlaid on the BPT diagram of these numerical photoionization models. The observed points do not follow ridges of constant metallicity. Instead, they appear in a region that corresponds to a range of metallicities of around 0.3 dex. The range of metallicity is similar to the range covered by Haro~14 on the observational BPT  (Fig.~\ref{Fig:bpt_sanchez}, top right panel) as well as in our metallicity estimate (Figs.~\ref{Fig:lgoh12_lgsfr} and \ref{fig:HIICM_Ha1}).

Haro~14 is just an example, but most galaxies have O3 and N2 spreading over an area of the BPT diagram that corresponds to \hii\ regions with a range in metallicities.  This need is not so clear in other cases like Mrk~32, which is the same conclusion reached when comparing the spread in O3 and N2 with observations (Fig.~\ref{Fig:bpt_sanchez}c).

%
\section{Discussion and Conclusions}
\label{Sect:Discussion}

Based on the analysis of IFU data from 14 dwarf star-forming galaxies, we find a clear anti-correlation between the H$\alpha$ flux in each spaxel and the value of N2. The anti-correlation is not due to noise in the spectra (Sect.~\ref{Sect:Data}). We attribute it to the existence of a relation between the local SFR (as traced by H$\alpha$) and the metallicity of the gas used to form stars (traced by N2). Each increase of the SFR by one order of magnitude is associated with a  drop in metallicity of 0.15\,dex (Eq.~[\ref{eq:slope}]). 

The anti-correlation is set by local variations of N2 and H$\alpha$, because it remains even if the large scale variation of these two quantities is subtracted out from the N2 and H$\alpha$ maps (Sect.~\ref{Sect:nolarge}). Thus, the anti-correlation is not produced by the radial drop in metallicity expected in spirals. Since the angular resolution of the individual maps is between 50 and 500 pc, the anti-correlation refers to integrated properties of entire \hii\ regions. It does not arise from the internal (sub-)structure of the individual \hii\ regions.

The relation between N2 and metallicity, and H$\alpha$ flux and SFR, are obtained from semi-empirical calibrations. The uncertainties inherent to the two calibrations may be correlated, thus faking a relation between metallicity and SFR. In order to discard this potential bias, 
a number of tests were carried out. We compute the metallicity using a technique which includes possible local variations in ionization parameter and N/O \citep[][]{2014MNRAS.441.2663P}, and the anti-correlation between metallicity and SFR remains (Sect.~\ref{sect:hiic}).
We also compare the position of the galaxies in the so-called BPT diagram with the position of many observed \hii\ regions having a large variety of metallicities  \citep[from][]{2015A&A...574A..47S,2012MNRAS.424.2316P,2013A&A...559A.114M}. Our galaxies  occupy an area corresponding to a range in metallicities of around 0.2~dex  (Sect.~\ref{Sect:sebastian}).
In addition, we compare the observations with photoionizatiom models \citep{2013ApJ...774..100K}, which suggest that a spread in metallicities is required to account for the location of the spaxels in the BPT plane.

We  studied plausible physical scenarios that create a local anti-correlation between metallicity and surface SFR. Thus, the observed anti-correlation is consistent with local infall of metal-poor gas, provided that the external gas mass is comparable to the gas mass previously existing in each spaxel (Sect.~\ref{sect:mixing}). Within this scenario, reproducing the observations requires some random variability in the mass of external gas contributing to each spaxel. It also demands the metallicity of the external gas to be much smaller than the solar metallicity. 
In addition, we explore the effect of self-enrichment of the observed \hii\ regions with metals produced in past star-formation episodes. In this case the anti-correlation emerges because those regions very active in the past are now quiescent and metallic, with the opposite happening with the ones that were less effective (Sect.~\ref{sect:selfcont}). If such self-enrichment is invoked as an explanation, then it comes with very intense outflows, having mass loading factors in excess of 10 (Eq.~[\ref{eq:massloading}]).
Even though we do not treat it explicitly in the main text, a combination of external metal-poor gas inflow plus self-enrichment should also render the observed anti-correlation.    

Numerical simulations of galaxies formed in a cosmological context show how this type of local anti-correlation naturally results from the accretion of cosmological metal-poor gas that triggers star-formation bursts \citep{2016MNRAS.457.2605C, 2014MNRAS.442.1830V}. Interpreting the observed relation in these terms is very appealing. However, it is not devoid of uncertainty since it involves either inefficient gas mixing in the galaxies showing the relation, or the  recent accretion of metal-poor gas. The gas in the disk is expected to mix in a timescale of the order of the rotational period or smaller \citep[e.g.,][]{2002ApJ...581.1047D, 2012ApJ...758...48Y,2015MNRAS.449.2588P}. Gas accretion, on the other hand, is theoretically predicted to fuel star formation in disk galaxies \citep[e.g.,][]{2009Natur.457..451D,2012RAA....12..917S}, but the observational evidence of this process is still rather indirect \citep{2014A&ARv..22...71S,2017ASSL..430...67S}. If the local anti-correlation between metallicity and SFR turns out to be universal, it would provide a strong support for the cosmological gas accretion predicted by these numerical models. 

Curiously, the slope of the relation between metallicity and log(SFR) is similar to the slope for the cosmological evolution of these two quantities. The mean gas metallicity of the star-forming galaxies decreases by 0.15 dex when their redshift changes from 0 and 1 \citep[][]{2007ASSP....3..435K}, while the SFR of the Universe increases by approximately one order of magnitude during this period \citep[e.g.,][]{2014ARA&A..52..415M}. The same slope is obtained from the variation with redshift of the metallicity and SFR in a group of relatively nearby low-mass irregular galaxies \citep[redshift $< 0.05$;][]{2013MNRAS.432.1217P}. The agreement may be coincidental, but we point it out because the change in the cosmic SFR probably follows the evolution of the cosmological gas accretion \citep[e.g.,][]{2013MNRAS.435..999D,2014A&ARv..22...71S}, and this cosmological process seems to produce an effect quantitatively similar to the local anti-correlation described in this work.

\section*{Acknowledgements}
We thank Sebasti\'an S\'anchez for providing some of the data used in Sect.~\ref{Sect:sebastian}, Enrique P\'erez-Montero for help with the use of HIIC, 
and Matteo~Miluzio for discussions and support during the early stages of the analysis. 
We also thank an anonymous referee whose comments help us improving the reliability of our argumentation. 
%
M.E.F. gratefully acknowledges the financial support of the {\em Funda\c c\~ao para a Ci\^encia e Tecnologia} (FCT -- Portugal), through the research grant SFRH/BPD/107801/2015.
This work has been partly funded by the Spanish Ministry of Economy and 
Competitiveness, projects {\em Estallidos} AYA2013--47742--C04--02--P and
AYA2016-79724-C4-2-P, as well as AYA2014-58861-C3-1-P.
%



\bibliographystyle{mnras}
\bibliography{metallicity,metallicity2} 

\appendix

\section{$M_{\it g2}/M_g$ when $M_{\it g2}$ varies from spaxel to spaxel}\label{appa}

\begin{figure}
\includegraphics[width=0.45\textwidth]{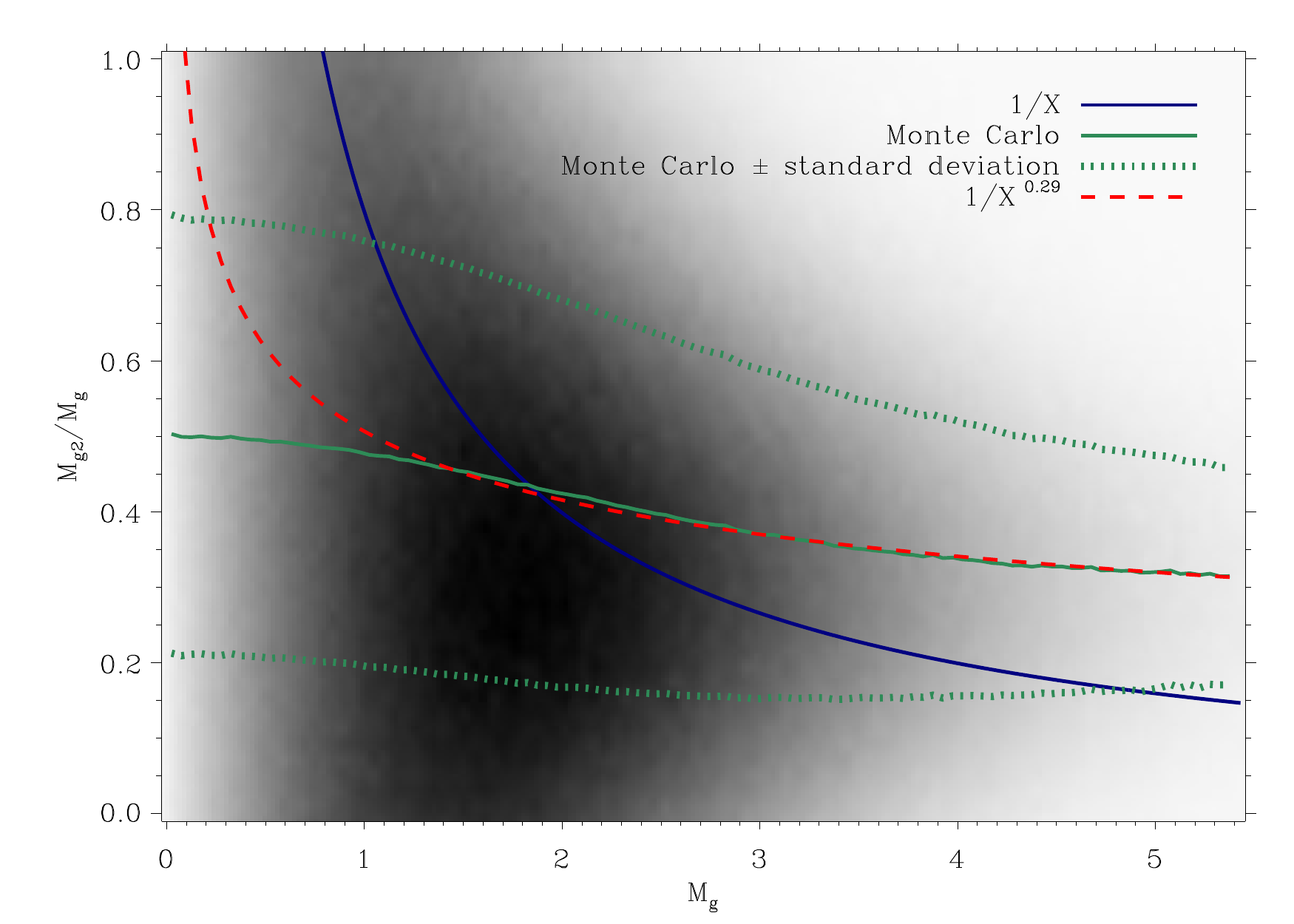}
\caption{$M_{g2}/M_g$ versus $M_g$ in a Monte Carlo simulation where both $M_{g1}$
and $M_{g2}$ are random variables extracted from the positive wing of a Gaussian PDF,
where the mean and the dispersion of $M_{g1}$ is 1.5 times larger than the mean and 
dispersion of $M_{g2}$. 
The mean $M_{g2}/M_g$ for each $M_g$ is shown as a solid green line,
with the 1-sigma dispersion around this mean represented as dotted green lines. The 
red dashed line shows a fit to this relation, which renders $\delta\simeq 0.3$. 
The curve corresponding to 
$\delta =1$ (i.e., $M_{g2}/M_g\propto 1/M_g$) is shown for reference as a solid blue 
line. Masses are given in arbitrary units.
}
\label{fig:monte_carlo}
\end{figure}

In Sect.~\ref{sect:mixing}, we explore the correlation between $Z$ and SFR 
resulting from the mixing of pre-existing gas with external metal-poor gas.
In the extreme case where the pre-existing gas-mass, $M_{g2}$, is the same 
for all the spaxels of the galaxy, then one expects a scaling of $Z$ 
as $1/M_g$, with $M_{g}$ the total gas mass in the spaxel. However, when $M_{g2}$
varies, the correlation of $Z$ with $M_g$ weakens since $M_g$ varies not
only with the mass of external metal-poor gas, $M_{g1}$, but also with 
$M_{g2}$ ($M_g=M_{g1}+M_{g2}$). This can be formulated in mathematical terms 
considering that $Z$ scales as $M_{g2}/M_g$ (Eqs.~[\ref{Eqn:formulaJorge}] and [\ref{eq:kslaw}]), 
so that when $M_{g2}$ varies,  then $M_{g2}/M_g$ varies as 
\begin{equation}
M_{g2}/M_g\propto M_g^{-\delta},
\label{eq:monte_carlo}
\end{equation}
with $\delta < 1$. The extreme case of constant $M_{g2}$ 
corresponds to $\delta=1$, whereas the case where the contribution of the external mass
is negligible (i.e., $M_{g1} \ll M_{g2}$) gives $\delta \sim 0$.
In order to explore the validity of the approximation in Eq.~(\ref{eq:monte_carlo}), 
we  carried out a series of Monte Carlo simulations that assume $M_{g1}$ and 
$M_{g2}$ to be two random variables, and then we study how $M_{g2}/M_g$
depends on $M_g$. Independently of the distribution function 
assumed for the variables, one finds a relation between the 
mean $M_{g2}/M_g$ for a given $M_g$ that can be parameterized as in
Eq.~(\ref{eq:monte_carlo}).
Figure~\ref{fig:monte_carlo} shows one example. In this particular case 
$M_{g1}$ and $M_{g2}$ are drawn from the positive wing of a Gaussian
probability distribution function (PDF), with the mean and the dispersion of $M_{g1}$ 1.5 times
larger than the mean and dispersion of $M_{g2}$. The resulting bidimensional PDF is displayed as 
an image in Fig.~\ref{fig:monte_carlo}. The mean $M_{g2}/M_g$ for each $M_g$ is shown as a solid 
green line, with the 1-sigma dispersion around this mean represented as dotted green lines. 
A fit to this relation is shown as the red dashed line, and it corresponds to
$\delta\simeq 0.3$. The curve representing the extreme case where $\delta =1$ 
(i.e., $M_{g2}/M_g\propto 1/M_g$) is shown as a solid blue line.

\bsp	
\label{lastpage}
\end{document}